\documentclass{emulateapj}
\usepackage{epsfig,natbib}
\usepackage{graphicx}
\usepackage{enumerate}
\usepackage{amsmath}
\usepackage{multirow}
\usepackage{bigdelim}
\usepackage{longtable}
\usepackage{threeparttable}
\usepackage{subfigure}
\usepackage{textcomp}
\usepackage{gensymb}
\usepackage{color}
\usepackage{morefloats}
\usepackage{xspace}
\usepackage[T1]{fontenc}

\newcommand{\ms}{\mbox{m\,s$^{-1}~$}}
\newcommand{\kms}{\mbox{km\,s$^{-1}~$}}

\newcommand{\msun}{$M_{\odot}$\xspace}

\newcommand{\rsun}{$R_{\odot}$\xspace}

\newcommand{\mearth}{$M_\earth$\xspace}

\newcommand{\rearth}{$R_\earth$\xspace}

\newcommand{\feh}{\ensuremath{[\mbox{Fe}/\mbox{H}]}\xspace}
\newcommand{\rphk}{\ensuremath{R'_{\mbox{\scriptsize HK}}}}
\newcommand{\lrphk}{\ensuremath{\log{\rphk}}\xspace}
\newcommand{\caii}{\ion{Ca}{2} H \& K\xspace}

\newcommand{\sini}{\ensuremath{\sin i}}

\newcommand{\vsinimin}{< 2}
\newcommand{\blank}{--}

\newcommand{\ktwoeoh}{K2-5\xspace}
\newcommand{\ktwoofe}{K2-3\xspace}
\newcommand{\ktwocib}{K2-8\xspace}
\newcommand{\ktwoceo}{K2-19\xspace}
\newcommand{\ktwohfo}{K2-35\xspace}
\newcommand{\ktwocdh}{K2-36\xspace}
\newcommand{\ktwocoe}{K2-16\xspace}
\newcommand{\ktwooih}{K2-24\xspace}
\newcommand{\ktwodcf}{K2-37\xspace}
\newcommand{\ktwobfc}{K2-38\xspace}
\newcommand{\ktwoihd}{K2-32\xspace}

\newcommand{\eoh}{201338508\xspace}
\newcommand{\ofe}{201367065\xspace}
\newcommand{\cib}{201445392\xspace}
\newcommand{\ceo}{201505350\xspace}
\newcommand{\hfo}{201549860\xspace}
\newcommand{\cdh}{201713348\xspace}
\newcommand{\coe}{201754305\xspace}
\newcommand{\oih}{203771098\xspace}
\newcommand{\dcf}{203826436\xspace}
\newcommand{\bfc}{204221263\xspace}
\newcommand{\ihd}{205071984\xspace}

\newcommand{\eohField}{C1}
\newcommand{\eohRA}{11:17:13}
\newcommand{\eohDec}{$-$\,01:52:41}
\newcommand{\eohPmRA}{\ensuremath{-10.3 \pm 4.4}}
\newcommand{\eohPmDec}{\ensuremath{+32.6 \pm 4.6}}

\newcommand{\eohTeff}{\ensuremath{3930 \pm 375}}
\newcommand{\eohlogg}{\ensuremath{ 4.71 \pm 0.21}}
\newcommand{\eohFeH}{\ensuremath{-0.33 \pm 0.19}}
\newcommand{\eohvsini}{\blank}
\newcommand{\eohRHKstar}{\blank}
\newcommand{\eohRstar}{\ensuremath{0.57 \pm 0.12}}
\newcommand{\eohMstar}{\ensuremath{0.61 \pm 0.13}}

\newcommand{\eohVmag}{\ensuremath{14.91 \pm 0.03}}
\newcommand{\eohKepmag}{\ensuremath{14.36}}

\newcommand{\eohJmag}{\ensuremath{12.45 \pm 0.03}}

\newcommand{\eohKmag}{\ensuremath{11.60 \pm 0.02}}

\newcommand{\ofeField}{C1}
\newcommand{\ofeRA}{11:29:20}
\newcommand{\ofeDec}{$-$\,01:27:17}
\newcommand{\ofePmRA}{\ensuremath{+88.3 \pm 2.0}}
\newcommand{\ofePmDec}{\ensuremath{-73.6 \pm 2.7}}

\newcommand{\ofeTeff}{\ensuremath{3896 \pm 189}}
\newcommand{\ofelogg}{\ensuremath{4.72 \pm 0.13}}
\newcommand{\ofeFeH}{\ensuremath{-0.32 \pm 0.13}}
\newcommand{\ofevsini}{\blank}
\newcommand{\ofeRHKstar}{\blank}
\newcommand{\ofeRstar}{\ensuremath{0.56 \pm 0.07}}
\newcommand{\ofeMstar}{\ensuremath{0.60 \pm 0.09}}

\newcommand{\ofeVmag}{\ensuremath{12.17 \pm 0.01}}
\newcommand{\ofeKepmag}{\ensuremath{11.57}}

\newcommand{\ofeJmag}{\ensuremath{\phn9.42 \pm 0.03}}

\newcommand{\ofeKmag}{\ensuremath{\phn8.56 \pm 0.02}}

\newcommand{\cibField}{C1}
\newcommand{\cibRA}{11:19:10}
\newcommand{\cibDec}{$-$\,00:17:04}
\newcommand{\cibPmRA}{\ensuremath{-34.7 \pm 4.9}}
\newcommand{\cibPmDec}{\ensuremath{-16.7 \pm 4.1}}

\newcommand{\cibTeff}{\ensuremath{4870 \pm 60}}
\newcommand{\ciblogg}{\ensuremath{4.52 \pm 0.10}}
\newcommand{\cibFeH}{\ensuremath{-0.02 \pm 0.04}}
\newcommand{\cibvsini}{\ensuremath{\vsinimin}}
\newcommand{\cibRHKstar}{\blank}
\newcommand{\cibSHKstar}{\ensuremath{0.33}}

\newcommand{\cibMstar}{\ensuremath{0.78 \pm 0.04}}
\newcommand{\cibRstar}{\ensuremath{0.74 \pm 0.04}}

\newcommand{\cibVmag}{\ensuremath{14.61 \pm 0.03}}
\newcommand{\cibKepmag}{\ensuremath{14.38}}

\newcommand{\cibJmag}{\ensuremath{12.83 \pm 0.03}}

\newcommand{\cibKmag}{\ensuremath{12.25 \pm 0.03}}

\newcommand{\ceoField}{C1}
\newcommand{\ceoRA}{11:39:50}
\newcommand{\ceoDec}{$+$\,00:36:13}
\newcommand{\ceoPmRA}{\ensuremath{-18.7 \pm 1.7}}
\newcommand{\ceoPmDec}{\phn\ensuremath{+4.5 \pm 2.0}}

\newcommand{\ceoTeff}{\ensuremath{5430 \pm 60}}
\newcommand{\ceologg}{\ensuremath{4.63 \pm 0.10}}
\newcommand{\ceoFeH}{\ensuremath{+0.10 \pm 0.04}}
\newcommand{\ceovsini}{\ensuremath{\vsinimin}}
\newcommand{\ceoRHKstar}{\ensuremath{-4.66}}

\newcommand{\ceoMstar}{\ensuremath{0.93 \pm 0.05}}
\newcommand{\ceoRstar}{\ensuremath{0.86 \pm 0.04}}

\newcommand{\ceoVmag}{\ensuremath{13.00 \pm 0.01}}
\newcommand{\ceoKepmag}{\ensuremath{12.81}}

\newcommand{\ceoJmag}{\ensuremath{11.60 \pm 0.02}}

\newcommand{\ceoKmag}{\ensuremath{11.16 \pm 0.03}}

\newcommand{\hfoField}{C1}
\newcommand{\hfoRA}{11:20:25}
\newcommand{\hfoDec}{$+$\,01:17:09}
\newcommand{\hfoPmRA}{\ensuremath{+10.4 \pm 6.4}}
\newcommand{\hfoPmDec}{\phn\ensuremath{-16.8 \pm 5.2}}

\newcommand{\hfoTeff}{\ensuremath{4680 \pm 60}}
\newcommand{\hfologg}{\ensuremath{4.56 \pm 0.10}}
\newcommand{\hfoFeH}{\ensuremath{+0.04 \pm 0.04}}
\newcommand{\hfovsini}{\ensuremath{3 \pm 1}}
\newcommand{\hfoRHKstar}{\blank}
\newcommand{\hfoSHKstar}{\ensuremath{0.33}}

\newcommand{\hfoMstar}{\ensuremath{0.76 \pm 0.04}}
\newcommand{\hfoRstar}{\ensuremath{0.72 \pm 0.04}}

\newcommand{\hfoVmag}{\ensuremath{14.35 \pm 0.06}}
\newcommand{\hfoKepmag}{\ensuremath{13.92}}

\newcommand{\hfoJmag}{\ensuremath{12.14 \pm 0.02}}

\newcommand{\hfoKmag}{\ensuremath{11.42 \pm 0.02}}

\newcommand{\cdhField}{C1}
\newcommand{\cdhRA}{11:17:48}
\newcommand{\cdhDec}{$+$\,03:51:59}
\newcommand{\cdhPmRA}{\ensuremath{-17.5 \pm 2.3}}
\newcommand{\cdhPmDec}{\ensuremath{+23.5 \pm 2.5}}

\newcommand{\cdhTeff}{\ensuremath{4924 \pm 60}}
\newcommand{\cdhlogg}{\ensuremath{4.65 \pm 0.10}}
\newcommand{\cdhFeH}{\ensuremath{-0.03 \pm 0.04}}
\newcommand{\cdhvsini}{\ensuremath{2 \pm 1}}
\newcommand{\cdhRHKstar}{\blank}
\newcommand{\cdhSHKstar}{\ensuremath{0.46}}

\newcommand{\cdhMstar}{\ensuremath{0.80 \pm 0.04}}
\newcommand{\cdhRstar}{\ensuremath{0.74 \pm 0.04}}

\newcommand{\cdhVmag}{\ensuremath{11.80 \pm 0.03}}
\newcommand{\cdhKepmag}{\ensuremath{11.53}}

\newcommand{\cdhJmag}{\ensuremath{10.03 \pm 0.02}}

\newcommand{\cdhKmag}{\ensuremath{\phn9.45 \pm 0.03}}

\newcommand{\coeField}{C1}
\newcommand{\coeRA}{11:40:23}
\newcommand{\coeDec}{$+$\,04:33:26}
\newcommand{\coePmRA}{\phn\ensuremath{-3.8 \pm 3.2}}
\newcommand{\coePmDec}{\ensuremath{+21.8 \pm 3.9}}

\newcommand{\coeTeff}{\ensuremath{4742 \pm 60}}
\newcommand{\coelogg}{\ensuremath{4.51 \pm 0.10}}
\newcommand{\coeFeH}{\ensuremath{-0.33 \pm 0.04}}
\newcommand{\coevsini}{\ensuremath{2 \pm 1}}
\newcommand{\coeRHKstar}{\blank}
\newcommand{\coeSHKstar}{\ensuremath{0.18}}

\newcommand{\coeMstar}{\ensuremath{0.68 \pm 0.03}}
\newcommand{\coeRstar}{\ensuremath{0.66 \pm 0.03}}

\newcommand{\coeVmag}{\ensuremath{14.67 \pm 0.04}}
\newcommand{\coeKepmag}{\ensuremath{14.30}}

\newcommand{\coeJmag}{\ensuremath{12.76 \pm 0.03}}

\newcommand{\coeKmag}{\ensuremath{12.09 \pm 0.02}}

\newcommand{\oihField}{C2}
\newcommand{\oihRA}{16:10:18}
\newcommand{\oihDec}{$-$\,24:59:25}
\newcommand{\oihPmRA}{\ensuremath{-60.6 \pm 2.5}}
\newcommand{\oihPmDec}{\ensuremath{-65.4 \pm 2.4}}

\newcommand{\oihTeff}{\ensuremath{5743 \pm 60}}
\newcommand{\oihlogg}{\ensuremath{4.29 \pm 0.08}}
\newcommand{\oihFeH}{\ensuremath{+0.42 \pm 0.04}}
\newcommand{\oihvsini}{\ensuremath{\vsinimin}}
\newcommand{\oihRHKstar}{\ensuremath{-5.15}}

\newcommand{\oihMstar}{\ensuremath{1.12 \pm 0.06}}
\newcommand{\oihRstar}{\ensuremath{1.21 \pm 0.12}}

\newcommand{\oihVmag}{\ensuremath{11.07 \pm 0.11}}
\newcommand{\oihKepmag}{\ensuremath{11.65}}

\newcommand{\oihJmag}{\ensuremath{\phn9.64 \pm 0.02}}

\newcommand{\oihKmag}{\ensuremath{\phn9.18 \pm 0.02}}

\newcommand{\dcfField}{C2}
\newcommand{\dcfRA}{16:13:48}
\newcommand{\dcfDec}{$-$\,24:47:13}
\newcommand{\dcfPmRA}{\phn\ensuremath{-9.4 \pm 1.9}}
\newcommand{\dcfPmDec}{\phn\ensuremath{+3.8 \pm 2.6}}

\newcommand{\dcfTeff}{\ensuremath{5413 \pm 60}}
\newcommand{\dcflogg}{\ensuremath{4.52 \pm 0.10}}
\newcommand{\dcfFeH}{\ensuremath{-0.03 \pm 0.04}}
\newcommand{\dcfvsini}{\ensuremath{\vsinimin}}
\newcommand{\dcfRHKstar}{\ensuremath{-4.85}}

\newcommand{\dcfMstar}{\ensuremath{0.90 \pm 0.05}}
\newcommand{\dcfRstar}{\ensuremath{0.85 \pm 0.04}}

\newcommand{\dcfVmag}{\ensuremath{12.52 \pm 0.06}}
\newcommand{\dcfKepmag}{\ensuremath{12.24}}

\newcommand{\dcfJmag}{\ensuremath{10.69 \pm 0.02}}

\newcommand{\dcfKmag}{\ensuremath{10.14 \pm 0.02}}

\newcommand{\bfcField}{C2}
\newcommand{\bfcRA}{16:00:08}
\newcommand{\bfcDec}{$-$\,23:11:21}
\newcommand{\bfcPmRA}{\ensuremath{-55.6 \pm 3.4}}
\newcommand{\bfcPmDec}{\ensuremath{-38.3 \pm 3.7}}

\newcommand{\bfcTeff}{\ensuremath{5757 \pm 60}}
\newcommand{\bfclogg}{\ensuremath{4.35 \pm 0.08}}
\newcommand{\bfcFeH}{\ensuremath{+0.28 \pm 0.04}}
\newcommand{\bfcvsini}{\ensuremath{\vsinimin}}
\newcommand{\bfcRHKstar}{\ensuremath{-5.13}}

\newcommand{\bfcMstar}{\ensuremath{1.07 \pm 0.05}}
\newcommand{\bfcRstar}{\ensuremath{1.10 \pm 0.09}}

\newcommand{\bfcVmag}{\ensuremath{11.39 \pm 0.03}}
\newcommand{\bfcKepmag}{\ensuremath{11.21}}

\newcommand{\bfcJmag}{\ensuremath{\phn9.91 \pm 0.02}}

\newcommand{\bfcKmag}{\ensuremath{\phn9.47 \pm 0.02}}

\newcommand{\ihdField}{C2}
\newcommand{\ihdRA}{16:49:42}
\newcommand{\ihdDec}{$-$\,19:32:34}
\newcommand{\ihdPmRA}{\ensuremath{-16.4 \pm 1.2}}
\newcommand{\ihdPmDec}{\ensuremath{-52.5 \pm 1.3}}

\newcommand{\ihdTeff}{\ensuremath{5315 \pm 60}}
\newcommand{\ihdlogg}{\ensuremath{4.43 \pm 0.10}}
\newcommand{\ihdFeH}{\ensuremath{+0.00 \pm 0.04}}
\newcommand{\ihdvsini}{\ensuremath{\vsinimin}}
\newcommand{\ihdRHKstar}{\ensuremath{-4.94}}

\newcommand{\ihdMstar}{\ensuremath{0.87 \pm 0.04}}
\newcommand{\ihdRstar}{\ensuremath{0.87 \pm 0.05}}

\newcommand{\ihdVmag}{\ensuremath{12.31 \pm 0.04}}
\newcommand{\ihdKepmag}{\ensuremath{12.01}}

\newcommand{\ihdJmag}{\ensuremath{10.40 \pm 0.02}}

\newcommand{\ihdKmag}{\ensuremath{\phn9.82 \pm 0.02}}

\newcommand{\eohbTt}{\ensuremath{808.8600 \pm 0.0048\xspace}}
\newcommand{\eohbPer}{\ensuremath{5.73594 \pm 0.00064\xspace}}
\newcommand{\eohbincdeg}{\ensuremath{87.35^{+1.88}_{-3.37}}\xspace}
\newcommand{\eohbRpRs}{\ensuremath{0.03031^{+0.00412}_{-0.00236}}\xspace}
\newcommand{\eohbRsa}{\ensuremath{0.0712^{+0.0487}_{-0.0159}}\xspace}
\newcommand{\eohbu}{\ensuremath{0.62 \pm 0.05\xspace}}
\newcommand{\eohbb}{\ensuremath{0.65^{+0.22}_{-0.41}}\xspace}
\newcommand{\eohbtdur}{\ensuremath{2.654^{+0.607}_{-0.268}}\xspace}
\newcommand{\eohbRp}{\ensuremath{1.91 \pm 0.44\xspace}}
\newcommand{\eohbrhostar}{\ensuremath{1.59^{+1.80}_{-1.25}}\xspace}
\newcommand{\eohba}{\ensuremath{0.0532 \pm 0.0038\xspace}}
\newcommand{\eohbSinc}{\ensuremath{24.1^{+18.4}_{-11.3}}\xspace}
\newcommand{\eohbTeq}{\ensuremath{565 \pm 84\xspace}}

\newcommand{\eohcTt}{\ensuremath{814.6010 \pm 0.0052\xspace}}
\newcommand{\eohcPer}{\ensuremath{10.93241 \pm 0.00134\xspace}}
\newcommand{\eohcincdeg}{\ensuremath{86.95 \pm 2.47\xspace}}
\newcommand{\eohcRpRs}{\ensuremath{0.03601^{+0.00805}_{-0.00475}}\xspace}
\newcommand{\eohcRsa}{\ensuremath{0.0639^{+0.0448}_{-0.0272}}\xspace}
\newcommand{\eohcu}{\ensuremath{0.62 \pm 0.05\xspace}}
\newcommand{\eohcb}{\ensuremath{0.83^{+0.09}_{-0.43}}\xspace}
\newcommand{\eohctdur}{\ensuremath{3.498^{+1.420}_{-0.580}}\xspace}
\newcommand{\eohcRp}{\ensuremath{2.26 \pm 0.62\xspace}}
\newcommand{\eohcrhostar}{\ensuremath{0.60^{+2.59}_{-0.48}}\xspace}
\newcommand{\eohca}{\ensuremath{0.0818 \pm 0.0059\xspace}}
\newcommand{\eohcSinc}{\ensuremath{10.2^{+7.8}_{-4.8}}\xspace}
\newcommand{\eohcTeq}{\ensuremath{456 \pm 68\xspace}}

\newcommand{\ofebTt}{\ensuremath{813.4173 \pm 0.0011\xspace}}
\newcommand{\ofebPer}{\ensuremath{10.05449 \pm 0.00026\xspace}}
\newcommand{\ofebincdeg}{\ensuremath{88.80^{+0.84}_{-1.12}}\xspace}
\newcommand{\ofebRpRs}{\ensuremath{0.03534^{+0.00286}_{-0.00153}}\xspace}
\newcommand{\ofebRsa}{\ensuremath{0.0391^{+0.0138}_{-0.0057}}\xspace}
\newcommand{\ofebu}{\ensuremath{0.60 \pm 0.05\xspace}}
\newcommand{\ofebb}{\ensuremath{0.54^{+0.23}_{-0.35}}\xspace}
\newcommand{\ofebtdur}{\ensuremath{2.726^{+0.252}_{-0.111}}\xspace}
\newcommand{\ofebRp}{\ensuremath{2.18 \pm 0.30\xspace}}
\newcommand{\ofebrhostar}{\ensuremath{3.12 \pm 1.87\xspace}}
\newcommand{\ofeba}{\ensuremath{0.0769 \pm 0.0039\xspace}}
\newcommand{\ofebSinc}{\ensuremath{10.9 \pm 3.7\xspace}}
\newcommand{\ofebTeq}{\ensuremath{463 \pm 39\xspace}}

\newcommand{\ofecTt}{\ensuremath{812.2812 \pm 0.0022\xspace}}
\newcommand{\ofecPer}{\ensuremath{24.64354 \pm 0.00117\xspace}}
\newcommand{\ofecincdeg}{\ensuremath{89.12^{+0.62}_{-0.86}}\xspace}
\newcommand{\ofecRpRs}{\ensuremath{0.03007^{+0.00304}_{-0.00203}}\xspace}
\newcommand{\ofecRsa}{\ensuremath{0.0237^{+0.0123}_{-0.0053}}\xspace}
\newcommand{\ofecu}{\ensuremath{0.60 \pm 0.05\xspace}}
\newcommand{\ofecb}{\ensuremath{0.65^{+0.20}_{-0.40}}\xspace}
\newcommand{\ofectdur}{\ensuremath{3.633^{+0.491}_{-0.191}}\xspace}
\newcommand{\ofecRp}{\ensuremath{1.85 \pm 0.27\xspace}}
\newcommand{\ofecrhostar}{\ensuremath{2.34^{+2.64}_{-1.67}}\xspace}
\newcommand{\ofeca}{\ensuremath{0.1399 \pm 0.0070\xspace}}
\newcommand{\ofecSinc}{\ensuremath{3.3 \pm 1.1\xspace}}
\newcommand{\ofecTeq}{\ensuremath{344 \pm 29\xspace}}

\newcommand{\ofedTt}{\ensuremath{826.2288 \pm 0.0034\xspace}}
\newcommand{\ofedPer}{\ensuremath{44.55983 \pm 0.00590\xspace}}
\newcommand{\ofedincdeg}{\ensuremath{89.38^{+0.43}_{-0.64}}\xspace}
\newcommand{\ofedRpRs}{\ensuremath{0.02453^{+0.00267}_{-0.00182}}\xspace}
\newcommand{\ofedRsa}{\ensuremath{0.0161^{+0.0093}_{-0.0038}}\xspace}
\newcommand{\ofedu}{\ensuremath{0.59 \pm 0.05\xspace}}
\newcommand{\ofedb}{\ensuremath{0.67^{+0.20}_{-0.40}}\xspace}
\newcommand{\ofedtdur}{\ensuremath{4.325^{+0.552}_{-0.256}}\xspace}
\newcommand{\ofedRp}{\ensuremath{1.51 \pm 0.23\xspace}}
\newcommand{\ofedrhostar}{\ensuremath{2.28^{+2.82}_{-1.70}}\xspace}
\newcommand{\ofeda}{\ensuremath{0.2076 \pm 0.0104\xspace}}
\newcommand{\ofedSinc}{\ensuremath{1.5 \pm 0.5\xspace}}
\newcommand{\ofedTeq}{\ensuremath{282 \pm 24\xspace}}

\newcommand{\cibbTt}{\ensuremath{813.6114 \pm 0.0030\xspace}}
\newcommand{\cibbPer}{\ensuremath{10.35239 \pm 0.00086\xspace}}
\newcommand{\cibbincdeg}{\ensuremath{86.42^{+2.56}_{-1.37}}\xspace}
\newcommand{\cibbRpRs}{\ensuremath{0.04457 \pm 0.00861\xspace}}
\newcommand{\cibbRsa}{\ensuremath{0.0710^{+0.0233}_{-0.0366}}\xspace}
\newcommand{\cibbu}{\ensuremath{0.70 \pm 0.05\xspace}}
\newcommand{\cibbb}{\ensuremath{0.88^{+0.04}_{-0.36}}\xspace}
\newcommand{\cibbtdur}{\ensuremath{3.362 \pm 0.885\xspace}}
\newcommand{\cibbRp}{\ensuremath{3.58 \pm 0.71\xspace}}
\newcommand{\cibbrhostar}{\ensuremath{0.49^{+3.85}_{-0.28}}\xspace}
\newcommand{\cibba}{\ensuremath{0.0856 \pm 0.0014\xspace}}
\newcommand{\cibbSinc}{\ensuremath{37.7 \pm 4.4\xspace}}
\newcommand{\cibbTeq}{\ensuremath{631 \pm 18\xspace}}

\newcommand{\cibcTt}{\ensuremath{813.0707 \pm 0.0033\xspace}}
\newcommand{\cibcPer}{\ensuremath{5.06416 \pm 0.00041\xspace}}
\newcommand{\cibcincdeg}{\ensuremath{86.70 \pm 2.48\xspace}}
\newcommand{\cibcRpRs}{\ensuremath{0.02990 \pm 0.00385\xspace}}
\newcommand{\cibcRsa}{\ensuremath{0.0743^{+0.0415}_{-0.0251}}\xspace}
\newcommand{\cibcu}{\ensuremath{0.70 \pm 0.05\xspace}}
\newcommand{\cibcb}{\ensuremath{0.78^{+0.12}_{-0.42}}\xspace}
\newcommand{\cibctdur}{\ensuremath{2.080^{+0.498}_{-0.253}}\xspace}
\newcommand{\cibcRp}{\ensuremath{2.41 \pm 0.33\xspace}}
\newcommand{\cibcrhostar}{\ensuremath{1.79^{+4.39}_{-1.32}}\xspace}
\newcommand{\cibca}{\ensuremath{0.0532 \pm 0.0009\xspace}}
\newcommand{\cibcSinc}{\ensuremath{97.8 \pm 11.4\xspace}}
\newcommand{\cibcTeq}{\ensuremath{801 \pm 23\xspace}}

\newcommand{\ceobTt}{\ensuremath{813.3837 \pm 0.0003\xspace}}
\newcommand{\ceobPer}{\ensuremath{7.91940 \pm 0.00005\xspace}}
\newcommand{\ceobincdeg}{\ensuremath{89.47 \pm 0.41\xspace}}
\newcommand{\ceobRpRs}{\ensuremath{0.07540^{+0.00060}_{-0.00043}}\xspace}
\newcommand{\ceobRsa}{\ensuremath{0.0540^{+0.0021}_{-0.0010}}\xspace}
\newcommand{\ceobu}{\ensuremath{0.48 \pm 0.03\xspace}}
\newcommand{\ceobb}{\ensuremath{0.17 \pm 0.12\xspace}}
\newcommand{\ceobtdur}{\ensuremath{3.502 \pm 0.063\xspace}}
\newcommand{\ceobRp}{\ensuremath{7.74 \pm 0.39\xspace}}
\newcommand{\ceobrhostar}{\ensuremath{1.91^{+0.12}_{-0.21}}\xspace}
\newcommand{\ceoba}{\ensuremath{0.0740 \pm 0.0012\xspace}}
\newcommand{\ceobSinc}{\ensuremath{125.9 \pm 14.4\xspace}}
\newcommand{\ceobTeq}{\ensuremath{854 \pm 24\xspace}}

\newcommand{\ceocTt}{\ensuremath{817.2755 \pm 0.0051\xspace}}
\newcommand{\ceocPer}{\ensuremath{11.90715 \pm 0.00150\xspace}}
\newcommand{\ceocincdeg}{\ensuremath{87.99^{+1.42}_{-1.99}}\xspace}
\newcommand{\ceocRpRs}{\ensuremath{0.04727^{+0.00568}_{-0.00352}}\xspace}
\newcommand{\ceocRsa}{\ensuremath{0.0553^{+0.0287}_{-0.0121}}\xspace}
\newcommand{\ceocu}{\ensuremath{0.64 \pm 0.05\xspace}}
\newcommand{\ceocb}{\ensuremath{0.63^{+0.20}_{-0.39}}\xspace}
\newcommand{\ceoctdur}{\ensuremath{4.371^{+0.939}_{-0.396}}\xspace}
\newcommand{\ceocRp}{\ensuremath{4.86^{+0.62}_{-0.44}}\xspace}
\newcommand{\ceocrhostar}{\ensuremath{0.79^{+0.87}_{-0.56}}\xspace}
\newcommand{\ceoca}{\ensuremath{0.0971 \pm 0.0016\xspace}}
\newcommand{\ceocSinc}{\ensuremath{73.1 \pm 8.4\xspace}}
\newcommand{\ceocTeq}{\ensuremath{745 \pm 21\xspace}}

\newcommand{\ceodTt}{\ensuremath{808.9207 \pm 0.0086\xspace}}
\newcommand{\ceodPer}{\ensuremath{2.50856 \pm 0.00041\xspace}}
\newcommand{\ceodincdeg}{\ensuremath{85.83^{+2.97}_{-4.74}}\xspace}
\newcommand{\ceodRpRs}{\ensuremath{0.01109 \pm 0.00116\xspace}}
\newcommand{\ceodRsa}{\ensuremath{0.1277^{+0.0586}_{-0.0254}}\xspace}
\newcommand{\ceodu}{\ensuremath{0.64 \pm 0.05\xspace}}
\newcommand{\ceodb}{\ensuremath{0.59^{+0.26}_{-0.38}}\xspace}
\newcommand{\ceodtdur}{\ensuremath{2.170 \pm 0.328\xspace}}
\newcommand{\ceodRp}{\ensuremath{1.14 \pm 0.13\xspace}}
\newcommand{\ceodrhostar}{\ensuremath{1.44^{+1.36}_{-0.97}}\xspace}
\newcommand{\ceoda}{\ensuremath{0.0344 \pm 0.0006\xspace}}
\newcommand{\ceodSinc}{\ensuremath{583.5 \pm 66.7\xspace}}
\newcommand{\ceodTeq}{\ensuremath{1252 \pm 36\xspace}}

\newcommand{\hfobTt}{\ensuremath{810.5871 \pm 0.0085\xspace}}
\newcommand{\hfobPer}{\ensuremath{2.39984 \pm 0.00039\xspace}}
\newcommand{\hfobincdeg}{\ensuremath{86.10^{+2.67}_{-4.43}}\xspace}
\newcommand{\hfobRpRs}{\ensuremath{0.01777^{+0.00234}_{-0.00166}}\xspace}
\newcommand{\hfobRsa}{\ensuremath{0.1237^{+0.0539}_{-0.0230}}\xspace}
\newcommand{\hfobu}{\ensuremath{0.72 \pm 0.05\xspace}}
\newcommand{\hfobb}{\ensuremath{0.56^{+0.26}_{-0.35}}\xspace}
\newcommand{\hfobtdur}{\ensuremath{2.064 \pm 0.308\xspace}}
\newcommand{\hfobRp}{\ensuremath{1.40 \pm 0.17\xspace}}
\newcommand{\hfobrhostar}{\ensuremath{1.73 \pm 1.31\xspace}}
\newcommand{\hfoba}{\ensuremath{0.0320 \pm 0.0005\xspace}}
\newcommand{\hfobSinc}{\ensuremath{217.4 \pm 25.5\xspace}}
\newcommand{\hfobTeq}{\ensuremath{979 \pm 29\xspace}}

\newcommand{\hfocTt}{\ensuremath{812.1158 \pm 0.0049\xspace}}
\newcommand{\hfocPer}{\ensuremath{5.60912 \pm 0.00071\xspace}}
\newcommand{\hfocincdeg}{\ensuremath{87.85^{+1.51}_{-2.25}}\xspace}
\newcommand{\hfocRpRs}{\ensuremath{0.02661^{+0.00407}_{-0.00266}}\xspace}
\newcommand{\hfocRsa}{\ensuremath{0.0575^{+0.0319}_{-0.0133}}\xspace}
\newcommand{\hfocu}{\ensuremath{0.72 \pm 0.05\xspace}}
\newcommand{\hfocb}{\ensuremath{0.66^{+0.20}_{-0.40}}\xspace}
\newcommand{\hfoctdur}{\ensuremath{2.050^{+0.343}_{-0.227}}\xspace}
\newcommand{\hfocRp}{\ensuremath{2.09^{+0.33}_{-0.24}}\xspace}
\newcommand{\hfocrhostar}{\ensuremath{3.16^{+3.79}_{-2.32}}\xspace}
\newcommand{\hfoca}{\ensuremath{0.0564 \pm 0.0009\xspace}}
\newcommand{\hfocSinc}{\ensuremath{70.1 \pm 8.2\xspace}}
\newcommand{\hfocTeq}{\ensuremath{737 \pm 22\xspace}}

\newcommand{\cdhbTt}{\ensuremath{809.4684 \pm 0.0017\xspace}}
\newcommand{\cdhbPer}{\ensuremath{1.42266 \pm 0.00005\xspace}}
\newcommand{\cdhbincdeg}{\ensuremath{87.75^{+1.62}_{-2.40}}\xspace}
\newcommand{\cdhbRpRs}{\ensuremath{0.01625^{+0.00093}_{-0.00060}}\xspace}
\newcommand{\cdhbRsa}{\ensuremath{0.1124^{+0.0201}_{-0.0091}}\xspace}
\newcommand{\cdhbu}{\ensuremath{0.69 \pm 0.05\xspace}}
\newcommand{\cdhbb}{\ensuremath{0.36 \pm 0.26\xspace}}
\newcommand{\cdhbtdur}{\ensuremath{1.206 \pm 0.078\xspace}}
\newcommand{\cdhbRp}{\ensuremath{1.32 \pm 0.09\xspace}}
\newcommand{\cdhbrhostar}{\ensuremath{6.57^{+1.89}_{-2.56}}\xspace}
\newcommand{\cdhba}{\ensuremath{0.0230 \pm 0.0004\xspace}}
\newcommand{\cdhbSinc}{\ensuremath{546.3 \pm 63.5\xspace}}
\newcommand{\cdhbTeq}{\ensuremath{1232 \pm 36\xspace}}

\newcommand{\cdhcTt}{\ensuremath{812.8422 \pm 0.0008\xspace}}
\newcommand{\cdhcPer}{\ensuremath{5.34059 \pm 0.00010\xspace}}
\newcommand{\cdhcincdeg}{\ensuremath{88.33^{+1.19}_{-1.63}}\xspace}
\newcommand{\cdhcRpRs}{\ensuremath{0.03468^{+0.00515}_{-0.00362}}\xspace}
\newcommand{\cdhcRsa}{\ensuremath{0.0405^{+0.0250}_{-0.0117}}\xspace}
\newcommand{\cdhcu}{\ensuremath{0.70 \pm 0.05\xspace}}
\newcommand{\cdhcb}{\ensuremath{0.72^{+0.16}_{-0.42}}\xspace}
\newcommand{\cdhctdur}{\ensuremath{1.267^{+0.301}_{-0.104}}\xspace}
\newcommand{\cdhcRp}{\ensuremath{2.80^{+0.43}_{-0.31}}\xspace}
\newcommand{\cdhcrhostar}{\ensuremath{9.96^{+17.74}_{-7.60}}\xspace}
\newcommand{\cdhca}{\ensuremath{0.0555 \pm 0.0009\xspace}}
\newcommand{\cdhcSinc}{\ensuremath{93.6 \pm 10.9\xspace}}
\newcommand{\cdhcTeq}{\ensuremath{793 \pm 23\xspace}}

\newcommand{\coebTt}{\ensuremath{811.6871 \pm 0.0038\xspace}}
\newcommand{\coebPer}{\ensuremath{7.61880 \pm 0.00087\xspace}}
\newcommand{\coebincdeg}{\ensuremath{87.97^{+1.47}_{-1.86}}\xspace}
\newcommand{\coebRpRs}{\ensuremath{0.02796^{+0.00363}_{-0.00259}}\xspace}
\newcommand{\coebRsa}{\ensuremath{0.0525^{+0.0268}_{-0.0127}}\xspace}
\newcommand{\coebu}{\ensuremath{0.71 \pm 0.05\xspace}}
\newcommand{\coebb}{\ensuremath{0.68^{+0.18}_{-0.43}}\xspace}
\newcommand{\coebtdur}{\ensuremath{2.487^{+0.382}_{-0.225}}\xspace}
\newcommand{\coebRp}{\ensuremath{2.02 \pm 0.24\xspace}}
\newcommand{\coebrhostar}{\ensuremath{2.25^{+2.90}_{-1.59}}\xspace}
\newcommand{\coeba}{\ensuremath{0.0667 \pm 0.0011\xspace}}
\newcommand{\coebSinc}{\ensuremath{44.5 \pm 5.2\xspace}}
\newcommand{\coebTeq}{\ensuremath{658 \pm 19\xspace}}

\newcommand{\coecTt}{\ensuremath{809.4800 \pm 0.0091\xspace}}
\newcommand{\coecPer}{\ensuremath{19.07863 \pm 0.00327\xspace}}
\newcommand{\coecincdeg}{\ensuremath{87.83 \pm 1.68\xspace}}
\newcommand{\coecRpRs}{\ensuremath{0.03526^{+0.01553}_{-0.00650}}\xspace}
\newcommand{\coecRsa}{\ensuremath{0.0439^{+0.0285}_{-0.0210}}\xspace}
\newcommand{\coecu}{\ensuremath{0.72 \pm 0.05\xspace}}
\newcommand{\coecb}{\ensuremath{0.86^{+0.07}_{-0.46}}\xspace}
\newcommand{\coectdur}{\ensuremath{3.859^{+1.640}_{-0.728}}\xspace}
\newcommand{\coecRp}{\ensuremath{2.54^{+1.12}_{-0.47}}\xspace}
\newcommand{\coecrhostar}{\ensuremath{0.61^{+3.71}_{-0.48}}\xspace}
\newcommand{\coeca}{\ensuremath{0.1229 \pm 0.0021\xspace}}
\newcommand{\coecSinc}{\ensuremath{13.1 \pm 1.5\xspace}}
\newcommand{\coecTeq}{\ensuremath{485 \pm 14\xspace}}

\newcommand{\oihbTt}{\ensuremath{905.7950 \pm 0.0007\xspace}}
\newcommand{\oihbPer}{\ensuremath{20.88508 \pm 0.00036\xspace}}
\newcommand{\oihbincdeg}{\ensuremath{88.95 \pm 0.62\xspace}}
\newcommand{\oihbRpRs}{\ensuremath{0.04409 \pm 0.00146\xspace}}
\newcommand{\oihbRsa}{\ensuremath{0.0388^{+0.0062}_{-0.0041}}\xspace}
\newcommand{\oihbu}{\ensuremath{0.56 \pm 0.03\xspace}}
\newcommand{\oihbb}{\ensuremath{0.47^{+0.16}_{-0.27}}\xspace}
\newcommand{\oihbtdur}{\ensuremath{5.881^{+0.269}_{-0.187}}\xspace}
\newcommand{\oihbRp}{\ensuremath{5.83 \pm 0.60\xspace}}
\newcommand{\oihbrhostar}{\ensuremath{0.74 \pm 0.28\xspace}}
\newcommand{\oihba}{\ensuremath{0.1542 \pm 0.0026\xspace}}
\newcommand{\oihbSinc}{\ensuremath{60.1 \pm 12.4\xspace}}
\newcommand{\oihbTeq}{\ensuremath{709 \pm 36\xspace}}

\newcommand{\oihcTt}{\ensuremath{915.6250 \pm 0.0005\xspace}}
\newcommand{\oihcPer}{\ensuremath{42.36342 \pm 0.00063\xspace}}
\newcommand{\oihcincdeg}{\ensuremath{89.43^{+0.26}_{-0.17}}\xspace}
\newcommand{\oihcRpRs}{\ensuremath{0.06147 \pm 0.00122\xspace}}
\newcommand{\oihcRsa}{\ensuremath{0.0224 \pm 0.0017\xspace}}
\newcommand{\oihcu}{\ensuremath{0.57 \pm 0.02\xspace}}
\newcommand{\oihcb}{\ensuremath{0.44^{+0.09}_{-0.18}}\xspace}
\newcommand{\oihctdur}{\ensuremath{7.058 \pm 0.179\xspace}}
\newcommand{\oihcRp}{\ensuremath{8.10 \pm 0.82\xspace}}
\newcommand{\oihcrhostar}{\ensuremath{0.94^{+0.27}_{-0.18}}\xspace}
\newcommand{\oihca}{\ensuremath{0.2471 \pm 0.0041\xspace}}
\newcommand{\oihcSinc}{\ensuremath{23.4 \pm 4.8\xspace}}
\newcommand{\oihcTeq}{\ensuremath{560 \pm 29\xspace}}

\newcommand{\dcfbTt}{\ensuremath{893.7013 \pm 0.0080\xspace}}
\newcommand{\dcfbPer}{\ensuremath{4.44117 \pm 0.00075\xspace}}
\newcommand{\dcfbincdeg}{\ensuremath{87.28^{+1.95}_{-3.37}}\xspace}
\newcommand{\dcfbRpRs}{\ensuremath{0.01728^{+0.00188}_{-0.00109}}\xspace}
\newcommand{\dcfbRsa}{\ensuremath{0.0878^{+0.0420}_{-0.0135}}\xspace}
\newcommand{\dcfbu}{\ensuremath{0.64 \pm 0.05\xspace}}
\newcommand{\dcfbb}{\ensuremath{0.55^{+0.27}_{-0.36}}\xspace}
\newcommand{\dcfbtdur}{\ensuremath{2.706^{+0.325}_{-0.222}}\xspace}
\newcommand{\dcfbRp}{\ensuremath{1.61 \pm 0.17\xspace}}
\newcommand{\dcfbrhostar}{\ensuremath{1.42 \pm 0.95\xspace}}
\newcommand{\dcfba}{\ensuremath{0.0511 \pm 0.0009\xspace}}
\newcommand{\dcfbSinc}{\ensuremath{213.3 \pm 27.8\xspace}}
\newcommand{\dcfbTeq}{\ensuremath{974 \pm 32\xspace}}

\newcommand{\dcfcTt}{\ensuremath{898.8603 \pm 0.0023\xspace}}
\newcommand{\dcfcPer}{\ensuremath{6.42904 \pm 0.00036\xspace}}
\newcommand{\dcfcincdeg}{\ensuremath{87.37^{+1.83}_{-2.50}}\xspace}
\newcommand{\dcfcRpRs}{\ensuremath{0.02955^{+0.00312}_{-0.00187}}\xspace}
\newcommand{\dcfcRsa}{\ensuremath{0.0739^{+0.0340}_{-0.0147}}\xspace}
\newcommand{\dcfcu}{\ensuremath{0.64 \pm 0.05\xspace}}
\newcommand{\dcfcb}{\ensuremath{0.62^{+0.21}_{-0.39}}\xspace}
\newcommand{\dcfctdur}{\ensuremath{3.127^{+0.457}_{-0.198}}\xspace}
\newcommand{\dcfcRp}{\ensuremath{2.75 \pm 0.27\xspace}}
\newcommand{\dcfcrhostar}{\ensuremath{1.13^{+1.07}_{-0.77}}\xspace}
\newcommand{\dcfca}{\ensuremath{0.0654 \pm 0.0011\xspace}}
\newcommand{\dcfcSinc}{\ensuremath{130.3 \pm 16.9\xspace}}
\newcommand{\dcfcTeq}{\ensuremath{861 \pm 28\xspace}}

\newcommand{\dcfdTt}{\ensuremath{907.2315 \pm 0.0031\xspace}}
\newcommand{\dcfdPer}{\ensuremath{14.09189 \pm 0.00135\xspace}}
\newcommand{\dcfdincdeg}{\ensuremath{88.34^{+1.18}_{-1.65}}\xspace}
\newcommand{\dcfdRpRs}{\ensuremath{0.02950 \pm 0.00351\xspace}}
\newcommand{\dcfdRsa}{\ensuremath{0.0382^{+0.0259}_{-0.0123}}\xspace}
\newcommand{\dcfdu}{\ensuremath{0.64 \pm 0.05\xspace}}
\newcommand{\dcfdb}{\ensuremath{0.76^{+0.14}_{-0.43}}\xspace}
\newcommand{\dcfdtdur}{\ensuremath{2.967^{+0.728}_{-0.261}}\xspace}
\newcommand{\dcfdRp}{\ensuremath{2.73 \pm 0.36\xspace}}
\newcommand{\dcfdrhostar}{\ensuremath{1.71^{+3.80}_{-1.34}}\xspace}
\newcommand{\dcfda}{\ensuremath{0.1103 \pm 0.0018\xspace}}
\newcommand{\dcfdSinc}{\ensuremath{45.7 \pm 6.0\xspace}}
\newcommand{\dcfdTeq}{\ensuremath{663 \pm 22\xspace}}

\newcommand{\bfcbTt}{\ensuremath{896.8786 \pm 0.0054\xspace}}
\newcommand{\bfcbPer}{\ensuremath{4.01593 \pm 0.00050\xspace}}
\newcommand{\bfcbincdeg}{\ensuremath{87.28^{+1.88}_{-3.08}}\xspace}
\newcommand{\bfcbRpRs}{\ensuremath{0.01281^{+0.00105}_{-0.00064}}\xspace}
\newcommand{\bfcbRsa}{\ensuremath{0.0993^{+0.0340}_{-0.0117}}\xspace}
\newcommand{\bfcbu}{\ensuremath{0.62 \pm 0.05\xspace}}
\newcommand{\bfcbb}{\ensuremath{0.48 \pm 0.30\xspace}}
\newcommand{\bfcbtdur}{\ensuremath{2.861 \pm 0.220\xspace}}
\newcommand{\bfcbRp}{\ensuremath{1.55 \pm 0.16\xspace}}
\newcommand{\bfcbrhostar}{\ensuremath{1.20^{+0.55}_{-0.70}}\xspace}
\newcommand{\bfcba}{\ensuremath{0.0506 \pm 0.0008\xspace}}
\newcommand{\bfcbSinc}{\ensuremath{465.9 \pm 80.1\xspace}}
\newcommand{\bfcbTeq}{\ensuremath{1184 \pm 51\xspace}}
\newcommand{\bfcbjitter}{\ensuremath{2.4^{+1.0}_{-0.7}}\xspace}
\newcommand{\bfcbgamma}{\ensuremath{-1.7 \pm 0.9\xspace}}
\newcommand{\bfcbdvdt}{\ensuremath{-37 \pm 11\xspace}}
\newcommand{\bfcbK}{\ensuremath{4.6 \pm 1.1\xspace}}
\newcommand{\bfcbMp}{\ensuremath{12.0 \pm 2.9\xspace}}
\newcommand{\bfcbrhop}{\ensuremath{17.5^{+8.5}_{-6.2}}\xspace}

\newcommand{\bfccTt}{\ensuremath{900.4752 \pm 0.0033\xspace}}
\newcommand{\bfccPer}{\ensuremath{10.56103 \pm 0.00090\xspace}}
\newcommand{\bfccincdeg}{\ensuremath{88.61^{+1.00}_{-1.67}}\xspace}
\newcommand{\bfccRpRs}{\ensuremath{0.02004^{+0.00236}_{-0.00135}}\xspace}
\newcommand{\bfccRsa}{\ensuremath{0.0381^{+0.0234}_{-0.0079}}\xspace}
\newcommand{\bfccu}{\ensuremath{0.61 \pm 0.05\xspace}}
\newcommand{\bfccb}{\ensuremath{0.64^{+0.23}_{-0.41}}\xspace}
\newcommand{\bfcctdur}{\ensuremath{2.533^{+0.312}_{-0.144}}\xspace}
\newcommand{\bfccRp}{\ensuremath{2.42 \pm 0.29\xspace}}
\newcommand{\bfccrhostar}{\ensuremath{3.06 \pm 2.71\xspace}}
\newcommand{\bfcca}{\ensuremath{0.0964 \pm 0.0016\xspace}}
\newcommand{\bfccSinc}{\ensuremath{128.3 \pm 22.1\xspace}}
\newcommand{\bfccTeq}{\ensuremath{858 \pm 37\xspace}}

\newcommand{\bfccK}{\ensuremath{2.8 \pm 1.3\xspace}}
\newcommand{\bfccMp}{\ensuremath{9.9 \pm 4.6\xspace}}
\newcommand{\bfccrhop}{\ensuremath{3.6^{+2.7}_{-1.9}}\xspace}

\newcommand{\ihdbTt}{\ensuremath{900.9258 \pm 0.0009\xspace}}
\newcommand{\ihdbPer}{\ensuremath{8.99218 \pm 0.00020\xspace}}
\newcommand{\ihdbincdeg}{\ensuremath{89.00^{+0.69}_{-0.90}}\xspace}
\newcommand{\ihdbRpRs}{\ensuremath{0.05635^{+0.00243}_{-0.00111}}\xspace}
\newcommand{\ihdbRsa}{\ensuremath{0.0526^{+0.0078}_{-0.0029}}\xspace}
\newcommand{\ihdbu}{\ensuremath{0.66 \pm 0.04\xspace}}
\newcommand{\ihdbb}{\ensuremath{0.33 \pm 0.22\xspace}}
\newcommand{\ihdbtdur}{\ensuremath{3.693^{+0.193}_{-0.105}}\xspace}
\newcommand{\ihdbRp}{\ensuremath{5.38 \pm 0.35\xspace}}
\newcommand{\ihdbrhostar}{\ensuremath{1.61^{+0.30}_{-0.54}}\xspace}
\newcommand{\ihdba}{\ensuremath{0.0808 \pm 0.0013\xspace}}
\newcommand{\ihdbSinc}{\ensuremath{82.9 \pm 10.6\xspace}}
\newcommand{\ihdbTeq}{\ensuremath{769 \pm 25\xspace}}

\newcommand{\ihdcTt}{\ensuremath{899.4306 \pm 0.0101\xspace}}
\newcommand{\ihdcPer}{\ensuremath{20.65614 \pm 0.00598\xspace}}
\newcommand{\ihdcincdeg}{\ensuremath{88.23^{+1.32}_{-2.68}}\xspace}
\newcommand{\ihdcRpRs}{\ensuremath{0.03636^{+0.01024}_{-0.00384}}\xspace}
\newcommand{\ihdcRsa}{\ensuremath{0.0420^{+0.0427}_{-0.0132}}\xspace}
\newcommand{\ihdcu}{\ensuremath{0.66 \pm 0.05\xspace}}
\newcommand{\ihdcb}{\ensuremath{0.74^{+0.18}_{-0.46}}\xspace}
\newcommand{\ihdctdur}{\ensuremath{5.024^{+2.307}_{-0.492}}\xspace}
\newcommand{\ihdcRp}{\ensuremath{3.48^{+0.97}_{-0.42}}\xspace}
\newcommand{\ihdcrhostar}{\ensuremath{0.60^{+1.24}_{-0.52}}\xspace}
\newcommand{\ihdca}{\ensuremath{0.1407 \pm 0.0024\xspace}}
\newcommand{\ihdcSinc}{\ensuremath{27.4 \pm 3.5\xspace}}
\newcommand{\ihdcTeq}{\ensuremath{583 \pm 19\xspace}}

\newcommand{\ihddTt}{\ensuremath{903.7846 \pm 0.0031\xspace}}
\newcommand{\ihddPer}{\ensuremath{31.71922 \pm 0.00236\xspace}}
\newcommand{\ihddincdeg}{\ensuremath{88.40^{+1.06}_{-0.65}}\xspace}
\newcommand{\ihddRpRs}{\ensuremath{0.04004^{+0.00279}_{-0.00474}}\xspace}
\newcommand{\ihddRsa}{\ensuremath{0.0355 \pm 0.0114\xspace}}
\newcommand{\ihddu}{\ensuremath{0.66 \pm 0.05\xspace}}
\newcommand{\ihddb}{\ensuremath{0.79^{+0.07}_{-0.37}}\xspace}
\newcommand{\ihddtdur}{\ensuremath{5.990 \pm 0.729\xspace}}
\newcommand{\ihddRp}{\ensuremath{3.75 \pm 0.40\xspace}}
\newcommand{\ihddrhostar}{\ensuremath{0.42^{+1.15}_{-0.22}}\xspace}
\newcommand{\ihdda}{\ensuremath{0.1873 \pm 0.0031\xspace}}
\newcommand{\ihddSinc}{\ensuremath{15.4 \pm 2.0\xspace}}
\newcommand{\ihddTeq}{\ensuremath{505 \pm 16\xspace}}

\newcommand{\montet}{M15}
\newcommand{\crossfield}{C15}
\newcommand{\petigura}{P15}
\newcommand{\armstrong}{A15}
\newcommand{\barros}{B15}

\newcommand{\monteteohbRp}{\ensuremath{1.92 {\pm} 0.23}}
\newcommand{\monteteohcRp}{\ensuremath{1.92 {\pm} 0.20}}
\newcommand{\montetofebRp}{\ensuremath{1.98 {\pm} 0.10}}
\newcommand{\montetofecRp}{\ensuremath{1.56 {\pm} 0.10}}
\newcommand{\crossfieldofebRp}{\ensuremath{2.14 {\pm} 0.27}}
\newcommand{\crossfieldofecRp}{\ensuremath{1.72 {\pm} 0.23}}
\newcommand{\crossfieldofedRp}{\ensuremath{1.52 {\pm} 0.21}}
\newcommand{\montetcibcRp}{\ensuremath{2.31 {\pm} 0.33}}
\newcommand{\montetcibbRp}{\ensuremath{2.97 {\pm} 0.51}}
\newcommand{\montetceobRp}{\ensuremath{7.11 {\pm} 0.81}}
\newcommand{\montetceocRp}{\ensuremath{4.31 {\pm} 0.49}}
\newcommand{\armstrongceobRp}{\ensuremath{7.23 {\pm} 0.54}}
\newcommand{\armstrongceocRp}{\ensuremath{4.21 {\pm} 0.31}}
\newcommand{\barrosceobRp}{\ensuremath{7.46 {\pm} 0.76}}
\newcommand{\barrosceocRp}{\ensuremath{4.51 {\pm} 0.47}}
\newcommand{\montethfocRp}{\ensuremath{2.20 {\pm} 0.40}}
\newcommand{\montetcoebRp}{\ensuremath{2.13 {\pm} 0.37}}
\newcommand{\montetcoecRp}{\ensuremath{2.14 {\pm} 0.41}}
\newcommand{\petiguraoihbRp}{\ensuremath{5.68 {\pm} 0.41}}
\newcommand{\petiguraoihcRp}{\ensuremath{7.82 {\pm} 0.72}}

\newcommand{\monteteohRstar}{\ensuremath{0.52 {\pm} 0.01}}
\newcommand{\montetofeRstar}{\ensuremath{0.52 {\pm} 0.02}}
\newcommand{\crossfieldofeRstar}{\ensuremath{0.56 {\pm} 0.07}}
\newcommand{\montetcibRstar}{\ensuremath{0.74^{+0.02}_{-0.03}}}
\newcommand{\montetceoRstar}{\ensuremath{0.81^{+0.09}_{-0.05}}}
\newcommand{\armstrongceoRstar}{\ensuremath{1.03 {\pm} 0.2}}
\newcommand{\barrosceoRstar}{\ensuremath{0.91 {\pm} 0.09}}
\newcommand{\montethfoRstar}{\ensuremath{0.69 {\pm} 0.02}}
\newcommand{\montetcoeRstar}{\ensuremath{0.64 {\pm} 0.03}}
\newcommand{\petiguraoihRstar}{\ensuremath{1.21 {\pm} 0.11}}

\newcommand{\PreviouslyValidated}{Valid.}
\newcommand{\NewlyValidated}{Valid.}
\newcommand{\PreviouslyConfirmed}{Conf.}
\newcommand{\NewlyConfirmed}{Conf.}

\newcommand{\eohbDisposition}{\PreviouslyValidated}
\newcommand{\eohcDisposition}{\PreviouslyValidated}
\newcommand{\ofebDisposition}{\PreviouslyConfirmed}
\newcommand{\ofecDisposition}{\PreviouslyValidated}
\newcommand{\ofedDisposition}{\PreviouslyValidated}
\newcommand{\cibbDisposition}{\PreviouslyValidated}
\newcommand{\cibcDisposition}{\NewlyValidated}
\newcommand{\ceobDisposition}{\PreviouslyConfirmed}
\newcommand{\ceocDisposition}{\PreviouslyConfirmed}
\newcommand{\ceodDisposition}{\NewlyValidated}
\newcommand{\hfobDisposition}{\NewlyValidated}
\newcommand{\hfocDisposition}{\NewlyValidated}
\newcommand{\cdhbDisposition}{\NewlyValidated}
\newcommand{\cdhcDisposition}{\NewlyValidated}
\newcommand{\coebDisposition}{\PreviouslyValidated}
\newcommand{\coecDisposition}{\PreviouslyValidated}
\newcommand{\oihbDisposition}{\PreviouslyConfirmed}
\newcommand{\oihcDisposition}{\PreviouslyConfirmed}
\newcommand{\dcfbDisposition}{\NewlyValidated}
\newcommand{\dcfcDisposition}{\NewlyValidated}
\newcommand{\dcfdDisposition}{\NewlyValidated}
\newcommand{\bfcbDisposition}{\NewlyConfirmed}
\newcommand{\bfccDisposition}{\NewlyConfirmed}
\newcommand{\ihdbDisposition}{\PreviouslyConfirmed}
\newcommand{\ihdcDisposition}{\NewlyValidated}
\newcommand{\ihddDisposition}{\NewlyValidated}

\newcommand{\eohbFPP}{$<$ 0.001} 
\newcommand{\ofebFPP}{$<$ 0.001} 
\newcommand{\ofecFPP}{$<$ 0.001} 
\newcommand{\ofedFPP}{$<$ 0.001} 
\newcommand{\cibbFPP}{0.013} 
\newcommand{\cibcFPP}{0.008} 
\newcommand{\ceobFPP}{$<$ 0.001} 
\newcommand{\ceocFPP}{$<$ 0.001} 
\newcommand{\ceodFPP}{$<$ 0.001} 
\newcommand{\hfobFPP}{$<$ 0.001} 
\newcommand{\hfocFPP}{$<$ 0.001} 
\newcommand{\cdhbFPP}{$<$ 0.001} 
\newcommand{\cdhcFPP}{$<$ 0.001} 
\newcommand{\coebFPP}{$<$ 0.001} 
\newcommand{\coecFPP}{0.002} 
\newcommand{\oihbFPP}{$<$ 0.001} 
\newcommand{\oihcFPP}{$<$ 0.001} 
\newcommand{\dcfbFPP}{0.009} 
\newcommand{\dcfcFPP}{$<$ 0.001} 
\newcommand{\dcfdFPP}{$<$ 0.001} 
\newcommand{\bfcbFPP}{$<$ 0.001} 
\newcommand{\bfccFPP}{$<$ 0.001} 
\newcommand{\ihdbFPP}{$<$ 0.001} 
\newcommand{\ihdcFPP}{0.022} 
\newcommand{\ihddFPP}{$<$ 0.001} 

\makeatletter

\newcommand{\Rmnum}[1]{\expandafter\@slowromancap\romannumeral #1@}

\newcommand{\MPS}{multi-planet systems\xspace}
\newcommand{\mytilde}{\raise.17ex\hbox{$\scriptstyle\mathtt{\sim}$}}
\newcommand{\specmatch}{\texttt{SpecMatch}\xspace}

\newcommand{\TERRA}{\texttt{TERRA}\xspace}
\newcommand{\VESPA}{\texttt{VESPA}\xspace}
\newcommand{\shk}{$S_\mathrm{HK}$\xspace}

\newcommand{\Teff}{$T_{\mathrm{eff}}$\xspace}  
\newcommand{\logg}{\ensuremath{\log g}\xspace} 
\newcommand{\vsini}{\ensuremath{v \sin i}\xspace} 
\newcommand{\gcc}{g\,cm$^{-3}$\xspace} 
\newcommand{\dvdt}{\ensuremath{dv/dt}\xspace} 

\newcommand{\Rp}{\ensuremath{R_P}\xspace}
\newcommand{\Teq}{$T_{\mathrm{eq}}$\xspace}

\newcommand{\Me}{\ensuremath{M_{\oplus}}\xspace} 
\newcommand{\Kepler}{\textit{Kepler}\xspace} 
\newcommand{\ktwo}{\textit{K2}\xspace}

\renewcommand{\Re}{\ensuremath{R_{\oplus}}\xspace}

\makeatother

\shortauthors{Sinukoff}
\shorttitle{Multi-planet Systems from \ktwo Campaigns Campaigns 1 \& 2}
\submitted{Accepted to ApJ}  
\begin{document}
\pagenumbering{arabic}


\title{Eleven Multi-planet Systems from \ktwo Campaigns 1 \& 2 \\ 
       and the Masses of Two Hot Super-Earths}
\author{
Evan Sinukoff\altaffilmark{1,11},
Andrew W.\ Howard\altaffilmark{1},
Erik A.\ Petigura\altaffilmark{2,12},
Joshua E.\ Schlieder\altaffilmark{3,13},
Ian J.\ M.\ Crossfield\altaffilmark{4,14},
David R.\ Ciardi\altaffilmark{5},
Benjamin J.\ Fulton\altaffilmark{1,15},
Howard Isaacson\altaffilmark{6},
Kimberly M.\ Aller\altaffilmark{1},
Christoph Baranec\altaffilmark{1}, 
Charles A.\ Beichman\altaffilmark{5},
Brad M.\ S.\ Hansen\altaffilmark{8},
Heather A.\ Knutson\altaffilmark{2},
Nicholas M.\ Law\altaffilmark{9},
Michael C.\ Liu\altaffilmark{1},
Reed Riddle\altaffilmark{10},
Courtney D.\ Dressing\altaffilmark{2,14}
}


\altaffiltext{1}{Institute for Astronomy, University of Hawai`i at M\={a}noa, Honolulu, HI 96822, USA} 
\altaffiltext{2}{California Institute of Technology, Pasadena, CA, USA}
\altaffiltext{3}{NASA Ames Research Center, Moffett Field, CA, USA}
\altaffiltext{4}{Lunar \& Planetary Laboratory, University of Arizona, 1629 E. University Blvd., Tucson, AZ, USA}
\altaffiltext{5}{NASA Exoplanet Science Institute, California Institute of Technology, 770 S. Wilson Ave., Pasadena, CA, USA}
\altaffiltext{6}{Astronomy Department, University of California, Berkeley, CA, USA}
\altaffiltext{7}{Institute for Astronomy, University of Hawai`i, Hilo, HI 96720, USA}
\altaffiltext{8}{Department of Physics \& Astronomy and Institute of Geophysics \& Planetary Physics, University of California Los Angeles, Los Angeles, CA 90095, USA}
\altaffiltext{9}{Department of Physics and Astronomy, University of North Carolina at Chapel Hill, Chapel Hill, NC 27599, USA}
\altaffiltext{10}{Division of Physics, Mathematics, and Astronomy, California Institute of Technology, Pasadena, CA 91125, USA}

\altaffiltext{11}{NSERC Postgraduate Research Fellow}
\altaffiltext{12}{Hubble Fellow}
\altaffiltext{13}{NASA Postdoctoral Program Fellow}
\altaffiltext{14}{NASA Sagan Fellow}
\altaffiltext{15}{NSF Graduate Research Fellow}

\begin{abstract}
We present a catalog of 11 multi-planet systems from Campaigns 1 and 2 of the \ktwo mission. We report the sizes and orbits of 26 planets split between seven 2-planet systems and four 3-planet systems.  These planets stem from a systematic search of the \ktwo photometry for all dwarf stars observed by \ktwo in these fields.  We precisely characterized the host stars with adaptive optics imaging and analysis of high-resolution optical spectra from Keck/HIRES and medium-resolution spectra from IRTF/SpeX.  We confirm two planet candidates by mass detection and validate the remaining 24 candidates to $>$99\% confidence.  Thirteen planets were previously validated or confirmed by other studies and 24 were previously identified as planet candidates. The planets are mostly smaller than Neptune (21/26 planets) as in the \Kepler mission and all have short periods ($P < 50$ d) due to the duration of the \ktwo photometry.  The host stars are relatively bright (most have $Kp < 12.5$ mag) and are amenable to follow-up characterization.  For \ktwobfc, we measured precise radial velocities using Keck/HIRES and provide initial estimates of the planet masses.  \ktwobfc{b} is a short-period super-Earth with a radius of \bfcbRp~\Re, a mass of \bfcbMp~\Me, and a high density consistent with an iron-rich composition. The outer planet \ktwobfc{c} is a lower density sub-Neptune-size planet with a radius of \bfccRp~\Re and a mass of \bfccMp~\Me that likely has a substantial envelope. This new planet sample demonstrates the capability of \ktwo to discover numerous planetary systems around bright stars.
\end{abstract}



\section{Introduction}
\label{sec:intro}




The prime \Kepler mission (2009--2013) demonstrated that compact, multi-planet systems are common \citep{Howard12,Fressin13,Petigura13}. Of the 4,000$+$ planet candidates from \Kepler, roughly 1500 are in systems with multiple candidates \citep{Mullally15}. Some exceptional systems include the high-multiplicity Kepler-11 \citep{Lissauer11} and Kepler-90 \citep{Schmitt14} systems that host six and seven planets, respectively, all within 1 AU. Another noteworthy system is Kepler-36 which hosts two planets with semi-major axes differing by 10\% but densities differing by a factor of eight \citep{Carter12}.

The ensemble properties of \Kepler multi-planet systems (``multis'') have provided key insights into the formation, evolution, and architectures of planetary systems \citep{Lissauer11b,Lissauer12,Lissauer14,Rowe14}. 
Most of the \Kepler multis have low ($\lesssim 3$\%) mutual inclinations \citep{Fang12}. Many multi-planet systems are dynamically packed, i.e.\ adding an additional planet on an intermediate orbit leads to dynamical instability \citep{Fang13}. While the distribution of orbital period ratios of \Kepler multis is roughly uniform, \cite{Fabrycky14} observed an excess of planet pairs with orbital period ratios exterior to first order mean motion resonance (MMR) and a deficit of planets lying interior to resonance.  This feature may be the outcome of eccentricity damping of resonant planet pairs by the protoplanetary disk \citep{Lithwick12,Batygin13}.


Systems with multiple transiting planets are particularly valuable because they are a clean sample with nearly zero false positive detections \citep{Lissauer12,Lissauer14}.  This is due to the low probability of having multiple stars with a false positive signals in the same photometric aperture, i.\ e.\ eclipsing binaries are distributed sparsely on the sky.

Given the photometric precision and four year baseline of the prime \Kepler mission, dynamical interactions between pairs of planets are often detected as transit timing variations (TTVs), which can constrain planet properties such as mass and eccentricity \citep{Holman05,Agol05}. The analysis of the Kepler-36 system by \cite{Carter12} demonstrated the power of TTV observations. They measured a mass of $4.45^{+0.33}_{-0.27}$~\Me for Kepler-36b, a planet with \Rp~=~$1.486\pm0.035$~\Re. Currently, Kepler-36b has the best-constrained mass of any exoplanet smaller than 2~\Re \citep{Dressing15}.

The prime \Kepler mission came to an end in 2013, following the failure of a second reaction wheel. Beginning in March, 2014, NASA began operating the telescope in a new mode called \ktwo \citep{Howell14}. During \ktwo operations, the spacecraft observes a different region of the ecliptic plane every $\sim85$~d. 

\Kepler planet catalogs \citep{Borucki11, Batalha13, Burke14, Rowe15, Mullally15}  spawned numerous statistical studies on planet occurrence, the distribution of planet sizes, and the diversity of system architectures. These studies deepened our understanding of planet formation and evolution. Continuing in this pursuit, \ktwo planet catalogs will provide a wealth of planets around bright stars that are particularly favorable for studying planet compositions---perhaps the best link to their formation histories. 

The first four \ktwo campaigns (C0--C3) plus an additional engineering test campaign have yielded over 230 planet candidates at the time of writing \footnote{NASA Exoplanet Archive, UT 28 April 2016, http://exoplanetarchive.ipac.caltech.edu.}. Moreover, $\sim 40$ of these planet candidates have been either statistically validated as planets at better than 99\% confidence or confirmed via radial velocity (RV) or TTV detection, including several noteworthy discoveries. Super-Earth HIP\,116454b, discovered in the \ktwo engineering test field, orbits a bright K dwarf. Its mass is well-constrained from follow-up RV measurements \citep{Vanderburg15}. From C1, \citet{Crossfield15} announced three super-Earths orbiting a nearby M0 dwarf, \ktwoofe.  \citet{Almenara15} and \citet{Dai16} detected the RV signature of the inner planet, which is consistent with a mostly rocky composition, although the water fraction could be as large as 60\%.   \citet{Foreman-Mackey15} discovered two planets of Neptune- and Saturn-size near a 3:2 MMR around \ktwoceo.  \citet{Armstrong15} used TTVs to constrain the mass of the larger outer planet and the masses of both planets were measured by \citet{Dai16} using RVs.  \citet{Vanderburg15b} reported a third Earth-size planet candidate, \ktwoceo{d}, at $P$=2.5 d.  The first \ktwo planet catalogs have already been assembled; \citet{Foreman-Mackey15} reported 36 planet candidates, 21 of which were validated at $>99\%$ confidence by \citet{Montet15}.  These include four validated multi-planet systems (\ktwoofe, \ktwoeoh, \ktwocoe, \ktwoceo) and one system, \ktwocib, with one validated planet and a second planet candidate.  More recently, \citet{Vanderburg15b} presented 234 planet candidates in C0--C3, including 20 systems with multiple planet candidates.

In this paper, we present 11 multi-planet systems with a total of 26 planets detected by our team in \ktwo photometry from Campaigns 1 and 2 (C1 and C2). We detected no multi-planet systems in Campaign 0. We adopt a ``confirmed'' disposition for planet candidates with detected RV or TTV signatures and, following \citet{Montet15}, a ``validated'' disposition for planet candidates found to have a false positive probability, FPP $<1\%$.  Under this definition, 13 of the 26 planet candidates are previously confirmed or validated, 11 are newly validated, and two are newly discovered and confirmed.  Most importantly, 9 of the 13 newly validated or confirmed planet candidates orbit stars $V\leq12.5$\,mag, amenable to RV follow-up. For one system, \ktwobfc, we obtained radial velocity measurements using Keck/HIRES to constrain planet masses. The remainder of this paper is organized as follows:  In Section~\ref{sec:obs} we describe the photometric detection of multis in \ktwo photometry along with follow-up observations that confirm the planets and characterize the their stellar hosts. Section~\ref{sec:stellar-properties} details the physical properties of stellar hosts. Section \ref{sec:validation} outlines our validation of each planet via AO images, archival images, as well as vetting of the light curves and spectra. In Section~\ref{sec:planet-properties}, we describe our light curve modeling and present derived planet properties. We present our RV measurements of \ktwobfc and the derived planet masses in Section~\ref{sec:RV}. In Section~\ref{sec:individual-systems}, we summarize the most noteworthy characteristics of each system, including additional findings of other studies where relevant. We discuss the likely compositions of the \ktwobfc planets, summarize the ensemble properties of our planet sample, and compare our results to other studies in Section~\ref{sec:discussion}. Our RVs, spectra, AO images and contrast curves will be uploaded to the ExoFOP-K2 website\footnote{https://cfop.ipac.caltech.edu/k2/}.  We note that throughout this paper, systems are ordered by EPIC number.

\section{Detection and Observations of \ktwo Multis}
\label{sec:obs}

\subsection{\ktwo Planet Search Program}
\label{sec:k2_program}

%

During the prime \Kepler\ mission, the project office selected nearly all of the stars that were observed.  This target list was dominated by a magnitude-limited set of F, G, K, and M dwarfs ($Kp < 16$) from which the major planet catalogs and occurrence analyses were derived. \ktwo is entirely community-driven with all targets selected from  Guest Observer proposals.  Our team has proposed large samples of G, K, and M dwarfs for every \ktwo Campaign (to date, Campaigns 0--10).  For the G and K dwarfs, our proposed sets of stars are magnitude-limited at $Kp < 13$ or 14 (depending on crowding) and $\sim$3,500--10,000 stars per Campaign have been selected for \ktwo observations.  During each \ktwo Campaign, the \Kepler\ telescope observes the selected stars nearly continuously for $\sim$75 d. 

This catalog of multi-planet systems is based on photometry collected by \ktwo during Campaign 1 (2014 May 30--2014 Aug 21) and Campaign 2 (2014 Aug 23--2014 Nov 13).  
The stars were part of \ktwo Guest Observer proposals led by I.\ Crossfield, R.\ Sanchis-Ojeda, A.\ Scholz, A.\ Sozzetti, P.\ Robertson, D.\ Stello, V.\ Sanchez Bejar, N.\ Deacon, B.-O.\ Demory (Campaign 1) and E.\ Petigura, R.\ Sanchis-Ojeda, and D.\ Stello (Campaign 2).
We searched for transiting planets in the photometry of all stars observed by \ktwo, not just those in the above programs.  

\subsection{\ktwo Photometry \& Transit Detection}
\label{sec:photometry}
During \ktwo observations, solar radiation pressure exerts a torque on the spacecraft that causes the telescope to roll around its boresight. Consequently, stars trace out small arcs of $\sim$1~pixel every $\sim$6~hr. As the stars sample different pixel phases, inter-pixel sensitivity variations cause their apparent brightnesses to change. Disentangling stellar variability from spacecraft systematics is non-trivial when working with \ktwo data.

We extracted the photometry from the \ktwo target pixel files, which are available at the Mikulski Archive for Space Telescopes (MAST)%
\footnote{https://archive.stsci.edu/k2/}. Our photometric extraction procedure is detailed in \cite{Crossfield15}. In brief, for a given target star we compute raw aperture photometry using a soft-edged circular aperture. For every frame in a \ktwo campaign, we solve for the roll angle between the target and an arbitrary reference frame using several hundred stars. We model the time- and roll-dependent variations using a Gaussian process, which are then subtracted from the raw photometry to produce calibrated photometry.  This process is repeated for different aperture sizes, and we adopt the aperture size which minimizes photometric noise on three-hour timescales.  Specifically, we use the median absolute deviation (MAD) of the three-hour Single Event Statistic (SES) as our noise metric.  We define the SES as the depth of a box-shaped dimming relative to the local photometric level.  Conceptually, this is similar to the three-hour Combined Differential Photometric Precision (CDPP-3) metric used by the {\em Kepler} project.  We compute a three-hour SES at every long cadence measurement as part of our transit search \footnote{As an example, to compute the SES on 1 hour timescales (corresponding to 2 Kepler long cadence measurements), we construct the following kernel, $g = \frac{1}{2}[0.5, 0.5, -1, -1, 0.5, 0.5]$, which is convolved with the measured photometry. See \citet{Petigura12} for further details.}. This method of aperture selection favors small apertures for faint stars (where background noise dominates) and large apertures for bright targets.

To search the calibrated photometry for planetary transits, we use the \TERRA algorithm \citep{Petigura13}. We have adapted \TERRA to search for multi-planet systems. When \TERRA identifies a candidate transit, it flags that star for additional analysis. \TERRA masks out the transit of the first candidate along with a buffer of $\Delta T$ on either side, where $\Delta T$ is the transit duration. \TERRA then repeats the transit search in the masked photometry. This process continues until no transits with signal-to-noise ratio (SNR) $>$ 8 are detected or when the number of iterations exceeds 5.

Table \ref{tb:hostpars} lists coordinates, proper motions, and multi-band photometry for the 11 stars around which we detect multiple transiting planets.

\begin{deluxetable*}{llccccccccc}
\tabletypesize{\footnotesize}
\tablecaption{\ktwo multi-planet host stars}
\tablehead{\ktwo Name & EPIC No. & Field & RA & Dec. & $\mu_{\mathrm{RA}}$\tablenotemark{a} & $\mu_{\mathrm{dec}}$\tablenotemark{a} & $V$\tablenotemark{b} & $Kp$\tablenotemark{a} & $J$\tablenotemark{a} & $Ks$\tablenotemark{a} \\ & & & (J2000) & (J2000) & (mas\,yr$^{-1}$) & (mas\,yr$^{-1}$) & (mag) & (mag) & (mag) & (mag)}
 \startdata
\ktwoeoh & \eoh & \eohField & \eohRA & \eohDec & \eohPmRA & \eohPmDec & \eohVmag & \eohKepmag & \eohJmag  & \eohKmag \\ 
\ktwoofe & \ofe & \ofeField & \ofeRA & \ofeDec & \ofePmRA & \ofePmDec & \ofeVmag & \ofeKepmag & \ofeJmag  & \ofeKmag \\  
\ktwocib & \cib & \cibField & \cibRA & \cibDec & \cibPmRA & \cibPmDec & \cibVmag & \cibKepmag & \cibJmag  & \cibKmag \\ 
\ktwoceo & \ceo & \ceoField & \ceoRA & \ceoDec & \ceoPmRA & \ceoPmDec & \ceoVmag & \ceoKepmag & \ceoJmag  & \ceoKmag \\ 
\ktwohfo & \hfo & \hfoField & \hfoRA & \hfoDec & \hfoPmRA & \hfoPmDec & \hfoVmag & \hfoKepmag & \hfoJmag  & \hfoKmag \\ 
\ktwocdh & \cdh & \cdhField & \cdhRA & \cdhDec & \cdhPmRA & \cdhPmDec & \cdhVmag & \cdhKepmag & \cdhJmag  & \cdhKmag \\ 
\ktwocoe & \coe & \coeField & \coeRA & \coeDec & \coePmRA & \coePmDec & \coeVmag & \coeKepmag & \coeJmag  & \coeKmag \\ 
\ktwooih & \oih & \oihField & \oihRA & \oihDec & \oihPmRA & \oihPmDec & \oihVmag & \oihKepmag & \oihJmag  & \oihKmag \\ 
\ktwodcf & \dcf & \dcfField & \dcfRA & \dcfDec & \dcfPmRA & \dcfPmDec & \dcfVmag & \dcfKepmag & \dcfJmag  & \dcfKmag \\ 
\ktwobfc & \bfc & \bfcField & \bfcRA & \bfcDec & \bfcPmRA & \bfcPmDec & \bfcVmag & \bfcKepmag & \bfcJmag  & \bfcKmag \\ 
\ktwoihd & \ihd & \ihdField & \ihdRA & \ihdDec & \ihdPmRA & \ihdPmDec & \ihdVmag & \ihdKepmag & \ihdJmag  & \ihdKmag 
\enddata
\tablenotetext{a}{From Ecliptic Plane Input Catalog (EPIC)}
\tablenotetext{b}{From AAVSO Photometric All-Sky Survey (APASS) 9th Data Release}
\label{tb:hostpars}
\end{deluxetable*}

\subsection{AO Imaging}
\label{ssec:ao}
We obtained near-infrared adaptive optics images of the 11 EPIC sources at the W.\ M.\ Keck Observatory on the nights of 
1 April 2015 UT (\ktwocib, \ktwocdh, \ktwoceo, \ktwohfo, \ktwooih, \ktwodcf, \ktwoihd), 
7 April 2015 UT (\ktwoofe, \ktwoeoh), and 
25 July 2015 UT (\ktwobfc) UT, and at Palomar Observatory on the night of 
29 May 2015 UT (\ktwoceo).  The observations were obtained with the $1024 \times 1024$ NIRC2 array at Keck Observatory behind the natural guide star AO system and the $1024 \times 1024$ PHARO array behind the PALM-3000 natural guide star system \citep{Dekany13}. In all cases, the target star was bright enough to be used as the guide star. NIRC2 has a pixel scale of 9.942 mas/pixel with a field of view of 10\arcsec; PHARO has a pixel scale of 25 mas/pixel with a field of view of $25\farcs6$. The observations were taken in either the $Ks$ or Br-$\gamma$ filters; Br-$\gamma$ has a narrower bandwidth (2.13--2.18 $\micron$), but a similar central wavelength (2.15 $\micron$) compared the $Ks$ filter (1.95--2.34 $\micron$; 2.15 $\micron$) and allows for longer integration times before saturation. For the Keck observations, a 3-point dither pattern was utilized to avoid the noisier lower left quadrant of the NIRC2 array; the 3-point dither pattern was observed three times for a total of 9 frames.  The Palomar observations were obtained with a 5-point dither pattern with 3 observations at each dither pattern position for a total of 15 frames. 

To optimize our use of NIRC2 and PHARO, we pre-screened three of the targets by acquiring visible-light adaptive optics images of \ktwoofe, \ktwoceo, and \ktwocdh on 8--9 March 2015 using the Robo-AO system \citep{Baranec13, Baranec14} on the 1.5\,m Telescope at Palomar Observatory. Observations comprise a sequence of full-frame-transfer EMCCD detector readouts at the maximum rate of 8.6 Hz for a total of 120\,s of integration time with a long-pass filter cutting on at 600\,nm, with longer wavelength sensitivity limited by the quantum efficiency of the silicon detector out to 1000\,nm. The individual 44\arcsec{}\,$\times$\,44\arcsec{} images are corrected for detector bias and flat-fielding effects before being combined using post-facto shift-and-add processing using the source as the tip-tilt star with 100\% frame selection to synthesize a long-exposure image \citep{Law14}. Sensitivity to faint stellar companions matched that of the high-performance detectable magnitude ratio in \citet{Law14}, typically $\Delta$\,mag = 5 at $5\sigma$ at $0\farcs5$. For these three sources, no stellar companions were detected.

\subsection{Spectroscopy}
\label{sec:spec}

\subsubsection{Keck/HIRES}
\begin{figure*}[!htb]
\includegraphics[width=\textwidth]{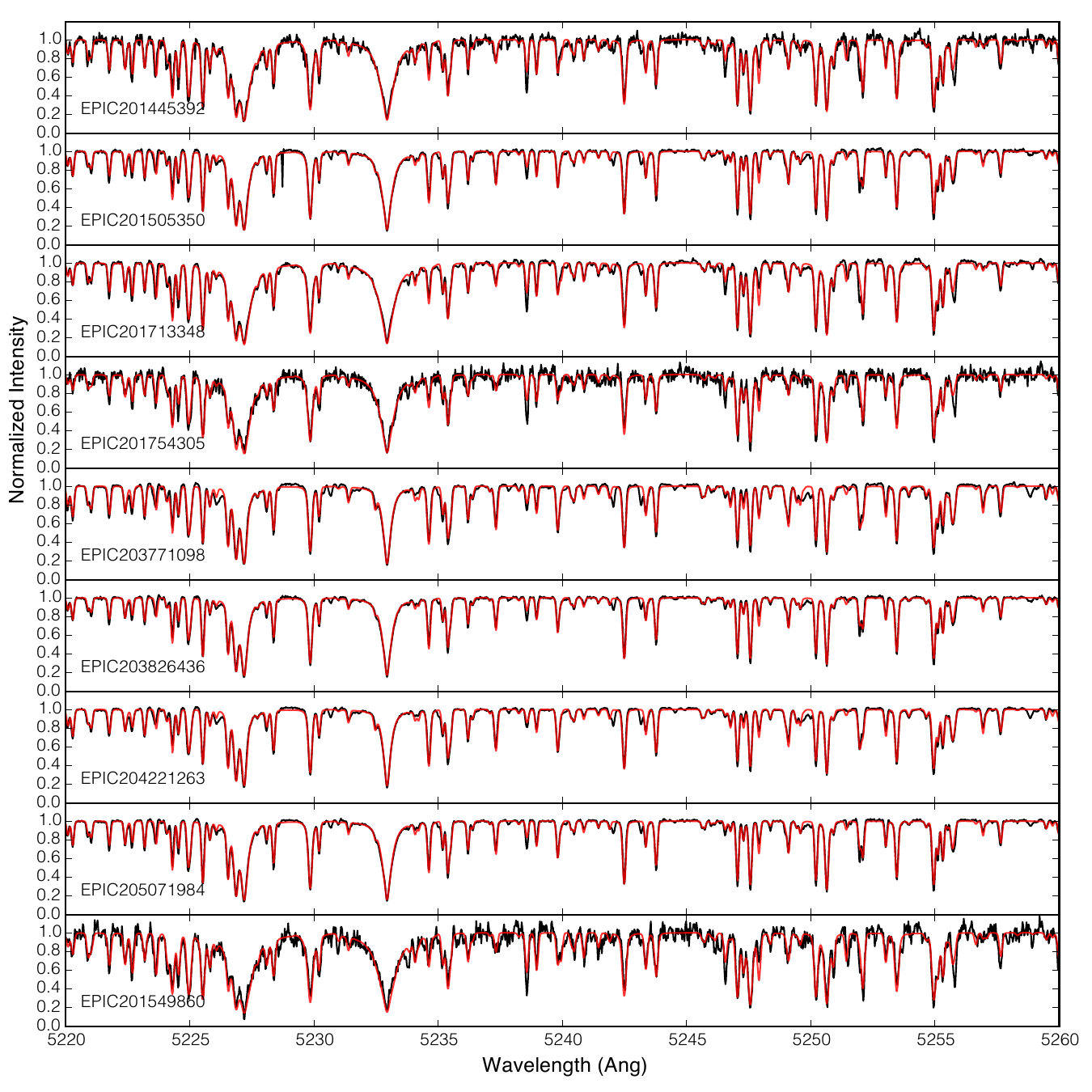}
\caption{A representative segment of our HIRES spectra spanning $\lambda$ = 5220--5260 \AA. The observed stellar spectra are shown in black and the best-fit \specmatch models \citep{Petigura15b} are overplotted in red. Note that this represents only about 10\% of the wavelength coverage modeled by \specmatch.}
\label{fig:specmatch}
\end{figure*}

We used HIRES \citep{Vogt94} at the W.\ M.\ Keck Observatory to measure high resolution optical spectra of all 11 stars except for the coolest and faintest star, \ktwoeoh.  Our observations followed standard procedures of the California Planet Search \citep[CPS;][]{Howard10}.  We used the ``C2'' decker ($0\farcs87$ $\times$ 14\arcsec{} slit) for a spectral resolution $R$ = 55,000 and subtracted the faint sky spectrum from the stellar spectrum. The HIRES exposure meter was set to achieve the desired SNR, which varied with stellar brightness. For our \ktwo follow-up program we generally obtain spectra of stars $V\le13.0$ mag having SNR = 45 per pixel at 550\,nm, while spectra of fainter stars (\ktwocib, \ktwohfo) have SNR = 32 per pixel. These exposure levels were chosen to be sufficient for determination of stellar parameters while keeping exposure times relatively short (1--10 min). Figure \ref{fig:specmatch} shows a wavelength segment of our HIRES spectra. Some of these spectra are higher SNR than prescribed because we obtained additional HIRES spectra for potential Doppler campaigns.

\subsubsection{IRTF/SpeX}
\label{sec:IRspec}
For two \ktwo multi-planet candidates with near-IR spectral types consistent with M dwarfs ($J-K_{s} \gtrsim 0.8$), we obtained spectra using the near-infrared cross-dispersed spectrograph SpeX \citep{Rayner03} on the 3.0-m NASA Infrared Telescope Facility (IRTF). These stars are \ktwoofe and \ktwoeoh. Our SpeX observations and analyses of \ktwoofe are described in detail in \citet{Crossfield15} and we adopt those results here. 

We observed \ktwoeoh on 2015 May 5 UT under clear
skies with an average seeing of 0\farcs5. We used SpeX in short cross-dispersed mode using the $0\farcs3 \times 15^{\prime\prime}$ slit which provides wavelength 
coverage from 0.68 to 2.5 $\mu$m at a resolution of $R \approx 2000$. The star was dithered to two 
positions along the slit following an ABBA pattern for sky subtraction. The \ktwoeoh observing sequence 
consisted of $8 \times 75$\,s exposures for a total integration time of 600\,s.  We also observed an A0 standard and flat and arc lamp 
exposures immediately after the target star for telluric correction and wavelength calibration.

\begin{figure}[!htb]
\epsscale{1.2}
\plotone{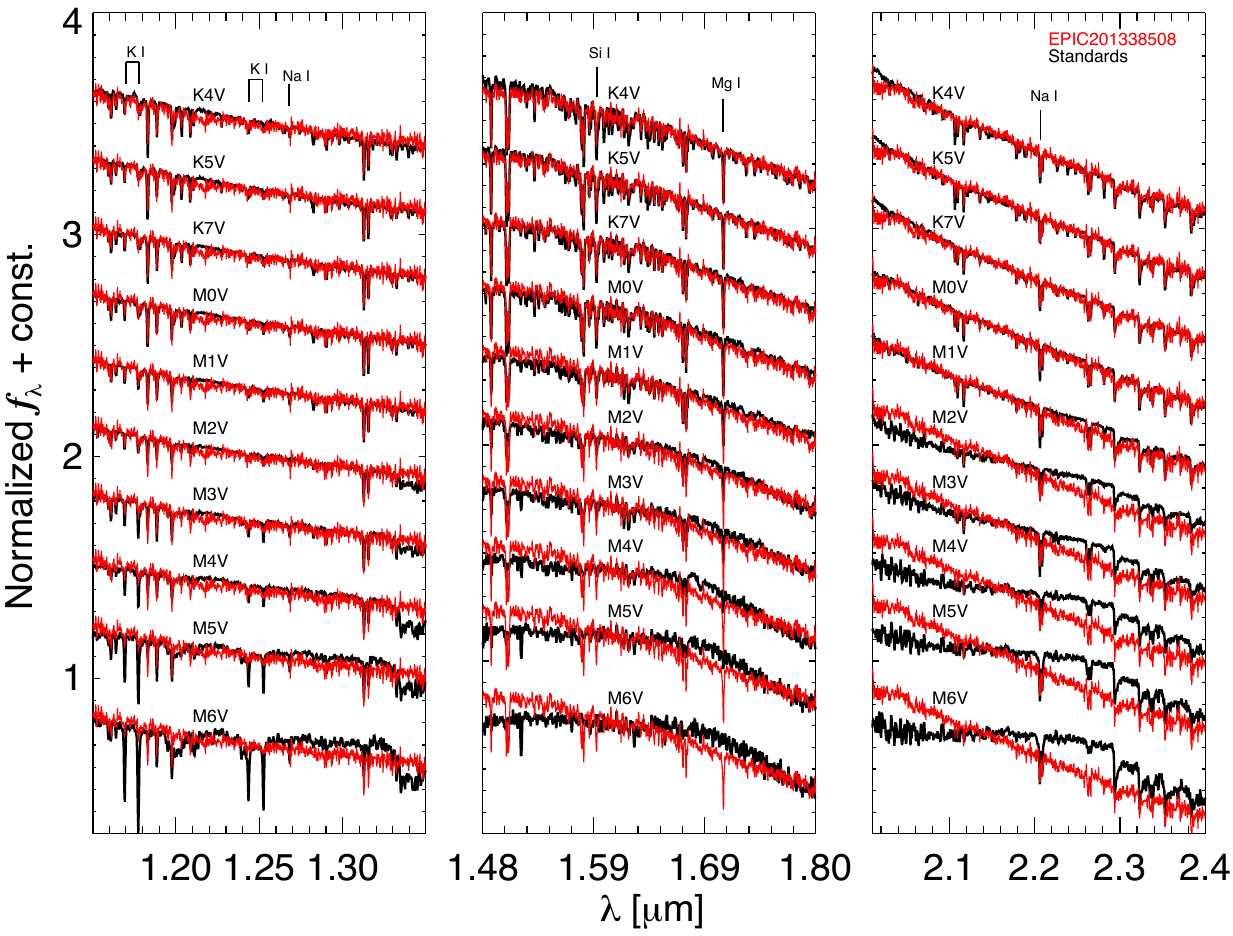}
\caption{$JHK$-band IRTF/SpeX spectra of \ktwoeoh compared to late-type standards from the IRTF spectral library. 
All spectra are normalized to the continuum in each of plotted regions. The star is a best visual match to spectral type $\sim$K7 across 
the three near-IR bands. This is consistent with the results from our analyses using spectroscopic indices.}
\label{fig:SpeX_fig1}
\end{figure}

The data were reduced using the SpeXTool package \citep{Vacca03, Cushing04}.  SpeXTool performs 
flat fielding, bad pixel removal, wavelength calibration, sky subtraction, spectral extraction and combination, 
telluric correction, flux calibration, and order merging. The final calibrated \ktwoeoh spectrum had $JHK$-band signal-to-noise ratios 
$\sim$ 50/75/60. The spectrum is compared to late-type 
standards from the IRTF Spectral Library\footnote{\url{http://irtfweb.ifa.hawaii.edu/\mytilde spex/IRTF\_Spectral\_Library/}}
\citep{Cushing05, Rayner09} in Figure \ref{fig:SpeX_fig1}.

\section{Host Star Characterization}
\label{sec:stellar-properties}
We used \specmatch \citep{Petigura15b} to determine stellar properties from our HIRES spectra for nine stars with spectral types of $\sim$K4 and earlier.  \specmatch estimates effective temperatures, surface gravities, metalicities, and rotational velocities by matching HIRES spectra to an interpolated library of model spectra from \citet{Coelho05}. These models are in good agreement with the spectra of well-characterized stars for \Teff $>$ 4700\,K. See \citet{Petigura15b} for details on \specmatch~ including demonstration that the uncertainties on \Teff, \logg, and \feh~ are 60\,K, 0.08--0.10\,dex, and 0.04\,dex, respectively. Figure \ref{fig:specmatch} shows the best-fit \specmatch model spectra for all nine stars with results.

 We estimated stellar masses and radii from spectroscopic parameters (\Teff, \logg, \feh) by fitting them to a grid of models from the Dartmouth Stellar Evolution Database \citep{Dotter08}. We used the \texttt{isochrones} Python package \citep{Morton15}, which interpolates the Dartmouth model grid (mass-age-\feh) and estimates uncertainties via the \texttt{emcee} Markov Chain Monte Carlo (MCMC) package \citep{Foreman-Mackey13}. This procedure gives mass and radius uncertainties as small as $\sim2\%$, not accounting for the intrinsic uncertainties of the Dartmouth models, which are most uncertain for cool stars. For $M_\star < 0.8$ \msun, \citet{Feiden12} find that most Dartmouth evolution models agree with observed stellar radii to within $\sim$4\%. Therefore, for the nine stars analyzed with \specmatch, we conservatively adopt minimum uncertainties for stellar mass and radius of 5\%. Our final stellar mass and radius uncertainties range from $\sim5-10\%$. 


We also used the HIRES spectra to measure stellar activity. The HIRES spectra span the \caii lines (3969 \AA, 3934 \AA) that are sensitive to chromospheric activity \citep{Wilson68}. Following \citet{Isaacson10}, we measured \shk indices---the 
ratio of flux in \caii line cores to flux in nearby continuum bands. These are converted into \lrphk values (tabulated in Table \ref{tb:specpars}), which account for differences in continuum flux levels with spectral type  \citep{Noyes84, Middelkoop82}. Since the conversion to \lrphk is only calibrated for stars with $B-V<0.9$ (\Teff $\sim$ 5000\,K), we provide only \lrphk values for stars \Teff $>$ 5000\,K and \shk values for the cooler stars. For reference, the Sun varies in the range \lrphk = $-4.85$ to $-5.05$ dex through the solar cycle \citep{Meunier10}.  \ktwoceo and EPIC~\cdh are moderately active (\lrphk = \ceoRHKstar\ dex and \shk = \cdhSHKstar, respectively), while the other GK dwarfs are inactive.




Stellar parameters for the two cooler stars (\ktwoofe, \ktwoeoh) are derived from infrared spectra discussed in Section \ref{sec:IRspec}. Determination of stellar parameters for \ktwoofe are detailed in \citet{Crossfield15}. Here we discuss characterization of \ktwoeoh using similar methods.

We used our SpeX spectrum to measure the near-IR H$_2$0-K2 index \citep{Rojas12} to estimate a spectral type for \ktwoeoh of K7.5 $\pm$ 0.5. This index-based measurement is consistent with the visual best 
match presented in Figure~\ref{fig:SpeX_fig1}. Following \citet{Crossfield15} and \citet{Petigura15}, we estimated
metallicity (\feh), effective temperature (\Teff),
radius ($R_\star$), and mass ($M_\star$) using the methods presented in \citet{Mann13_temp} and \cite{Mann13_metal}. 
Metallicity is estimated using spectroscopic index and equivalent width based methods \citep{Rojas12, Terrien12, Mann13_metal} that were 
calibrated using a sample  of M dwarfs having wide, co-moving FGK companions with well determined [Fe/H]. We use IDL software made publicly available 
by A. Mann\footnote{\url{https://github.com/awmann/metal}} to calculate the $H$- and $K$-band metallicities of \ktwoeoh. We average the $H$ and $K$ 
metallicities and add the measurement and systematic uncertainties in quadrature to arrive at the final value of $\mathrm{[Fe/H] = -0.33 \pm 0.20}$\,dex. This star is metal poor. 
Effective temperature, radius, and mass are calculated using temperature sensitive spectroscopic indices in the $JHK$-bands \citep{Mann13_temp} 
and empirical relations calibrated using nearby, bright M dwarfs with interferometrically measured radii \citep{Boyajian12}. We calculated \Teff in 
the $JHK$-bands and averaged the results. Conservative \Teff uncertainties were estimated by adding in quadrature the RMS scatter in the $JHK$-band values 
and the systematic errors in the empirical fits for each band \citep{Mann13_temp}. The stellar radius and mass were computed using publicly available software from 
A. Mann\footnote{\url{https://github.com/awmann/Teff\_rad\_mass\_lum}}. The resulting fundamental parameters for \ktwoeoh are: 
\Teff $= 3930 \pm 375$\,K, $R_\star$ = $0.57 \pm 0.12$ $R_{\odot}$,
and $M_\star$ = $0.61 \pm 0.13$ $M_{\odot}$. 

\ktwoeoh was presented as a multi-planet system in \citet{Montet15} where their fundamental parameters were estimated 
using broadband photometry and model fits. Our spectroscopic parameter estimates are consistent within uncertainties. 

Table \ref{tb:specpars} lists spectroscopically derived stellar parameters.

\begin{deluxetable*}{llccccccc}[h]
\tabletypesize{\footnotesize}
\tablecaption{Spectroscopic Stellar Properties}
\tablehead{\ktwo Name & EPIC Number & \Teff & $\log{g}$ & \feh & \vsini & \lrphk & $M_\star$ & $R_\star$ \\ 
& & (K) & (cgs) & (dex) & (\kms) & (dex) & (\msun) & (\rsun)}
 \startdata
\ktwoeoh & \eoh & \eohTeff & \eohlogg & \eohFeH & \eohvsini & \eohRHKstar & \eohMstar & \eohRstar \\ 
\ktwoofe & \ofe & \ofeTeff & \ofelogg & \ofeFeH & \ofevsini & \ofeRHKstar & \ofeMstar & \ofeRstar \\ 
\ktwocib & \cib & \cibTeff & \ciblogg & \cibFeH & \cibvsini & \cibRHKstar & \cibMstar & \cibRstar \\ 
\ktwoceo & \ceo & \ceoTeff & \ceologg & \ceoFeH & \ceovsini & \ceoRHKstar & \ceoMstar & \ceoRstar \\  
\ktwohfo & \hfo & \hfoTeff & \hfologg & \hfoFeH & \hfovsini & \hfoRHKstar & \hfoMstar & \hfoRstar \\  
\ktwocdh & \cdh & \cdhTeff & \cdhlogg & \cdhFeH & \cdhvsini & \cdhRHKstar & \cdhMstar & \cdhRstar \\  
\ktwocoe & \coe & \coeTeff & \coelogg & \coeFeH & \coevsini & \coeRHKstar & \coeMstar & \coeRstar \\  
\ktwooih & \oih & \oihTeff & \oihlogg & \oihFeH & \oihvsini & \oihRHKstar & \oihMstar & \oihRstar \\ 
\ktwodcf & \dcf & \dcfTeff & \dcflogg & \dcfFeH & \dcfvsini & \dcfRHKstar & \dcfMstar & \dcfRstar \\ 
\ktwobfc & \bfc & \bfcTeff & \bfclogg & \bfcFeH & \bfcvsini & \bfcRHKstar & \bfcMstar & \bfcRstar \\ 
\ktwoihd & \ihd & \ihdTeff & \ihdlogg & \ihdFeH & \ihdvsini & \ihdRHKstar & \ihdMstar & \ihdRstar 
\enddata
\tablenotetext{a}{For \ktwoeoh and \ktwoofe, \Teff, \feh,  $M_\star$, $R_\star$ are derived using spectroscopic indices of \citep{Mann13_temp} and empirical relations of \citep{Boyajian12}.  For the other stars \Teff, $\log{g}$, \feh, and \vsini are derived using \specmatch \citep{Petigura15b}, \lrphk is derived using the recipe of \citet{Isaacson10}, $M_\star$ and $R_\star$ are derived using the \texttt{isochrones} Python package \citep{Morton15}.}
\tablenotetext{b}{We only list \lrphk for stars with \Teff $>$ 5000\,K, for which this activity metric is well-calibrated. \shk values for cooler stars \ktwocib, \ktwohfo, \ktwocdh, and \ktwocoe are \cibSHKstar, \hfoSHKstar, \cdhSHKstar, and \coeSHKstar, respectively. Spectra of \ktwoeoh and \ktwoofe come from observations with IRTF/SpeX and do not contain \caii lines.}
\label{tb:specpars}
\end{deluxetable*}

\section{Validation of Planet Candidates}
\label{sec:validation}

There are several potential astrophysical events whose light curves can be confused with transiting planets. One example is a blended eclipsing binary (EB) system, either bound to the primary or in the background of  the target star's photometric aperture. Some of these astrophysical false positives can be distinguished from planet transits by secondary eclipses (SEs), but SEs do not always occur and are often undetectably small. Even if the primary star does host a planet, blending of other stars within the photometric aperture can dilute the transit depth causing the planet radius to be underestimated \citep{Ciardi15}. Follow-up observations are crucial for identifying any sources blended within the same 4\arcsec{} \Kepler pixels.

All of the \MPS presented in this catalog passed a series of complementary vetting tests:  First, from the \ktwo light curves, we identified eclipsing binaries (EBs) via their characteristic ``V-shaped'' dimming profiles and secondary eclipses. We also searched for nearby companions in AO images and archival images. In addition, we searched for multiple sets of stellar lines in the high resolution optical spectra.  We also estimate false positive probabilities (FPPs) of each planet candidate, which are constrained by these follow-up observations.

Even without AO imaging or spectroscopy to screen for these blends, the FPPs for \MPS are intrinsically lower compared to systems with a single planet candidate \citep{Lissauer12, Lissauer14}.  For the prime \Kepler mission the FPP for a single planet candidate system is $\sim 10\%$ \citep{Morton11, Fressin13}, but is reduced by factors of $\sim 25$ and $\sim 100$ for systems with one and two additional planet candidates respectively \citep{Lissauer12}.  These FP rates apply to the prime \Kepler mission and cannot be blindly applied to \ktwo, which has a different degree of source crowding as well as different photometric noise properties, target selection criteria, and vetting procedures, all of which factor into the FP rate and ``multiplicity boost'' estimation.  

In this section, we summarize the results of our AO and archival image searches, spectroscopic validation efforts, and FPP assessment.  We estimate multiplicity boosts for \ktwo fields C1 and C2 using available \ktwo planet candidate catalogs.  

\subsection{AO imaging}
\label{sec:AOimaging}
\begin{figure*}
\includegraphics[width=\textwidth]{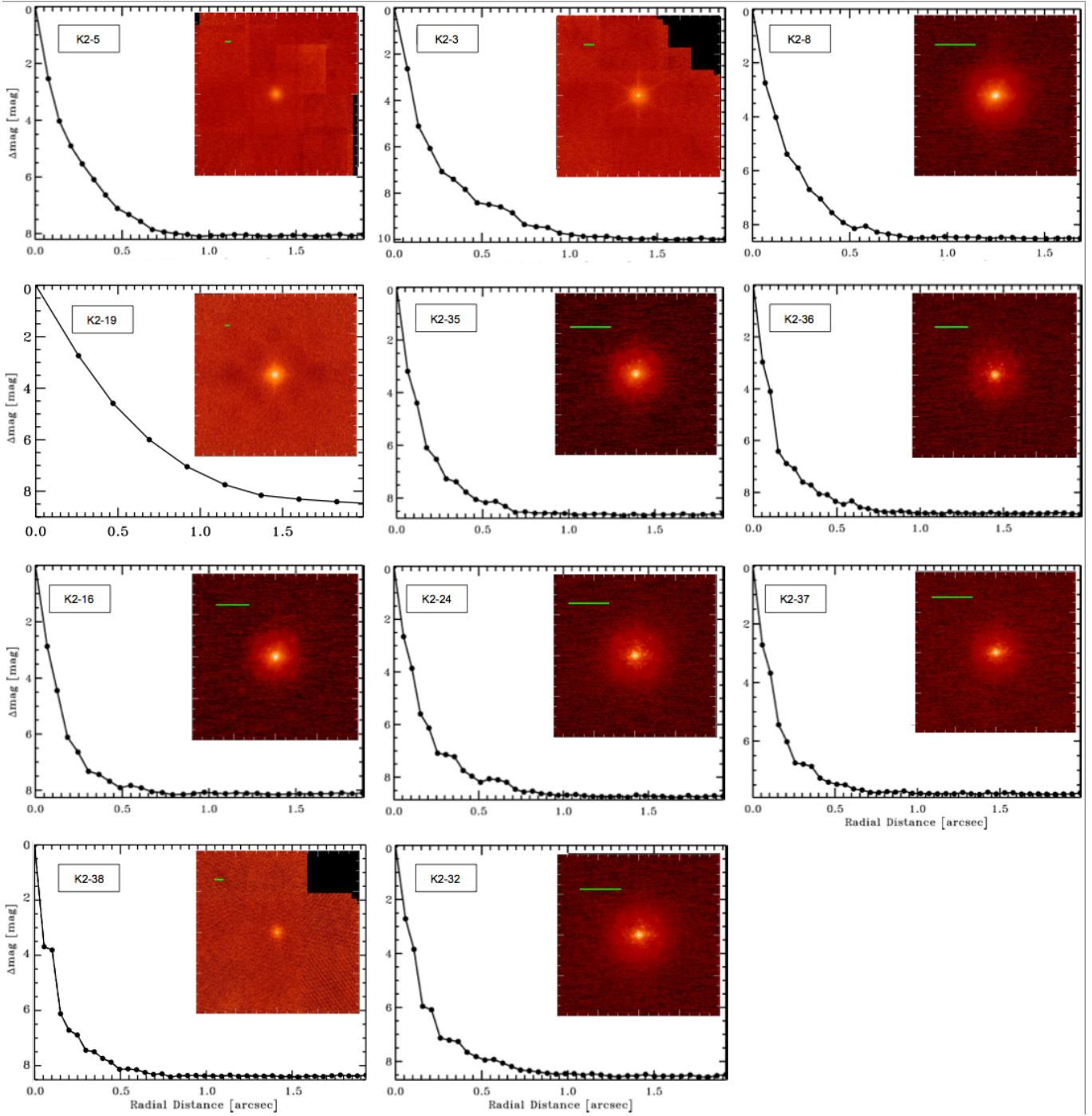}
\caption{AO images and contrast curves for all multi-planet hosts. Targets were imaged using Keck/NIRC2 AO, with the exception of \ktwoceo, which was observed with Palomar/PHARO. Green horizontal lines correspond to 1\arcsec in each field. Dotted black lines indicate where companions would be detectable with 5-$\sigma$ confidence. No companions were detected near any of the 11 stars.}
\label{fig:AO}
\end{figure*}

For each target star, our AO images were combined using a median average.  Typical final FWHM resolutions were 4--6 pixels for a resolution of $\approx 0\farcs05$ with Keck/NIRC2 and $\approx 0\farcs1$ for Palomar/PHARO.  For every target considered here, no other stars were detected within the fields of view of the cameras. For each final combined image,  we estimated the sensitivities by injecting fake sources with a signal-to-noise ratio of 5 at distances of $N$ $\times$ FWHM from the central source, where $N$ is an integer. The 5$\sigma$ sensitivities, as a function of radius from the stars, are shown in Figure \ref{fig:AO} along with a full field of view combined image.  Typical sensitivities yield contrasts of 2--3 mag within 1 FWHM of the target star and contrasts 4--6 mag within 3--4 FWHM.  In the ``flat'' ($>6$ FWHM) of the image, the typical contrasts were 8--9 mag fainter than the target star.

\subsection{Archival imaging}
We also searched for neighboring stars at separations beyond the edges of our AO images (typically 10\arcsec), but within the \ktwo photometric apertures (typically 10--15\arcsec). We downloaded 60\arcsec\,x\,60\arcsec{} $r_{\rm P1}$-band images from the Pan-STARRS1 3$\pi$ survey \citep{Kaiser10}, surrounding each of the EPIC target stars. The images have a plate scale of $0\farcs25$ per pixel and average seeing-limited resolution of $\sim$1\arcsec. The target stars are generally saturated in these images but background sources can be easily identified down to a limiting magnitude of $r_{\rm P1}\approx23$ mag. We used the magnitude zero points in the FITS headers and performed our own aperture photometry on the images to estimate the magnitudes of nearby sources.  Secondary sources are detected within the \ktwo aperture of four systems.  For three of these systems (\ktwoceo, \ktwodcf, \ktwoihd) secondary sources were bright enough to produce observed transit depths.  In these three cases, we regenerated the light curves using smaller apertures that excluded those other stars and verified that the transit signals remained.  We note that all listed transit parameters derive from light curves produced with the original (larger) apertures because of reduced photometric noise.  All secondary sources are sufficiently faint such that dilution corrections would have negligible effects on measured transit depths  --- correction factors would be more than an order of magnitude less than uncertainties on $R_p/R_\star$.  An analysis of each EPIC target is given below.    


\textbf{\ktwoeoh:} No sources fall within the 12\arcsec{} aperture to a limiting magnitude of $r_{\rm P1}\approx23$ mag. The nearest bright source is 15\arcsec{} to the NE with $r_{\rm P1}$=18.9 mag.

\textbf{\ktwoofe:} No stellar sources fall within the 16\arcsec{} aperture to a limiting magnitude of $r_{\rm P1}\approx22$ mag. The nearest star detected in the Pan-STARRS1 imaging is 26\arcsec{} to the NE with $r_{\rm P1}$=17.2 mag.

\textbf{\ktwocib:} No sources fall within the 12\arcsec{} aperture to a limiting magnitude of $r_{\rm P1}\approx22$ mag. No stars or galaxies brighter than $r_{\rm P1}$=21.6 mag fall within 30\arcsec{} of the target.

\textbf{\ktwoceo:} One faint star falls within the 12\arcsec{} aperture 10.7\arcsec{} to the NWW with $r_{\rm P1}$=20.7 mag. We estimate that this source is contributing 0.6 ppt to the \ktwo photometry. An eclipse of the secondary source would not be deep enough to produce the observed transits of \ktwoceo b or c.  Moreover, \citet{Narita15} measure a contrast of $\sim$ 0.1 ppt in H-band and detect the transits of \ktwoceo b and c when the faint star lies outside the photometric aperture, localizing them to the primary.  However, an eclipse of the secondary could produce the observed 0.1 ppt transits of \ktwoceo d.  We re-extracted the photometry using an 8\arcsec{} aperture, small enough to exclude the faint nearby source. The transit signals of all three planets were detected and their depths were consistent with those measured using the original (larger) aperture. The transits of all three planets are therefore localized to the bright star of interest.  Dilution correction factors are negligible compared to measurement uncertainties on $R_p/R_\star$, so we do not apply them.

\textbf{\ktwohfo:} No sources fall within the 12\arcsec{} aperture to a limiting magnitude of $r_{\rm P1}$ $\approx$ 22 mag. There are two nearby stars just outside the \ktwo aperture. One star is 22\arcsec{} to the WNW with $r_{\rm P1}$ = 15.1 mag, and the other is 27\arcsec{} to the NNW with $r_{\rm P1}$ = 14.3 mag.

\textbf{\ktwocdh:} There are two very bright sources 24\arcsec{} to the NE and 24\arcsec{} to the SE of \ktwocdh but both fall outside the 12\arcsec{} \ktwo aperture. These sources are saturated in the Pan-STARRS1 data but are of comparable brightness to \ktwocdh.

\textbf{\ktwocoe:} No sources fall within the 8\arcsec{} \ktwo aperture to a limiting magnitude of $r_{\rm P1}\approx22$ mag. The nearest detected source is 29\arcsec{} to the W with $r_{\rm P1}$=20.3 mag.

\textbf{\ktwooih:} One faint star falls within the 12\arcsec{} \ktwo aperture 6\arcsec{} to the W of the target. This star is badly blended with the wings of the saturated PSF of \ktwooih so reliable photometry can not be extracted. However, we estimate that the star is no brighter than $r_{\rm P1}\approx18.5$ mag, which would contribute only 0.8 ppt to the light in the \ktwo aperture and could not be the source of the 2 ppt and 4 ppt transits of \ktwooih b and c, respectively.  We do not apply a dilution correction to the measured transit depths because it would have negligible effect.

\textbf{\ktwodcf:} Two other stars fall within the 12\arcsec{} aperture. One star, $r_{\rm P1}$=19.1 mag, is located 9.4\arcsec{} to the ESE and the other, $r_{\rm P1}$=19.8 mag, is 8.3\arcsec{} to the WNW. Combined, the contaminating sources contribute 3 ppt to the flux in the \ktwo aperture.   We regenerated the photometry using an 8\arcsec{} aperture that excluded the two other stars, and the transits were still visible, confirming that the primary star is being transited.  We do not correct the transit depths for dilution as this would have negligible effect.

\textbf{\ktwobfc:} No stars brighter then $r_{\rm P1}\approx20.3$ mag fall within the 12\arcsec{} \ktwo aperture. The nearest comparably bright source is a $r_{\rm P1}=19.1$ mag star 29\arcsec{} to the NW.

\textbf{\ktwoihd:} Several faint sources fall within the 16\arcsec{} \ktwo aperture. The brightest of these is 15\arcsec{} to the south with $r_{\rm P1}$=18.9 mag. This contaminating source contributes 2 ppt to the \ktwo aperture flux and is bright enough to account for the transit depths of planets c and d but not planet b. We re-extracted the photometry using an 8\arcsec{} aperture, small enough to exclude the other nearby stars and the transits were still visible. The transits are therefore localized to the target of interest.  Changes in transit depths caused by dilution from the secondary sources are negligible.

\subsection{Spectroscopic vetting}
\label{sec:Reamatch}
We searched the HIRES spectra for multiple sets of stellar lines using the algorithm of \cite{Kolbl15}. The algorithm is sensitive to blends from secondary stars in the 0.87$''$ $\times$ 14$''$ HIRES slit that have effective temperatures \Teff = 3400--6100\, brightness ratios $\gtrsim$1\% in $V$ and $R$ bands, and differ in radial velocity by $\gtrsim$10\,km\,s$^{-1}$ from the primary star. For the 10 targets with HIRES spectra, no spectroscopic blends were detected. 
 
\subsection{False Positive Assessment}
\label{sec:VESPA}
We estimate the False Positive Probability (FPP) of each planet candidate signal using the Python package \VESPA \citep{Morton15}.  We supply \VESPA with the phase-folded \ktwo light curve, photometry from APASS, 2MASS, and WISE, stellar parameters derived from spectroscopy (\S \ref{sec:stellar-properties}), the contrast curve from AO imaging (\S \ref{sec:AOimaging}), and the maximum allowed contrast and velocity offset determined by our spectroscopic vetting (\S \ref{sec:Reamatch}).  Given these constraints, \VESPA estimates the FPP --- the likelihood that the transit signal was produced by a true planet around the target star and not by an eclipsing binary, hierarchical triple system, or non-associated star with a transiting planet.  The FPPs returned by \VESPA are listed in Table \ref{tb:highlights} as well as planet candidate dispositions. Dispositions take into account this study as well as other previously published studies. 

For planet candidates that are not ``confirmed'' by mass detection, we rely on computing a FPP as a means of validation.  Following \citet{Montet15}, we assign a ``validated'' disposition to planet candidates with FPP $<$1\%.  All planet candidates are validated by VESPA to better than 99\% confidence, except for \ktwocib{b} and \ktwoihd{c}, which have FPPs of 1.3\% and 2.2\% respectively. However, these estimates neglect the ``multiplicity boost'' --- an additional factor of confidence in the planet hypothesis gained from the detection of multiple planet candidates in these systems. For the prime \Kepler mission, \citet{Lissauer12} estimate that the a priori FPP is $\sim50\times-100\times$ lower for systems with three or more planet candidates and $\sim25\times$ lower for systems with two planet candidates compared to those with one planet candidate. These were derived using two different methods, each of which assumes that false positives (FPs) are randomly distributed among Kepler targets, and that the presence of FPs and detectable planet signals are uncorrelated.  

We apply the same methods to \ktwo planets.  Following \citet{Lissauer12}, for a system with two planet candidates, if $P_1$ is the probability of a candidate's planethood before considering multiplicity, then the probability of planethood after accounting for multiplicity is

\begin{equation}
\label{eqn:boost}
P_2 \approx \frac{X_2P_1}{X_2P_1 + (1-P_1)}, 
\end{equation} 
where $X_2$ is the ``multiplicity boost'' for systems of two planet candidates.  \ktwoihd{c} is part of a three-candidate system but, for argument sake, it is sufficient to assume that the multiplicity boost for three-candidate systems will be at least as large as that for two-candidate systems. \citet{Lissauer12} estimate $X_2$ using two different methods:  

The first method compares the fraction of Kepler targets with planet candidates ($F_{\mathrm{cand}}\sim1/150$) to the fraction of planet candidate hosts with more than one planet candidate ($F_{\mathrm{multi}}\sim1/6$).  If planets and FPs were randomly distributed among Kepler targets, the detection rate of Kepler multis would be much lower ($F_{\mathrm{multi}}\sim F_{\mathrm{cand}}$).   Assuming that FPs are randomly distributed and that planets are not, \citet{Lissauer12} estimate $X_2 \sim F_{\mathrm{multi}}/F_{\mathrm{cand}} = 25$.  We make the same assumptions to estimate $X_2$ for \ktwo fields C1 and C2 using the catalog of \citet{Vanderburg15b}, who adopt a transit detection threshold of SNR $>$ 9.  We assume that the target sample in a given \ktwo field consists of all objects denoted as a ``STAR'' in the EPIC catalog.  Combining C1 and C2, we compute $F_{\mathrm{cand}} \sim (116/32264) = 0.4\%$ and $F_{\mathrm{multi}}\sim (10/116) = 8.6\%$.   These suggest $X_2 \sim 24$, similar to \Kepler.  Substituting $X_2$ = 24 into Equation \ref{eqn:boost} and setting $P_1$ according to our \VESPA constraints give corrected FPPs, ($1-P_2$) = 0.06\% and 0.09\% for \ktwocib{b} and \ktwoihd{c} respectively. We repeated these estimates for C1 and C2 independently, with similar results.  We also applied these methods to our own catalog of $\sim$100 planet candidates in C1 and C2 detected by TERRA (Crossfield et al., submitted), requiring SNR $>$ 12 and three transits. This yields $X_2\sim 34$ and corrected FPPs (1-$P_2$) = 0.04\% and 0.07\% for \ktwocib{b} and \ktwoihd{c} respectively.  The multiplicity boosts estimated using either catalog are an order of magnitude larger than those needed to validate \ktwocib{b} and \ktwoihd{c} to better than 99\% confidence.

As an additional check, we estimate the multiplicity boost using a second method of \citet{Lissauer12}.  This method assumes that some fraction of candidates $F_{\mathrm{true}}$ are true planets in order to estimate the expected fraction of multi-candidate systems that have at least one FP.  Note that $F_{\mathrm{true}}$ is denoted as $P$ in \citet{Lissauer12} Equations 2 and 4.  Those two equations are used to estimate the expected number of FPs in two-candidate systems based on $F_{\mathrm{true}}$ as well as the total numbers of observed targets and planet candidates.  Subsequently dividing by the number of candidates in two-planet systems yields the fraction of candidates in two-candidate systems expected to be true planets.  Using this method, \citet{Lissauer12} estimated $X_2\sim 25$ for the prime Kepler mission, consistent with the first method above. We apply this same method to our own catalog of \ktwo planet candidates --- unlike \citet{Vanderburg15b}, we compute FPPs for all candidates, most of which have been vetted via spectroscopy and high-resolution imaging.  By integrating over all FPPs, we estimate $F_{\mathrm{true}} \sim 90\%$ and $\sim 60\%$ and $X_2 \sim 70$ and $\sim 20$ for C1 and C2 respectively.  These are similar to the prime Kepler mission ($F_{\mathrm{true}}$ $\sim 90\%$).  Plugging these $X_2$ values into Equation \ref{eqn:boost} and setting $P_1$ according to our VESPA constraints yields corrected FPPs, ($1-P_2$) = 0.02\% and 0.12\% for \ktwocib{b} and \ktwoihd{c}.  For this method, the multiplicity boost is still effective at validating these two planet candidates to FPP $<$ 1\% as long as $F_{\mathrm{true}}$ $\gtrsim$ 15\%.

In summary, the multiplicity boosts estimated via both methods, when combined with \VESPA constraints, are large enough to validate \ktwocib{b} and \ktwoihd{c} to much better than 99\%.

\section{Derived Planet Properties}
\label{sec:planet-properties}

Our light curve analysis follows \citet{Crossfield15}, which we summarize here. Starting with the detrended light curves from \TERRA (Section \ref{sec:photometry}), we perform a sliding median subtraction to removes variability on several-day timescales, including stellar modulation. We fit \texttt{JKTEBOP} transit models \citep{Southworth04, Southworth11} to the light curves, using the \texttt{emcee} MCMC package \citep{Foreman-Mackey13} to generate posterior probability distributions for the transit model parameter. We use the best-fitting transit depth, phase, and orbital period from \TERRA for the initial model guess. 

We assume circular orbits and adopt a linear limb-darkening model, imposing a Gaussian prior on the limb darkening coefficient $u$. The mean of this Gaussian is selected by interpolating the limb darkening tables of \citet{Claret12, Claret13} to our spectroscopically measured \Teff and \logg. The standard deviation was taken to be 0.05. We tested standard deviations of 0.1 and also tried propagating our \Teff and \logg uncertainties through the interpolation procedure, but our results were insensitive to the chosen method. The use of a quadratic limb darkening model also resulted in negligible changes to the posteriors.  

Detrended light curves, fitted transit models, and derived planet parameters are presented for all 11 systems in Figures \ref{fig:eoh_LC}--\ref{fig:ihd_LC} and Tables \ref{tb:eoh}--\ref{tb:ihd} in the Appendix. Derived parameters include orbital distance, $a$, incident flux $S_\oplus$, and equilibrium temperatures, \Teq. The discoveries of \ktwoofe and \ktwooih planets are reported in \citet{Crossfield15} and \citet{Petigura2037}, based on the same data products and analysis methods presented here. We include them in our catalog for completeness.  All stellar and planet parameters will be provided in an online supplementary table. 

\begin{figure}[h]
\includegraphics[width=\columnwidth]{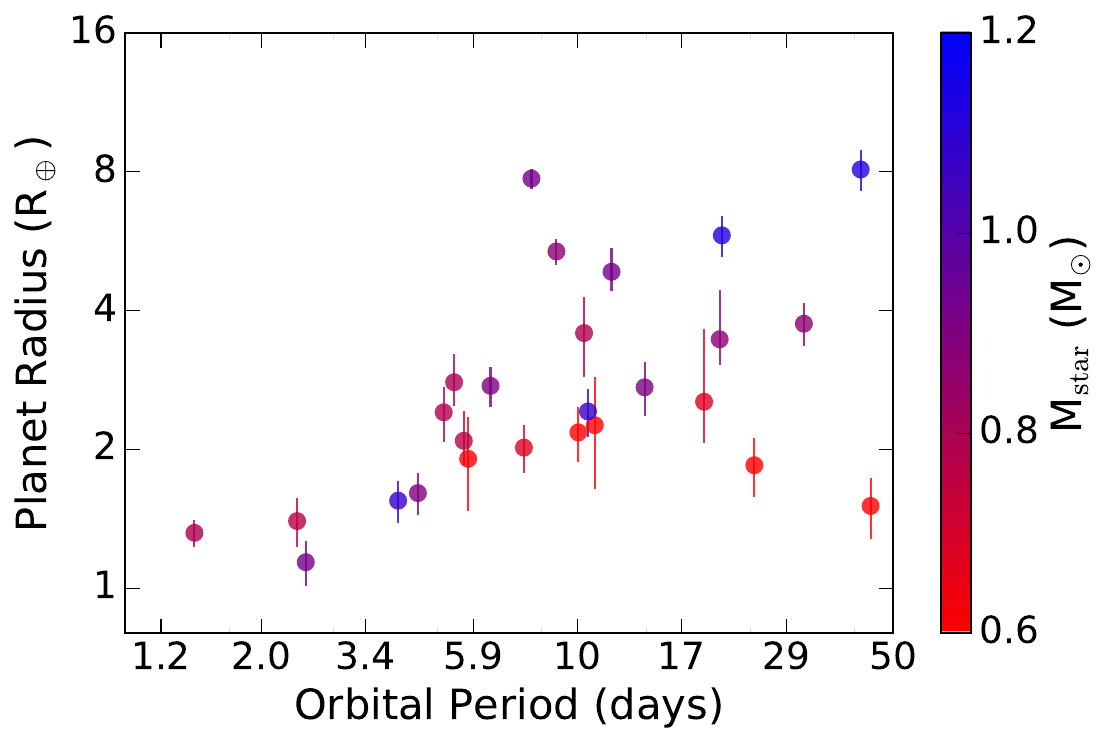} 
\caption{Radii and orbital periods of all 26 planets detected in 11 multi-planet systems in \ktwo Campaigns 1 and 2. The points are colored according to host star mass, with redder colors corresponding to less massive stars. Twenty-one of the 26 planets are likely smaller than 4 \rearth.} 
\label{fig:Rp_vs_P}
\end{figure}

Table \ref{tb:highlights} lists key parameters for each multi-planet system.  
The 26 planets are plotted in radius versus orbital period in Figure \ref{fig:Rp_vs_P}.  The points are colored according to host stellar mass. Twenty-one of the planets are likely smaller than Neptune ($R_p < 3.8$ \rearth).  

\begin{figure}[ht]
\includegraphics[width=\columnwidth]{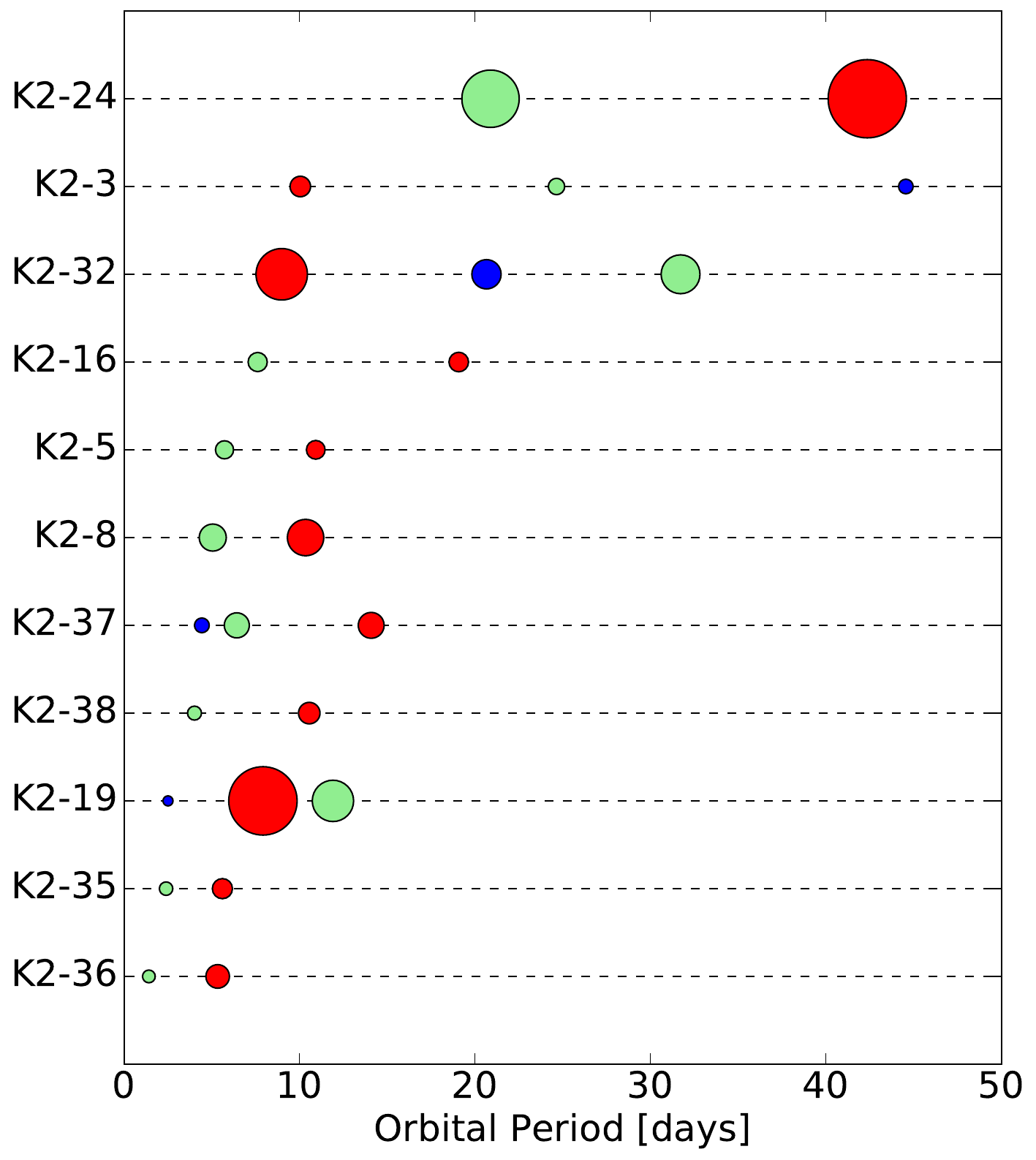} 
\caption{Architecture of the 11 \ktwo \MPS in this study. Each row shows the planets in a particular system ordered top to bottom by decreasing orbital period of the inner planet. The symbol sizes are proportional to planet sizes. The largest planet in each system is colored red, the second largest planet is green and the third largest planet (if present) is blue.}
\label{fig:architectures}
\end{figure}

Figure \ref{fig:architectures} displays the architectures of all systems. Systems are ordered top to bottom by decreasing orbital period of the inner planet. The largest planet in each system is colored red, the second largest planet is green and, the third largest planet (if present) is blue. This ranking scheme considers posterior medians and does not account for uncertainties, thus providing the most likely ranking. In six out of seven systems having only planets $R_p < 3$ \rearth, planet size increases with $P$.   

\begin{deluxetable*}{llccccccrc}
\tabletypesize{\footnotesize}
\centering
\tablecaption{Summary of \ktwo multi-planet systems}
\tablehead{\ktwo & EPIC No. & \Teff & $Kp$ & $T_0$ & $P$ & $R_p$ & \Teq & FPP & Dispo-\\
  Name & & (K) & (mag) & (BJD$_{\mathrm{TDB}}$--2456000) & (d) & (\rearth) & (K) &  & sition\tablenotemark{a}}
\startdata
\ktwoeoh      & \eoh     & \eohTeff & \eohKepmag &         &  &  & & & \\
\ktwoeoh{b} & \eoh{b} &          &            & \eohbTt & \eohbPer & \eohbRp & \eohbTeq & \eohbFPP & \eohbDisposition \\
\ktwoeoh{c} & \eoh{c} &          &            & \eohcTt & \eohcPer & \eohcRp & \eohcTeq &  \eohbFPP & \eohcDisposition \vspace*{0.1in} \\
\ktwoofe      & \ofe       & \ofeTeff & \ofeKepmag &         &  &  & & & \\
\ktwoofe{b}  & \ofe{b} &          &            & \ofebTt & \ofebPer & \ofebRp & \ofebTeq &  \ofebFPP & \ofebDisposition \\
\ktwoofe{c}  & \ofe{c} &          &            & \ofecTt & \ofecPer & \ofecRp & \ofecTeq &  \ofecFPP & \ofecDisposition \\
\ktwoofe{d}  & \ofe{d} &          &            & \ofedTt & \ofedPer & \ofedRp & \ofedTeq &  \ofedFPP & \ofedDisposition \vspace*{0.1in} \\
\ktwocib       & \cib    & \cibTeff & \cibKepmag &         &  &  & & & \\
\ktwocib{b}  & \cib{b} &          &            & \cibbTt & \cibbPer & \cibbRp & \cibbTeq &  \cibbFPP & \cibbDisposition\tablenotemark{b} \\
\ktwocib{c}  & \cib{c} &          &            & \cibcTt & \cibcPer & \cibcRp & \cibcTeq &  \cibcFPP & \cibcDisposition \vspace*{0.1in} \\
\ktwoceo     & \ceo    & \ceoTeff & \ceoKepmag &         &  &  & & & \\
\ktwoceo{b} & \ceo{b} &          &            & \ceobTt & \ceobPer & \ceobRp & \ceobTeq &  \ceobFPP & \ceobDisposition \\
\ktwoceo{c} & \ceo{c} &          &            & \ceocTt & \ceocPer & \ceocRp & \ceocTeq &  \ceocFPP & \ceocDisposition \\
\ktwoceo{d} & \ceo{d} &          &            & \ceodTt & \ceodPer & \ceodRp & \ceodTeq &  \ceodFPP & \ceodDisposition \vspace*{0.1in} \\
\ktwohfo      & \hfo    & \hfoTeff & \hfoKepmag &         &  &  & & & \\
\ktwohfo{b}  & \hfo{b} &          &            & \hfobTt & \hfobPer & \hfobRp & \hfobTeq &  \hfobFPP & \hfobDisposition \\
\ktwohfo{c}  & \hfo{c} &          &            & \hfocTt & \hfocPer & \hfocRp & \hfocTeq &  \hfocFPP & \hfocDisposition \vspace*{0.1in} \\
\ktwocdh     & \cdh    & \cdhTeff & \cdhKepmag &         &  &  & & & \\
\ktwocdh{b} & \cdh{b} &          &            & \cdhbTt & \cdhbPer & \cdhbRp & \cdhbTeq &  \cdhbFPP & \cdhbDisposition \\
\ktwocdh{c} & \cdh{c} &          &            & \cdhcTt & \cdhcPer & \cdhcRp & \cdhcTeq &  \cdhcFPP & \cdhcDisposition \vspace*{0.1in} \\
\ktwocoe     & \coe    & \coeTeff & \coeKepmag &         &  &  & & & \\
\ktwocoe{b} & \coe{b} &          &            & \coebTt & \coebPer & \coebRp & \coebTeq &  \coebFPP & \coebDisposition \\
\ktwocoe{c} & \coe{c} &          &            & \coecTt & \coecPer & \coecRp & \coecTeq &  \coecFPP & \coecDisposition \vspace*{0.1in} \\
\ktwooih      & \oih    & \oihTeff & \oihKepmag &         &  &  & & & \\
\ktwooih{b}  & \oih{b} &          &            & \oihbTt & \oihbPer & \oihbRp & \oihbTeq &  \oihbFPP & \oihbDisposition \\
\ktwooih{c}  & \oih{c} &          &            & \oihcTt & \oihcPer & \oihcRp & \oihcTeq &  \oihcFPP & \oihcDisposition \vspace*{0.1in} \\
\ktwodcf      & \dcf    & \dcfTeff & \dcfKepmag &         &  &  & & & \\
\ktwodcf{b}  & \dcf{b} &          &            & \dcfbTt & \dcfbPer & \dcfbRp & \dcfbTeq &  \dcfbFPP & \dcfbDisposition \\
\ktwodcf{c}  & \dcf{c} &          &            & \dcfcTt & \dcfcPer & \dcfcRp & \dcfcTeq &  \dcfcFPP & \dcfcDisposition \\
\ktwodcf{d}  & \dcf{d} &          &            & \dcfdTt & \dcfdPer & \dcfdRp & \dcfdTeq &  \dcfdFPP & \dcfdDisposition \vspace*{0.1in} \\
\ktwobfc      & \bfc    & \bfcTeff & \bfcKepmag &         &  &  & & & \\
\ktwobfc{b}  & \bfc{b} &          &            & \bfcbTt & \bfcbPer & \bfcbRp & \bfcbTeq &  \bfcbFPP & \bfcbDisposition \\
\ktwobfc{c}  & \bfc{c} &          &            & \bfccTt & \bfccPer & \bfccRp & \bfccTeq &  \bfccFPP & \bfccDisposition \vspace*{0.1in} \\
\ktwoihd      & \ihd    & \ihdTeff & \ihdKepmag &         &  &  & & & \\
\ktwoihd{b}  & \ihd{b} &          &            & \ihdbTt & \ihdbPer & \ihdbRp & \ihdbTeq &  \ihdbFPP & \ihdbDisposition  \\
\ktwoihd{c}  & \ihd{c} &          &            & \ihdcTt & \ihdcPer & \ihdcRp & \ihdcTeq &  \ihdcFPP & ~\ihdcDisposition\tablenotemark{b}\\
\ktwoihd{d}  & \ihd{d} &          &            & \ihddTt & \ihddPer & \ihddRp & \ihddTeq &  \ihddFPP & \ihddDisposition 
\enddata
\tablenotetext{}{$T_0$ = mid-transit time, \Teq = equilibrium temperature assuming albedo = 0.3.}
\tablenotetext{a}{Conf. =   
planet candidate confirmed by RV and/or TTV detections. Cand. = planet candidate (FPP $>$ 1\%).  Valid. =  Statistically validated planet candidate, $>$99\% confidence (FPP $<$ 1\%).  Dispositions take into account this study as well as other previously published studies.}
\tablenotetext{b}{Although the FPPs of \ktwocib{b} and \ktwoihd{c} exceed our 1\% threshold for a ``validated'' disposition, FPP values do not account for the ``multiplicity boost'' (reduction in FPP) resulting from the presence of additional planet candidates around the same star.  We estimate that the multiplicity boosts for C1 and C2 are large enough by an order of magnitude to validate \ktwocib{b} and \ktwoihd{c} to better than 99\% confidence (See \S\ref{sec:VESPA}).}
\label{tb:highlights}
\end{deluxetable*}

\section{Masses of \ktwobfc Super-Earths}
\label{sec:RV}

\subsection{Doppler Measurements}

In an initial campaign with Keck/HIRES, we obtained 14 radial velocity (RV) measurements of \ktwobfc  between 24 June 2015 UT and 3 October 2015 UT.  These observations followed the standard procedures of the California Planet Search \citep[CPS;][]{Howard10}.  We used the  ``C2'' decker (0.87\arcsec{} $\times$ 14\arcsec{} slit) with a cell of molecular iodine gas placed in front of the spectrometer slit to imprint a dense set of molecular absorption lines on the stellar spectrum, subjected to the same instrumental effects.  Exposure times were typically $\sim$20 min and were determined by an exposure meter that terminated exposures when an SNR per pixel of 160 in the continuum near 550\,nm was reached. The iodine lines serve as a wavelength reference and calibration for the point spread function (PSF) over the entire spectral formal.  We also gathered an iodine-free spectrum with the ``B3'' decker ($0\farcs57$ $\times$ 14\arcsec{} slit).  RVs were  determined by forward-modeling the iodine-free spectrum, a high-resolution/high-SNR spectrum of the iodine transmission, and the instrumental response \citep{Marcy92, Valenti95, Butler96, Howard09}. Our measured RVs are listed in Table \ref{tb:RVdata}. Individual RV measurement uncertainties are in the range 1.3--1.8\,\ms.


\begin{figure*}[h]
\subfigure{
\includegraphics[width=\textwidth,trim={0 0 0 0}, clip,]{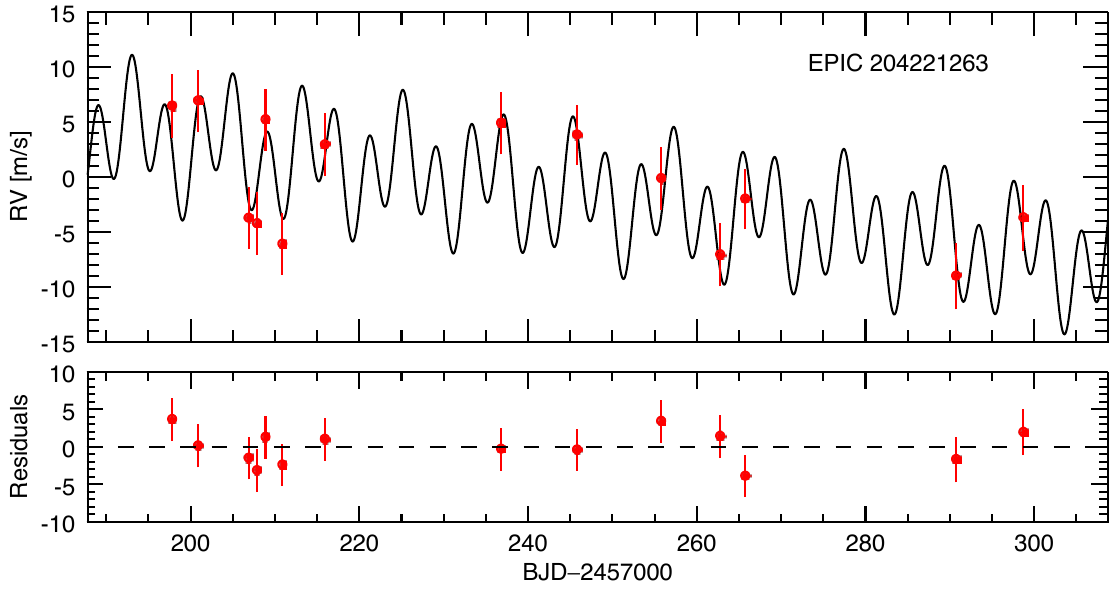} 
} \\
\subfigure{
\includegraphics[width=\textwidth,trim={0 0 0 0}, clip,]{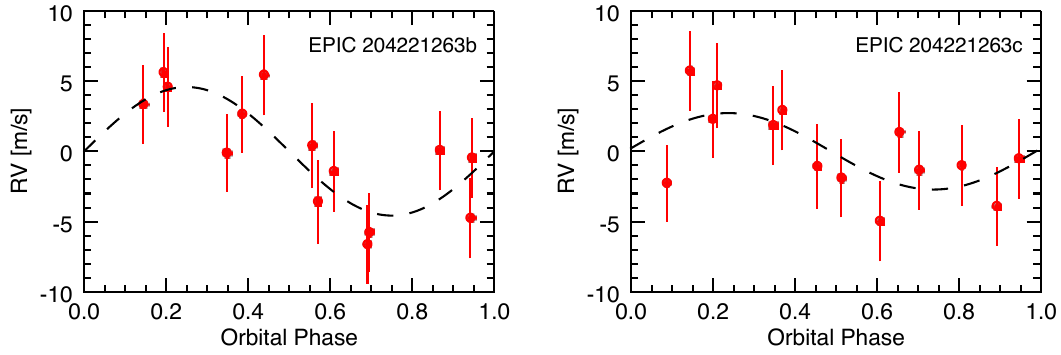} 
} 
\caption{\textit{Top}: RV time series observed with Keck/HIRES (red points), and the best two-planet fit, which includes a significant constant acceleration, \dvdt, evidence of a third bound companion at larger orbital distances. RV error bars represent the quadrature sum of individual measurement uncertainty and the best-fit jitter (2.4 \ms). \textit{Bottom}: RV time series of planets b (left) and c (right), folded at the orbital period of each planet with the linear trend and the Keplerian signal from the other planet subtracted. Transits occur at an orbital phase of 0.5.}
\label{fig:RV}
\end{figure*}

Figure \ref{fig:RV} shows the measured RV time series for \ktwobfc.  The star has low astrophysical jitter and we were able to make initial mass measurements of the two planets.  We fit a two-planet model using the IDL package \texttt{RVLIN} \citep{Wright09}.  
Our RV model assumes two planets with circular orbits, with the orbital periods and phases fixed to the values measured from transits. 
We used a likelihood function constructed as in \cite{Howard14}.  The model has five free parameters including the RV semi-amplitude of planets b and c, $K_b$ and $K_c$, a constant RV offset, $\gamma$, and a constant radial acceleration (RV changing linearly with time), \dvdt. We also include an RV ``jitter" parameter,  $\sigma_{\mathrm{jitter}}$, to account for additional Doppler noise, which might have astrophysical or instrumental origins.  We estimate a jitter of \bfcbjitter \ms that is consistent with expectations for old, solar-type star \citep{Wright05,Isaacson10}.

The RV time series has a negative slope, suggesting that we are seeing a small orbital segment of a third companion with a wider orbital separation. To test this hypothesis, we compared models with and without the constant RV acceleration parameter using the Bayesian Information Criterion \citep[BIC;][]{Schwarz78, Liddle07}.  
Comparing the the best-fitting models, we find $\mathrm{BIC}_{\mathrm{\dvdt}} - \mathrm{BIC}_{\mathrm{\dvdt=0}} =-5.6$, indicating that the model that includes constant acceleration is strongly preferred. There is very likely to be a third, more distant companion in this system, and we constrain its properties in Section \ref{sec:companion}. Additional RV measurements in early 2016, when the target is next observable, will provide a stronger test of the long-term trend.     


Using the same RV model and likelihood function, we performed an MCMC analysis of the RVs to determine parameter uncertainties.  We used \texttt{emcee} \citep{Foreman-Mackey13} and adopted Gaussian priors on the orbital periods and phases, with means and widths matching the posteriors derived from fitting the light curve (Section \ref{sec:planet-properties}). We adopted uninformed priors for $K_b$, $K_c$, $\sigma_{\mathrm{jitter}}$, $\gamma$, and \dvdt.  By allowing the model to explore unphysical solutions with $K<0$, we did not bias the analysis to positive planet masses. We used the best fits from \texttt{RVLIN} to initialize the process. We discarded the first 500 steps. Every 2000 MCMC steps thereafter, we computed the Gelman-Rubin statistic \citep[GRS,][]{Gelman92} to assess convergence. We adopted convergence criterion GRS $<$ 1.03 and generated posteriors once this condition was satisfied.

Table \ref{tb:RVfit} lists measured RV semi-amplitudes, masses, and bulk densities of \ktwobfc b and c. We list posterior medians, with quoted uncertainties being 15.87 and 84.13 percentiles. For the inner planet, we measure a mass of \bfcbMp \mearth. Combining this our planet radius measurement gives a bulk density of \bfcbrhop \gcc.
The mass and bulk density of the outer planet are \bfccMp \mearth and \bfccrhop \gcc, respectively.  We discuss possible compositions for both planets in Section \ref{sec:discussion}.

\begin{deluxetable}{lrr}[h]
\tabletypesize{\footnotesize}
\tablecaption{Relative radial velocities, \ktwobfc}
\tablehead{BJD$-$2457000 & Radial Velocity & Uncertainty \\ & (\ms) & (\ms)}
\startdata
$197.81009$ & $     6.49$ & $     1.49$ \\
$200.88524$ & $     6.96$ & $     1.42$ \\
$206.87767$ & $    -3.70$ & $     1.36$ \\
$207.86695$ & $    -4.20$ & $     1.46$ \\
$208.87926$ & $     5.25$ & $     1.40$ \\
$210.87654$ & $    -6.06$ & $     1.34$ \\
$215.91724$ & $     2.97$ & $     1.46$ \\
$236.80323$ & $     4.93$ & $     1.40$ \\
$245.80773$ & $     3.87$ & $     1.26$ \\
$255.77713$ & $    -0.10$ & $     1.44$ \\
$262.77317$ & $    -7.03$ & $     1.51$ \\
$265.74243$ & $    -1.96$ & $     1.27$ \\
$290.74044$ & $    -8.97$ & $     1.75$ \\
$298.71937$ & $    -3.67$ & $     1.76$
\enddata
\tablenotetext{}{}
\label{tb:RVdata}
\end{deluxetable}

\begin{deluxetable}{lrcrl}[h]
\tabletypesize{\footnotesize}
\tablecaption{RV model, \ktwobfc}
\tablehead{Parameter & Planet b & & Planet c & Units}
\startdata
$K$ & \bfcbK & & \bfccK & m\,s$^{-1}$ \\
$M_p$ & \bfcbMp & & \bfccMp & $M_{\oplus}$ \\
$\rho_p$ & \bfcbrhop & & \bfccrhop & g\,cm$^{-3}$ \\
$\sigma_\mathrm{jit}$ & & \bfcbjitter & & \ms \\
$\gamma$ & & \bfcbgamma & & \ms \\
$\dvdt$ &  & \bfcbdvdt & & \ms{yr$^{-1}$} 
\enddata
\tablenotetext{}{$K$ = RV semi-amplitude, $M_p$ = planet mass, $\rho_p$ = planet density, $\sigma_\mathrm{jit}$ = RV ``jitter'', $\gamma$ = constant RV offset, $\dvdt$ = constant RV acceleration}
\label{tb:RVfit}
\end{deluxetable}

The marginalized posterior distribution for the linear trend gives \dvdt = \bfcbdvdt \ms\,yr$^{-1}$. This linear trend contributes a change in RV of $\sim$10 \ms over the $\sim$100-day time baseline of our RV campaign. This suggests a Keplerian signal with semi-amplitude $K\gtrsim5$\,\ms, and $P\gtrsim200$\,d ($a\gtrsim0.7$\,AU). 

\subsection{Constraints on an additional body}
\label{sec:companion}
\begin{figure}[h]
\includegraphics[width=\columnwidth]{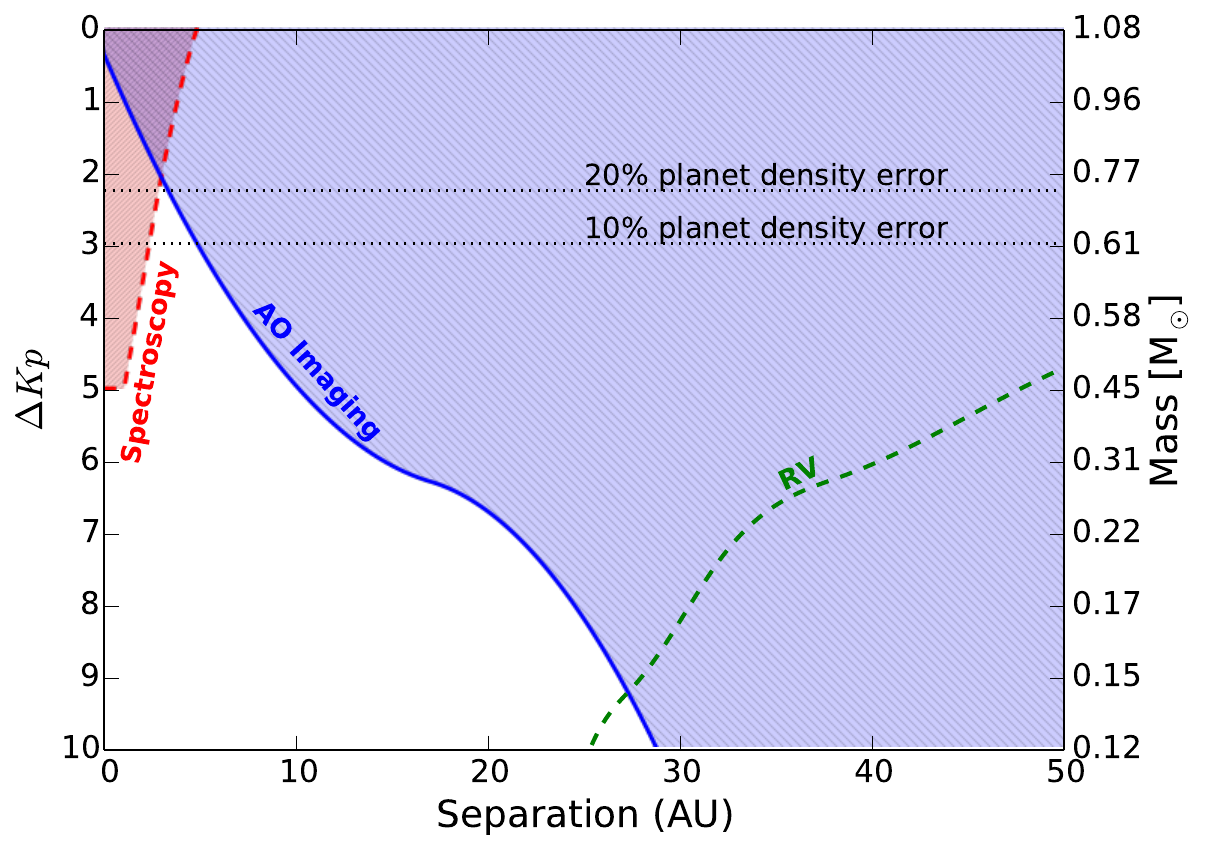} 
\caption{Constraints on the properties of an unseen companion.  The axes denote a parameter space of brightness contrast of the companion with \ktwobfc (alternatively the mass of the companion) and the orbital separation between the two bodies. AO imaging excludes companions in the hatched blue region.  The dashed red line shows the limits of our  search for secondary lines in the high-resolution optical spectrum from HIRES. The dashed green line (lower right) corresponds to the masses and orbital semi-major axes consistent with the measured linear RV trend, assuming a circular, edge-on orbit and a companion mass much lower than that of the primary star (Equation \ref{eqn:companion-mass}). The horizontal dashed lines represent companion contrasts at which the dilution of the observed transit depths would cause planet densities to be overestimated by 10\% and 20\%. AO imaging and spectroscopy rule out companions that would cause systematic errors of $>20\%$ in planet density with high confidence (see Section \ref{sec:companion} for discussion).}
\label{fig:companion-constraint}
\end{figure}

We considered the possibility that the source of the linear RV trend is a companion star that contributes enough light to the \ktwo photometry to significantly dilute the observed transit depths. In this scenario we would underestimate the planet radii and overestimate of planet densities \citep{Ciardi15}. We assessed this possibility using our AO images and HIRES spectra.  We confined the companion properties to a small domain of companion mass (or contrast with the primary) and orbital separation. These constraints are summarized in Figure \ref{fig:companion-constraint} and suggest a low likelihood that the companion is bright enough to affect our measured density by more than 20\%. The non-detection of secondary lines in the HIRES spectrum allows us to exclude stars close in proximity and mass to the primary, specifically $\Delta{Kp}\leq 5$ and $\Delta{\mathrm{RV}} \gtrsim 10$ \kms (red dashed line). The plotted boundary (dashed red line) assumes that the HIRES spectrum was acquired at an orbital phase of maximum $\Delta{\mathrm{RV}}$. Our Keck/NIRC2 AO contrast curve (blue solid line) extends the exclusion region to fainter companions at larger separations. Horizontal dotted lines show stellar companions that would cause planet densities to be overestimated by 10\% and 20\%.  There is a small window of unvetted parameter space, spanning companion masses $\sim$0.6--0.7 \msun, orbital separations $\sim$4--5\,AU, which would cause planet densities to be overestimated by 10--20\% (planet radii underestimated by 3\% and 6\%, respectively).  This potential underestimate is smaller than our measurement uncertainties. There are a few noteworthy caveats: If the AO imaging happened to take place when the projected separation was small, then a brighter companion could go undetected.  The same applies if the HIRES spectrum was taken when the difference between the RVs of the primary star and its companion was low or if the orbit is near face-on (misaligned with the transiting planets). Note that while a near face-on orbit would limit spectroscopic constraints, it would maximize detectability by AO imaging.

For the above analysis, we used $riJHK$ photometric calibrations of \citet{Kraus07} to convert angular separation to orbital distance ($\sim170$\,pc) and to convert contrasts in the NIRC2-AO bandpass to contrasts in the Kepler bandpass and to companion masses. Following \citet{Winn10}, if we assume the companion has a circular orbit and mass $M_{2} \ll M_\star$, then our measured $dv/dt$ = \bfcbdvdt \ms\,yr$^{-1}$ implies

\begin{equation}
M_2\sini\sim0.2\,M_{\mathrm{Jup}}\left(\frac{a}{1\,\mathrm{AU}}\right)^2,
\label{eqn:companion-mass}
\end{equation}
where $M_{\mathrm{Jup}}$ is the mass of Jupiter. If the companion is 1\,$M_{\mathrm{Jup}}$ it would be located at $\sim$ 2 AU. The constraints given by Equation \ref{eqn:companion-mass} are also shown in Figure \ref{fig:companion-constraint} (green dashed line).  However, we stress that these assume the companion has a circular orbit in the same plane as the planets b and c, and a mass much lower than that of the primary star. Therefore, it does not decrease the likelihood of close-in companions bright enough to significantly dilute the transit depth.

\section{Individual systems}
\label{sec:individual-systems}
Here we summarize some of the most important characteristics of each system determined from this study, as well as other studies where relevant. 

\textbf{\ktwoeoh} is a $\sim$K7.5V star, hosting at least two planets with $P$ = 5.7 and 10.9 d and $R_p$ = \eohbRp \rearth and \eohcRp \rearth, respectively (Figure \ref{fig:eoh_LC}, Table \ref{tb:eoh}). Both planets were reported by \citet{Foreman-Mackey15}, and \citet{Vanderburg15b} and validated by \citet{Montet15}. The large uncertainties on planet radii ($\sim 20-30\%$) relative to the rest of our targets is due to the modest SNR of our IRTF-SpeX stellar spectrum (see Section \ref{sec:IRspec}). 

\textbf{\ktwoofe} is a nearby, $\sim$M0V star hosting three super-Earths with $P$ = 10.1, 24.6, and 44.6 d and $R_p$ = \ofebRp, \ofecRp, and \ofedRp \rearth, respectively (Figure \ref{fig:ofe_LC}, Table \ref{tb:ofe}). All three planets were first reported in \citet{Crossfield15}. Spitzer observations independently confirmed transits of each planet \citet{Beichman16}.  \citet{Almenara15} and \citet{Dai16} measured masses of $8.4 \pm 2.1$ \mearth and $8.1^{+2.0}_{-1.9}$ \mearth respectively for the inner planet, \ktwoofe{b}, which suggests it is composed mostly of rock, but possibly as much as 60\% water.  The outer planet receives $\sim$50\% more stellar flux than Earth receives from the Sun, with an equilibrium temperature, \Teq$\sim$282\,K.      

\textbf{\ktwocib} is a $\sim$K3V star hosting two planets with $P$ = 5.1 and 10.4 d, near a 2:1 MMR. These planets have radii of \cibcRp\xspace and \cibbRp\,\rearth, respectively (Figure \ref{fig:cib_LC}, Table \ref{tb:cib}). Both planets were reported in \citet{Foreman-Mackey15} and \citet{Vanderburg15b}.   \citet{Montet15} estimated FPPs using \VESPA. In their study, only the outer planet candidate (K2-8b) was given a ``validated'' disposition.  They listed the inner planet as a ``planet candidate'' because the FPP exceeded 1\%, dominated by the 1.9\% probability of a background eclipsing binary (BEB). They looked for companions using images from data release nine of the Sloan Digital Sky Survey (SDSS), which has a 1.4\arcsec~PSF \citep{Ahn12}. Our NIRC2 AO imaging excludes the possibility of BEBs several times closer to the star and we compute a lower FPP of 0.8\%.  Surprisingly, for the outer planet candidate validated by \citet{Montet15} (FPP=0.2\%), we compute a higher FPP of 1.3\%, despite better constraints from AO imaging and spectroscopy.  The reason for this discrepancy is unclear but possibly the result of having slightly different stellar parameter constraints or photometry.  Nevertheless, \VESPA does not account for the multiplicity of this system, which would reduce our FPP estimate below 1\% (See ¤4.4).  Therefore, we assign validated dispositions to both \ktwocib{b} and \ktwocib{c}.

\textbf{\ktwoceo} is a $\sim$G9V star hosting three planets. The star is magnetically active; we measure \lrphk = \ceoRHKstar\,dex and the light curve exhibits quasi-periodic variations in brightness by $\sim1\%$ over 15--20 d. The inner planet, $P=2.5$ d, is near Earth-size with $R_p$ = \ceodRp \rearth.  The outer two planets are larger and near 3:2 MMR having $P$ = 7.9 and 11.9 d and $R_p$ = \ceobRp \rearth and \ceocRp \rearth, respectively (Figure \ref{fig:ceo_LC}, Table \ref{tb:ceo}). The two outer planets were reported as planet candidates by \citet{Foreman-Mackey15} and first confirmed by \citet{Armstrong15}, who used ground based telescopes to detect additional transits and measure hour-long TTVs for the middle planet.  A similar study by \citet{Narita15} found consistent TTVs and precisely characterized the star via AO imaging with Subaru-HiCIAO and high resolution spectroscopy with Subaru-HDS.  \citet{Barros15} simultaneously modeled \ktwo and ground-based photometry and RVs from the SOPHIE spectrograph of $\sim$16\,\ms precision. They find that planet b has mass $44 \pm 12$ \mearth and radius $7.46 \pm 0.76$ \rearth, while planet c has mass $15.9^{+7.7}_{-2.8}$ \mearth and radius $4.51 \pm 0.47$ \rearth. \citet{Dai16} measured $M_b = 28.5^{+5.4}_{-5.0}$ \mearth and $M_c = 25.6 \pm 7.1$ \mearth. Planet d was first reported as a planet candidate by \citet{Vanderburg15b}.  The 2.5-day period of the transits is $\sim$10x the spacecraft thruster firing period of 6 hours.  However, in our planet search and light curve fitting, we omit all photometry collected during a thruster firing.  Thus, we are confident that thruster firings are not the source of the transit.  As an additional test, we visually inspected the complete set of photometric measurements phased at the transit period and found there was no excess of thruster firings during the transits.  Thus, even if these data points were included in our fitting, they would not significantly bias the derived planet properties. We also verified that the photometric scatter during the transit phase is not systematically different than the out-of-transit phase.  

\textbf{\ktwohfo} is a bright $\sim$K4V star hosting two close-in super-Earths with $P$ = 2.4 and 5.6 d.  The planets are \hfobRp \rearth and \hfocRp \rearth respectively (Figure \ref{fig:hfo_LC}, Table \ref{tb:hfo}).  The outer planet was reported as a planet candidate by \citet{Foreman-Mackey15} and \citet{Montet15}.  Both were listed as planet candidates by \citet{Vanderburg15b}. 

\textbf{\ktwocdh} is a bright $\sim$K2V star hosting two hot super-Earths with $P$ = 1.4 and 5.3 d, first reported as planet candidates by \citet{Vanderburg15b}.  The inner planet, $R_p$ = \cdhbRp \rearth, has \Teq$\sim$1200\,K and the outer planet, $R_p=$ \cdhcRp \rearth has \Teq $\sim$ 800\,K (Figure \ref{fig:cdh_LC}, Table \ref{tb:cdh}).  The light curve shows 1--2\% modulation with a period of $\sim$10 d and the spectrum has strong \caii emission lines (\shk=\cdhSHKstar), indicating that the star is magnetically active.  

\textbf{\ktwocoe} is a faint, $\sim$K3V star having two planets with $P$ =7.6 and 19.1 d and $R_p$ = \coebRp \rearth and \coecRp \rearth, respectively (Figure \ref{fig:coe_LC}, Table \ref{tb:coe}). These planets are near 5:2 MMR.  They were first detected by \citet{Foreman-Mackey15} and validated by \citet{Montet15}.  Both planets were detected by \citet{Vanderburg15b}.

\textbf{\ktwooih} is a bright $\sim$G9V star with two cool, sub-Saturn-size planets near 2:1 MMR, $P$ = 20.9 and 42.4 d, $R_p$ = \oihbRp\,\rearth and \oihcRp\,\rearth (Figure \ref{fig:oih_LC}, Table \ref{tb:oih}). Using Keck/HIRES RVs, \citet{Petigura2037} measured masses of $21 \pm 5.4$\,\mearth and $27 \pm 6.9$\,\mearth and densities of $0.63 \pm 0.25$\,\gcc and $0.31 \pm 0.12$\,\gcc for inner and outer planets, respectively.  \citet{Dai16} measured masses of $19.8^{+4.5}_{-4.4}$ \mearth and $26.0^{+5.8}_{-6.1}$ \mearth respectively. The transit signals of both planets were also detected by \citet{Vanderburg15b}.

\textbf{\ktwodcf} is a bright $\sim$G3V star with three small, tightly packed planets having $P$ = 9.0 d, 24.6 d, and 44.6 d and $R_p$ = \dcfbRp, \rearth \dcfcRp \rearth and \dcfdRp \rearth, respectively (Figure \ref{fig:dcf_LC}, Table \ref{tb:dcf}). These were reported as planet candidates by \citet{Vanderburg15b}.

\textbf{\ktwobfc} is a bright $\sim$G2V star, with two hot super-Earths, $P$ = 4.0 and 10.6 d and $R_p$ = \bfcbRp \rearth and \bfccRp \rearth, respectively (Figure \ref{fig:bfc_LC}, Table \ref{tb:bfc}). We measure planet masses of \bfcbMp \mearth and \bfccMp \mearth and densities of \bfcbrhop \gcc and \bfccrhop \gcc (Table \ref{tb:RVfit}). These indicate that the inner planet is likely rocky and possibly iron-rich, while the outer planet is likely to have an envelope of low-density volatiles (Section \ref{sec:compositions}). A linear RV trend also suggests a third companion at larger orbital distances (Section \ref{sec:RV}).  None of these planets were previously reported.

\textbf{\ktwoihd} is a bright $\sim$G9V star with three planets, $P$ = 9.0, 24.6, and 44.6 d, $R_p$ = \ihdbRp, \rearth \ihdcRp, \rearth and \ihddRp \rearth, respectively (Figure \ref{fig:ihd_LC}, Table \ref{tb:ihd}).   \citet{Vanderburg15b} reported these as planet candidates. The outer two planets are near 3:2 MMR.  \citet{Dai16} confirmed the inner planet b, using RVs to measure a mass of $21.1 \pm 5.9$ \mearth.  For planet c, \VESPA returns a FPP of 2.2\%, which does not meet our criterion for a ``validated'' disposition (FPP $<$ 1\%).  However, \VESPA does not account for the ``multiplicity boost'', which is more than the factor of 2.2 necessary to reduce the FPP of \ktwoihd{c} below 1\% (See discussion in \S\ref{sec:VESPA}).  Therefore we deem all three planet candidates in this system ``validated''.

\section{Discussion \& Conclusions}
\label{sec:discussion}

We have detected, validated, and characterized 11 multi-planet systems comprised of 26 planets in \ktwo fields C1 and C2.  Seven of these systems have two detected planets and four of them have three detected planets, the majority of which are smaller than Neptune. Moreover, seven of the stars have $Kp<13$ and are amenable to RV follow-up to measure planet masses and densities. 
This study is distinguished from previous \ktwo catalogs because it focuses on multi-planet systems with intrinsically low false positive probabilities and we have characterized each host star with high contrast imaging and spectroscopy.  
We detected the RV signatures of \ktwobfc{b} and \ktwobfc{c}, allowing us to constrain their masses and densities and infer their bulk compositions. 

\subsection{Compositions of \ktwobfc Super-Earths}
\label{sec:compositions}

\begin{figure}[ht]
\centering
\includegraphics[width=\columnwidth]{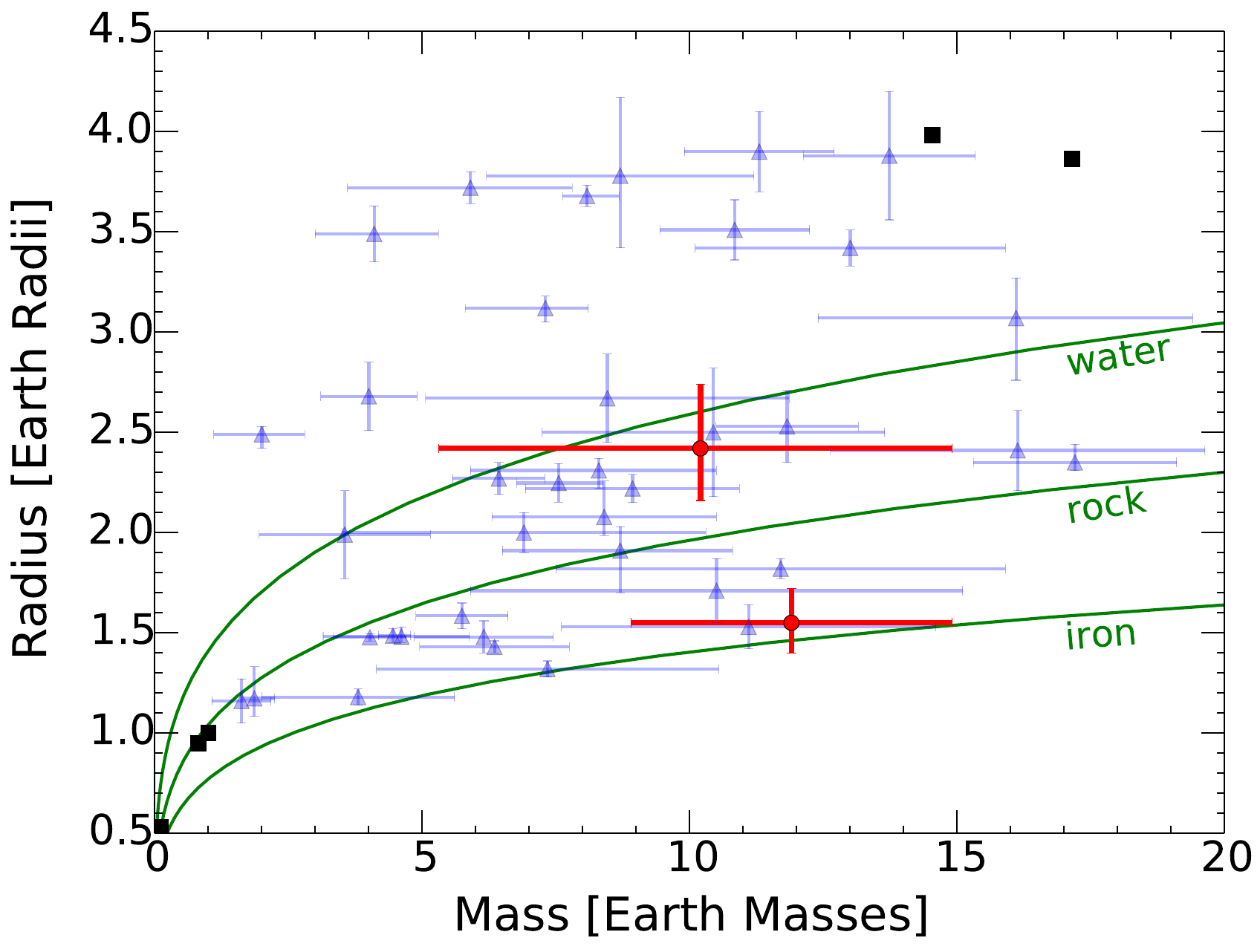} 
\caption{Radii and masses of all confirmed planets whose mass and radius are measured to better than 50\% (2$\sigma$) precision (blue triangles). Solar System planets are represented as black squares. Red circles indicate our measurements of \ktwobfc{b} and {c} (top and bottom points, respectively). Green curves show the expected planet mass-radius curves for pure iron, rock, and water compositions according to models by \citet{Zeng13}. \ktwobfc{b} likely has a large iron fraction and could be the densest planet detected to date. The composition of \ktwobfc{c} is less certain, but the planet likely possesses an outer envelope comprised of low-density volatiles.}  
\label{fig:MR}
\end{figure}

Figure \ref{fig:MR} shows the mass-radius and density-radius distributions of all planets with $R_p$ < 4.0 \rearth whose mass and radius are measured to better than 50\%  precision ($2\sigma$) either by RVs or TTVs \footnote{NASA Exoplanet Archive, UT 13 November 2015, http://exoplanetarchive.ipac.caltech.edu}. Solar System planets are included as well as theoretical mass-radius relations for pure iron, rock, and water compositions, based on models by \citet{Zeng13}. The red points in Figure \ref{fig:MR} show our mass and radius constraints of \ktwobfc{b} and \ktwobfc{c}. 
   
The measured mass and radius of \ktwobfc{b} are consistent with a rocky or iron-rich composition, matching \ktwoofe{d} and KOI-94b within uncertainties. Comparing mass and radius estimates to compositional models of \citet{Zeng13} gives a 97.7\% probability that the planet is denser than pure rock. With a bulk density of \bfcbrhop \gcc, \ktwobfc{b} could be the densest planet discovered to date, but additional RV measurements are needed to confirm this. While our $1\sigma$ measurement errors do not rule out densities exceeding that of pure iron, we can reject such compositions on the basis that there are incompatible with planet formation theory and the low abundances of heavy elements in planet-forming disks. Iron-rich planets (e.g., Mercury) might result from collisional stripping of the rocky mantle of a larger, differentiated, planet. Simulations by \citet{Marcus10} suggest that collisional stripping is unlikely to produce super-Earths with iron mass fractions exceeding $\sim70\%$. The initial assembly of an iron-rich core might be expedited by photophoretic segregation of metals and silicates in the inner protoplanetary disk, which preferentially drives the rocky material outward \citep{Wurm13}. With an equilibrium temperature, \Teq$\sim$1200\,K, \ktwobfc{b} is perhaps a remnant core of a larger planet whose atmosphere was removed by photoevaporation. In such a scenario, the precursor could have been a gas giant that formed beyond the snow line and migrated inwards. Indeed, such a massive core would have rapidly accreted nebular gas, if still present. Alternatively, if the planet assembled in-situ, photoevaporation might have been less important to its present composition; nebular gas might have dispersed before the core was massive enough to accrete, or atmospheric accretion could have been limited by the creation of a gap in the disk \citep{Hansen12}.   

The mass and radius of the outer planet, \ktwobfc{c}, are consistent with many other planets, including GJ 1214b, Kepler-68b, Kepler-96b, Kepler102-e, Kepler-106c, HD 97658b, and HIP 116454b. Its equilibrium temperature of \bfccTeq\,K is intermediate to those of the other planets, which have \Teq spanning $\sim$550--1150\,K. This planet is unlikely to have experienced significant atmospheric photoevaporation. We measure the planet's mass to $\sim$50\% (2$\sigma$) precision, which allows for a range of possible compositions --- even with smaller measurement uncertainties, planet compositions in this region of the mass-radius diagram are highly degenerate \citep{Seager07, Adams08, Valencia13}. \ktwobfc{c} likely contains an outer envelope comprised of low-density volatiles. It could have a small rocky core, with an extended H/He envelope or steam atmosphere.  Alternatively, since the measured density is consistent with pure water, the planet could be a ``water-world'', with a core rich in water ice and interior to a mostly steam atmosphere.  A more precise mass is needed to meaningfully constrain core to envelope mass ratios and possible mass fractions of rock, water, and H/He. Due to the mass-radius degeneracies between water-worlds and rocky cores with extended H/He atmospheres, the atmospheric composition must be measured by other means (e.g. transmission spectroscopy) in order to distinguish between these two different archetypes.  

\subsection{Orbital Stability}
We analytically assess the orbital stability of each system by comparing orbital separations of each planet pair to their mutual Hill radii, which is the length scale applicable to dynamical interactions: 

\begin{equation}
R_H = \left[\frac{M_\mathrm{in}+ M_\mathrm{out}}{3M_\star}\right]^{1/3}\frac{\left(a_\mathrm{in}+ a_\mathrm{out}\right)}{2}.
\label{eqn:MHill}
\end{equation}
Here, $M_\mathrm{in}$ and $M_\mathrm{out}$ are the masses of the inner and outer planets and $a_\mathrm{in}$ and $a_\mathrm{out}$ are their respective orbital distances from the host star.  
For planets $R_p$ = 1.4--4.0 \rearth, we use the power-law scaling $M_p=2.69R_p^{0.93}$ from  \citet{Weiss14}to convert radii to masses. 
For planets $R_p > 4.0$ \rearth we use $M_p=1.6R_p^{1.8}$ \citep{Wolfgang15}. The one exception is \ktwobfc, for which we use our measured RV masses (Section \ref{sec:RV}). We compute orbital separations in units of $R_H$:

\begin{equation}
\Delta = \frac{a_\mathrm{out}- a_\mathrm{in}}{R_H}.
\label{eqn:delta}
\end{equation}

For a two-planet system, if $\Delta < 2\sqrt{3}$ then even circular orbits are likely to be unstable on short timescales \citep{Gladman93}.  All of our \ktwo multis have $\Delta > 5$, so we have no reason to suspect that their orbits are unstable. While there are no such analytic criterion to assess the orbital stability of three planet systems, \citet{Fabrycky14} suggest $\Delta_\mathrm{in} + \Delta_\mathrm{out} > 18$ as a conservative requirement.  All four of the triple-planet systems presented here satisfy this criterion. 


\subsection{Orbital Resonances}

\begin{figure*}[ht]
\centering
\includegraphics[width=15cm]{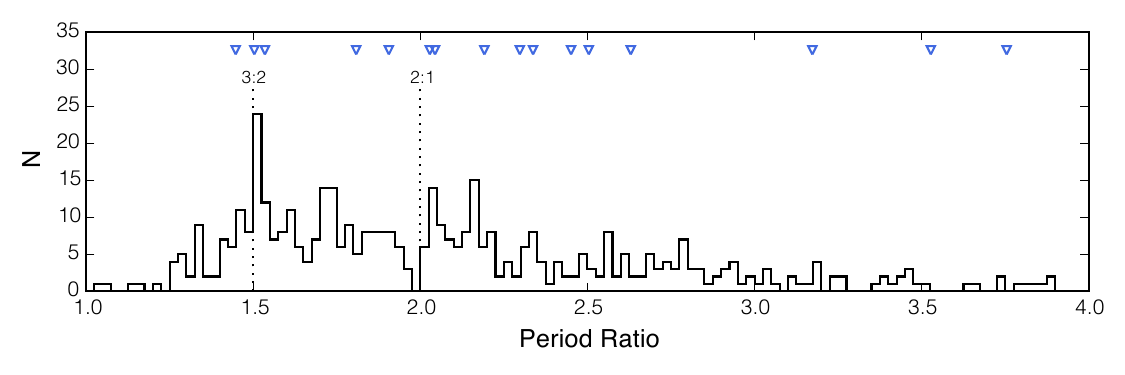} 
\caption{Histogram of the distribution of period ratios for planets from the prime \textit{Kepler} mission \citep{Fabrycky14}. In systems with three or more transiting planets, all pairs of planets are considered, not only adjacent pairs. In order to make a more direct comparison to the population probed by \ktwo, we have only shown \textit{Kepler} planets having $P < 50$ d. The period ratios of \ktwo planet pairs presented in this work are shown using blue triangles. The \ktwo distribution of period ratios is qualitatively similar to the period ratios from the Kepler prime mission.} 
\label{fig:Pratio}
\end{figure*}

In multi-planet systems, the distribution of planet orbital period ratios contains important clues regarding their formation and evolution. \citet{Fabrycky14} found that the distribution of period ratios among Kepler multis was fairly uniform. They noted, however, a slight overabundance of planet pairs just outside first order mean-motion resonances, and an underabundance of pairs just inside. \citet{Lithwick12} and \citet{Batygin13} interpreted this feature as a natural outcome of resonant pairs of planets that experience eccentricity damping. Figure \ref{fig:Pratio} shows the distribution of period ratios for planet pairs discovered during the prime Kepler mission. In order to make a more direct comparison to our \ktwo planets, we have restricted to orbital periods $<$ 50 d. We have indicated the period ratios for the planet pairs presented in this paper. While there are too few planet pairs for a detailed comparison, the distribution of period ratios is qualitatively similar: fairly uniform with a few planet pairs lying just out side the 2:1 and 3:2 MMR.

Two planet pairs, \ktwoceo bc and \ktwoihd cd, are just wide of 3:2 MMR, having $P_\mathrm{rel}$ = 1.5036 and 1.5351, respectively. Two pairs, \ktwooih bc and \ktwocib bc, orbit just outside of 2:1 resonance ($P_\mathrm{rel}$ = 2.0284, 2.0441). The TTV signals of these planet pairs will be significantly enhanced by their near-resonant orbits \citep{Holman05, Lithwick12}.

\subsection{Comparison with other studies}

Table \ref{tb:compare} compares our measured planet radii and host star radii with those published in other studies. All measurements agree within $1\sigma$.  \citet{Montet15} derive the radii of all 11 stars from photometry, yet their quoted uncertainties are often comparable to or smaller than our spectroscopic constraints. For example, for the two reddest stars (\ktwoofe, \ktwoeoh) they estimate uncertainties $\lesssim2\%$. They interpolate Dartmouth stellar evolution models, which, as the authors acknowledge, might systematically underestimate M-dwarf radii by as much as 15\% \citep{Montet15b, Newton15}. Moreover, below $\sim$0.8\msun, the scatter relative to precisely measured stellar radii is $\sim$4\% \citep{Feiden12}. For the nine hotter stars, uncertainties from our stellar characterization algorithm Specmatch are well calibrated large samples of exquisitely characterized stars \citep{Petigura15b}.  Therefore, we believe that our typical $\sim$5--10\% uncertainties are appropriate. Although the photometric derived uncertainties of \citet{Montet15} agree with our measurements, as both studies acknowledge, one should be cautious of adopting them for other analysis.  


\begin{deluxetable*}{lccccc}[h!]
\centering
\tabletypesize{\footnotesize}
\tablecaption{Comparison with other studies}
\tablehead{EPIC & $R_p$ (this study) & $R_p$ (other) & $R_\star$ (this study) & $R_\star$ (other) & Reference\tablenotemark{a}}
\startdata
\eoh{b} & \eohbRp & \monteteohbRp & \eohRstar & \monteteohRstar & \montet \\
\eoh{c} & \eohcRp & \monteteohcRp & \eohRstar & \monteteohRstar & \montet \\
\ofe{b} & \ofebRp & \montetofebRp, \crossfieldofebRp & \ofeRstar & \montetofeRstar, \crossfieldofeRstar & \montet, \crossfield \\
\ofe{c} & \ofecRp & \montetofecRp, \crossfieldofecRp & \ofeRstar & \montetofeRstar, \crossfieldofeRstar & \montet, \crossfield \\
\ofe{d} & \ofedRp & \crossfieldofedRp & \ofeRstar & \crossfieldofeRstar & \crossfield \\
\cib{b} & \cibbRp & \montetcibbRp & \cibRstar & \montetcibRstar & \montet \\
\cib{c} & \cibcRp & \montetcibcRp & \cibRstar & \montetcibRstar & \montet \\ 
\ceo{b} & \ceobRp & \montetceobRp, \armstrongceobRp, \barrosceobRp  & \ceoRstar & \montetceoRstar, \armstrongceoRstar, \barrosceoRstar & \montet, \armstrong, \barros \\
\ceo{c} & \ceocRp & \montetceocRp, \armstrongceocRp, \barrosceocRp & \ceoRstar & \montetceoRstar, \armstrongceoRstar, \barrosceoRstar & \montet, \armstrong, \barros \\
\hfo{c} & \hfocRp & \montethfocRp & \hfoRstar & \montethfoRstar & \montet \\
\coe{b} & \coebRp & \montetcoebRp & \coeRstar & \montetcoeRstar & \montet \\
\coe{c} & \coecRp & \montetcoecRp & \coeRstar & \montetcoeRstar & \montet \\
\oih{b} & \oihbRp & \petiguraoihbRp & \oihRstar & \petiguraoihRstar & \petigura \\
\oih{c} & \oihcRp & \petiguraoihcRp & \oihRstar & \petiguraoihRstar & \petigura 
\enddata
\tablenotetext{a}{Citation key: M15 = \cite{Montet15}, C15 = \cite{Crossfield15}, A15 = \cite{Armstrong15}, B15 = \cite{Barros15}, P15 = \cite{Petigura2037}}
\label{tb:compare}
\end{deluxetable*}

\acknowledgements

We thank Sam Grunblatt,  Matthew Hosek Jr., John Livingston, and Geoff Marcy for helpful discussions. We thank Lauren Weiss and Lea Hirsch for their help with observing with Keck-HIRES.  E.~S. is supported by a post-graduate scholarship from the Natural Sciences and Engineering Research Council of Canada (NSERC).  E.~A.~P.\ acknowledges support from a Hubble Fellowship grant HST-HF2-51365.001-A awarded by the Space Telescope Science Institute, which is operated by the Association of Universities for Research in Astronomy, Inc. for NASA under contract NAS 5-26555.  A.~W.~H.\ acknowledges support for our \ktwo team through a NASA Astrophysics Data Analysis Program grant.  A.~W.~H.\ and I.~J.~M.~C.\ acknowledge support from the \ktwo Guest Observer Program. E.~D.~L. received funding from the European Union Seventh Framework Programme (FP7/2007-2013) under grant agreement number 313014 (ETAEARTH). The research of J.E.S was supported by an appointment to the NASA Postdoctoral Program at NASA Ames Research Center, administered by Oak Ridge Associated Universities through a contract with NASA.B.~J.~F.\ acknowledges support from a National Science Foundation Graduate Research Fellowship under grant No. 2014184874.  This research used resources of the National Energy Research Scientific Computing Center, a DOE Office of Science User Facility  supported by the Office of Science of the U.S. Department of Energy  under Contract No. DE-AC02-05CH11231. This work made use of the SIMBAD database (operated at CDS, Strasbourg, France), NASA's Astrophysics Data System Bibliographic Services, and data products from the Two Micron All Sky Survey (2MASS), the APASS database, the SDSS-III project, and the Digitized Sky Survey. Some of the data presented in this paper were obtained from the Mikulski Archive for Space Telescopes (MAST). Support for MAST for non-HST data is provided by the NASA Office of Space Science via grant NNX09AF08G and by other grants and contracts. This research was made possible through the use of the AAVSO Photometric All-Sky Survey (APASS), funded by the Robert Martin Ayers Sciences Fund. This study benefits from use of the Robo-AO system, which was developed by collaborating partner institutions, the California Institute of Technology and the Inter-University Centre for Astronomy and Astrophysics, and with the support of the National Science Foundation under Grant Nos. AST-0906060, AST-0960343 and AST-1207891, the Mt. Cuba Astronomical Foundation and by a gift from Samuel Oschin. Ongoing science operation support of Robo-AO is provided by the California Institute of Technology and the University of Hawai`i. C.~B. acknowledges support from the Alfred P. Sloan Foundation. Some of the data presented herein were obtained at the W.~M.~Keck Observatory (which is operated as a scientific partnership among Caltech, UC, and NASA).  The authors wish to recognize and acknowledge the very significant cultural role and reverence that the summit of Maunakea has always had within the indigenous Hawaiian community.  We are most fortunate to have the opportunity to conduct observations from this mountain.

{\it Facility:} \facility{\Kepler}, \facility{\ktwo}, \facility{Keck-I (HIRES)}, \facility{Keck-II (NIRC2)}, \facility{IRTF (Spex)}, \facility{Palomar:Hale (PALM-3000/PHARO)}, \facility{PO:1.5m (Robo-AO)}

\bibliographystyle{apj}
\bibliography{references}

\begin{thebibliography}{98}
\expandafter\ifx\csname natexlab\endcsname\relax\def\natexlab#1{#1}\fi

\bibitem[{{Adams} {et~al.}(2008){Adams}, {Seager}, \&
  {Elkins-Tanton}}]{Adams08}
{Adams}, E.~R., {Seager}, S., \& {Elkins-Tanton}, L. 2008, \apj, 673, 1160

\bibitem[{{Agol} {et~al.}(2005){Agol}, {Steffen}, {Sari}, \&
  {Clarkson}}]{Agol05}
{Agol}, E., {Steffen}, J., {Sari}, R., \& {Clarkson}, W. 2005, \mnras, 359, 567

\bibitem[{{Ahn} {et~al.}(2012){Ahn}, {Alexandroff}, {Allende Prieto},
  {Anderson}, {Anderton}, {Andrews}, {Aubourg}, {Bailey}, {Balbinot}, {Barnes},
  \& et~al.}]{Ahn12}
{Ahn}, C.~P., {et~al.} 2012, \apjs, 203, 21

\bibitem[{{Almenara} {et~al.}(2015){Almenara}, {Astudillo-Defru}, {Bonfils},
  {Forveille}, {Santerne}, {Albrecht}, {Barros}, {Bouchy}, {Delfosse},
  {Demangeon}, {Diaz}, {H{\'e}brard}, {Mayor}, {Neves}, {Rojo}, {Santos}, \&
  {W{\"u}nsche}}]{Almenara15}
{Almenara}, J.~M., {et~al.} 2015, \aap, 581, L7

\bibitem[{{Armstrong} {et~al.}(2015){Armstrong}, {Santerne}, {Veras}, {Barros},
  {Demangeon}, {Lillo-Box}, {McCormac}, {Osborn}, {Tsantaki}, {Almenara},
  {Barrado}, {Boisse}, {Bonomo}, {Brown}, {Bruno}, {Rey Cerda}, {Courcol},
  {Deleuil}, {D{\'{\i}}az}, {Doyle}, {H{\'e}brard}, {Kirk}, {Lam}, {Pollacco},
  {Rajpurohit}, {Spake}, \& {Walker}}]{Armstrong15}
{Armstrong}, D.~J., {et~al.} 2015, \aap, 582, A33

\bibitem[{{Baranec} {et~al.}(2013){Baranec}, {Riddle}, {Law}, {Ramaprakash},
  {Tendulkar}, {Bui}, {Burse}, {Chordia}, {Das}, {Davis}, {Dekany}, {Kasliwal},
  {Kulkarni}, {Morton}, {Ofek}, \& {Punnadi}}]{Baranec13}
{Baranec}, C., {et~al.} 2013, Journal of Visualized Experiments, 72, e50021

\bibitem[{{Baranec} {et~al.}(2014){Baranec}, {Riddle}, {Law}, {Ramaprakash},
  {Tendulkar}, {Hogstrom}, {Bui}, {Burse}, {Chordia}, {Das}, {Dekany},
  {Kulkarni}, \& {Punnadi}}]{Baranec14}
---. 2014, \apjl, 790, L8

\bibitem[{{Barros} {et~al.}(2015){Barros}, {Almenara}, {Demangeon}, {Tsantaki},
  {Santerne}, {Armstrong}, {Barrado}, {Brown}, {Deleuil}, {Lillo-Box},
  {Osborn}, {Pollacco}, {Abe}, {Andre}, {Bendjoya}, {Boisse}, {Bonomo},
  {Bouchy}, {Bruno}, {Cerda}, {Courcol}, {D{\'{\i}}az}, {H{\'e}brard}, {Kirk},
  {Lachuri{\'e}}, {Lam}, {Martinez}, {McCormac}, {Moutou}, {Rajpurohit},
  {Rivet}, {Spake}, {Suarez}, {Toublanc}, \& {Walker}}]{Barros15}
{Barros}, S.~C.~C., {et~al.} 2015, \mnras, 454, 4267

\bibitem[{{Batalha} {et~al.}(2013){Batalha}, {Rowe}, {Bryson}, {Barclay},
  {Burke}, {Caldwell}, {Christiansen}, {Mullally}, {Thompson}, {Brown},
  {Dupree}, {Fabrycky}, {Ford}, {Fortney}, {Gilliland}, {Isaacson}, {Latham},
  {Marcy}, {Quinn}, {Ragozzine}, {Shporer}, {Borucki}, {Ciardi}, {Gautier},
  {Haas}, {Jenkins}, {Koch}, {Lissauer}, {Rapin}, {Basri}, {Boss}, {Buchhave},
  {Carter}, {Charbonneau}, {Christensen-Dalsgaard}, {Clarke}, {Cochran},
  {Demory}, {Desert}, {Devore}, {Doyle}, {Esquerdo}, {Everett}, {Fressin},
  {Geary}, {Girouard}, {Gould}, {Hall}, {Holman}, {Howard}, {Howell},
  {Ibrahim}, {Kinemuchi}, {Kjeldsen}, {Klaus}, {Li}, {Lucas}, {Meibom},
  {Morris}, {Pr{\v s}a}, {Quintana}, {Sanderfer}, {Sasselov}, {Seader},
  {Smith}, {Steffen}, {Still}, {Stumpe}, {Tarter}, {Tenenbaum}, {Torres},
  {Twicken}, {Uddin}, {Van Cleve}, {Walkowicz}, \& {Welsh}}]{Batalha13}
{Batalha}, N.~M., {et~al.} 2013, \apjs, 204, 24

\bibitem[{{Batygin} \& {Morbidelli}(2013)}]{Batygin13}
{Batygin}, K., \& {Morbidelli}, A. 2013, \aj, 145, 1

\bibitem[{{Beichman} {et~al.}(2016){Beichman}, {Livingston}, {Werner},
  {Gorjian}, {Krick}, {Deck}, {Knutson}, {Wong}, {Petigura}, {Christiansen},
  {Ciardi}, {Greene}, {Schlieder}, {Line}, {Crossfield}, {Howard}, \&
  {Sinukoff}}]{Beichman16}
{Beichman}, C., {et~al.} 2016, \apj, 822, 39

\bibitem[{{Borucki} {et~al.}(2011){Borucki}, {Koch}, {Basri}, {Batalha},
  {Brown}, {Bryson}, {Caldwell}, {Christensen-Dalsgaard}, {Cochran}, {DeVore},
  {Dunham}, {Gautier}, {Geary}, {Gilliland}, {Gould}, {Howell}, {Jenkins},
  {Latham}, {Lissauer}, {Marcy}, {Rowe}, {Sasselov}, {Boss}, {Charbonneau},
  {Ciardi}, {Doyle}, {Dupree}, {Ford}, {Fortney}, {Holman}, {Seager},
  {Steffen}, {Tarter}, {Welsh}, {Allen}, {Buchhave}, {Christiansen}, {Clarke},
  {Das}, {D{\'e}sert}, {Endl}, {Fabrycky}, {Fressin}, {Haas}, {Horch},
  {Howard}, {Isaacson}, {Kjeldsen}, {Kolodziejczak}, {Kulesa}, {Li}, {Lucas},
  {Machalek}, {McCarthy}, {MacQueen}, {Meibom}, {Miquel}, {Prsa}, {Quinn},
  {Quintana}, {Ragozzine}, {Sherry}, {Shporer}, {Tenenbaum}, {Torres},
  {Twicken}, {Van Cleve}, {Walkowicz}, {Witteborn}, \& {Still}}]{Borucki11}
{Borucki}, W.~J., {et~al.} 2011, \apj, 736, 19

\bibitem[{{Boyajian} {et~al.}(2012){Boyajian}, {von Braun}, {van Belle},
  {McAlister}, {ten Brummelaar}, {Kane}, {Muirhead}, {Jones}, {White},
  {Schaefer}, {Ciardi}, {Henry}, {L{\'o}pez-Morales}, {Ridgway}, {Gies}, {Jao},
  {Rojas-Ayala}, {Parks}, {Sturmann}, {Sturmann}, {Turner}, {Farrington},
  {Goldfinger}, \& {Berger}}]{Boyajian12}
{Boyajian}, T.~S., {et~al.} 2012, \apj, 757, 112

\bibitem[{{Burke} {et~al.}(2014){Burke}, {Bryson}, {Mullally}, {Rowe},
  {Christiansen}, {Thompson}, {Coughlin}, {Haas}, {Batalha}, {Caldwell},
  {Jenkins}, {Still}, {Barclay}, {Borucki}, {Chaplin}, {Ciardi}, {Clarke},
  {Cochran}, {Demory}, {Esquerdo}, {Gautier}, {Gilliland}, {Girouard}, {Havel},
  {Henze}, {Howell}, {Huber}, {Latham}, {Li}, {Morehead}, {Morton}, {Pepper},
  {Quintana}, {Ragozzine}, {Seader}, {Shah}, {Shporer}, {Tenenbaum}, {Twicken},
  \& {Wolfgang}}]{Burke14}
{Burke}, C.~J., {et~al.} 2014, \apjs, 210, 19

\bibitem[{{Butler} {et~al.}(1996){Butler}, {Marcy}, {Williams}, {McCarthy},
  {Dosanjh}, \& {Vogt}}]{Butler96}
{Butler}, R.~P., {Marcy}, G.~W., {Williams}, E., {McCarthy}, C., {Dosanjh}, P.,
  \& {Vogt}, S.~S. 1996, \pasp, 108, 500

\bibitem[{{Carter} {et~al.}(2012){Carter}, {Agol}, {Chaplin}, {Basu},
  {Bedding}, {Buchhave}, {Christensen-Dalsgaard}, {Deck}, {Elsworth},
  {Fabrycky}, {Ford}, {Fortney}, {Hale}, {Handberg}, {Hekker}, {Holman},
  {Huber}, {Karoff}, {Kawaler}, {Kjeldsen}, {Lissauer}, {Lopez}, {Lund},
  {Lundkvist}, {Metcalfe}, {Miglio}, {Rogers}, {Stello}, {Borucki}, {Bryson},
  {Christiansen}, {Cochran}, {Geary}, {Gilliland}, {Haas}, {Hall}, {Howard},
  {Jenkins}, {Klaus}, {Koch}, {Latham}, {MacQueen}, {Sasselov}, {Steffen},
  {Twicken}, \& {Winn}}]{Carter12}
{Carter}, J.~A., {et~al.} 2012, Science, 337, 556

\bibitem[{{Ciardi} {et~al.}(2015){Ciardi}, {Beichman}, {Horch}, \&
  {Howell}}]{Ciardi15}
{Ciardi}, D.~R., {Beichman}, C.~A., {Horch}, E.~P., \& {Howell}, S.~B. 2015,
  \apj, 805, 16

\bibitem[{{Claret} {et~al.}(2012){Claret}, {Hauschildt}, \& {Witte}}]{Claret12}
{Claret}, A., {Hauschildt}, P.~H., \& {Witte}, S. 2012, \aap, 546, A14

\bibitem[{{Claret} {et~al.}(2013){Claret}, {Hauschildt}, \& {Witte}}]{Claret13}
---. 2013, \aap, 552, A16

\bibitem[{{Coelho} {et~al.}(2005){Coelho}, {Barbuy}, {Mel{\'e}ndez},
  {Schiavon}, \& {Castilho}}]{Coelho05}
{Coelho}, P., {Barbuy}, B., {Mel{\'e}ndez}, J., {Schiavon}, R.~P., \&
  {Castilho}, B.~V. 2005, \aap, 443, 735

\bibitem[{{Crossfield} {et~al.}(2015){Crossfield}, {Petigura}, {Schlieder},
  {Howard}, {Fulton}, {Aller}, {Ciardi}, {L{\'e}pine}, {Barclay}, {de Pater},
  {de Kleer}, {Quintana}, {Christiansen}, {Schlafly}, {Kaltenegger}, {Crepp},
  {Henning}, {Obermeier}, {Deacon}, {Weiss}, {Isaacson}, {Hansen}, {Liu},
  {Greene}, {Howell}, {Barman}, \& {Mordasini}}]{Crossfield15}
{Crossfield}, I.~J.~M., {et~al.} 2015, \apj, 804, 10

\bibitem[{{Cushing} {et~al.}(2005){Cushing}, {Rayner}, \& {Vacca}}]{Cushing05}
{Cushing}, M.~C., {Rayner}, J.~T., \& {Vacca}, W.~D. 2005, \apj, 623, 1115

\bibitem[{{Cushing} {et~al.}(2004){Cushing}, {Vacca}, \& {Rayner}}]{Cushing04}
{Cushing}, M.~C., {Vacca}, W.~D., \& {Rayner}, J.~T. 2004, \pasp, 116, 362

\bibitem[{{Dai} {et~al.}(2016){Dai}, {Winn}, {Albrecht}, {Arriagada},
  {Bieryla}, {Butler}, {Crane}, {Hirano}, {Johnson}, {Kiilerich}, {Latham},
  {Narita}, {Nowak}, {Palle}, {Ribas}, {Rogers}, {Sanchis-Ojeda}, {Shectman},
  {Teske}, {Thompson}, {Van Eylen}, {Vanderburg}, {Wittenmyer}, \&
  {Yu}}]{Dai16}
{Dai}, F., {et~al.} 2016, ArXiv e-prints

\bibitem[{{Dekany} {et~al.}(2013){Dekany}, {Roberts}, {Burruss}, {Bouchez},
  {Truong}, {Baranec}, {Guiwits}, {Hale}, {Angione}, {Trinh}, {Zolkower},
  {Shelton}, {Palmer}, {Henning}, {Croner}, {Troy}, {McKenna}, {Tesch},
  {Hildebrandt}, \& {Milburn}}]{Dekany13}
{Dekany}, R., {et~al.} 2013, \apj, 776, 130

\bibitem[{{Dotter} {et~al.}(2008){Dotter}, {Chaboyer}, {Jevremovi{\'c}},
  {Kostov}, {Baron}, \& {Ferguson}}]{Dotter08}
{Dotter}, A., {Chaboyer}, B., {Jevremovi{\'c}}, D., {Kostov}, V., {Baron}, E.,
  \& {Ferguson}, J.~W. 2008, \apjs, 178, 89

\bibitem[{{Dressing} {et~al.}(2015){Dressing}, {Charbonneau}, {Dumusque},
  {Gettel}, {Pepe}, {Collier Cameron}, {Latham}, {Molinari}, {Udry}, {Affer},
  {Bonomo}, {Buchhave}, {Cosentino}, {Figueira}, {Fiorenzano}, {Harutyunyan},
  {Haywood}, {Johnson}, {Lopez-Morales}, {Lovis}, {Malavolta}, {Mayor},
  {Micela}, {Motalebi}, {Nascimbeni}, {Phillips}, {Piotto}, {Pollacco},
  {Queloz}, {Rice}, {Sasselov}, {S{\'e}gransan}, {Sozzetti}, {Szentgyorgyi}, \&
  {Watson}}]{Dressing15}
{Dressing}, C.~D., {et~al.} 2015, \apj, 800, 135

\bibitem[{{Fabrycky} {et~al.}(2014){Fabrycky}, {Lissauer}, {Ragozzine}, {Rowe},
  {Steffen}, {Agol}, {Barclay}, {Batalha}, {Borucki}, {Ciardi}, {Ford},
  {Gautier}, {Geary}, {Holman}, {Jenkins}, {Li}, {Morehead}, {Morris},
  {Shporer}, {Smith}, {Still}, \& {Van Cleve}}]{Fabrycky14}
{Fabrycky}, D.~C., {et~al.} 2014, \apj, 790, 146

\bibitem[{{Fang} \& {Margot}(2012)}]{Fang12}
{Fang}, J., \& {Margot}, J.-L. 2012, \apj, 761, 92

\bibitem[{{Fang} \& {Margot}(2013)}]{Fang13}
---. 2013, \apj, 767, 115

\bibitem[{{Feiden} \& {Chaboyer}(2012)}]{Feiden12}
{Feiden}, G.~A., \& {Chaboyer}, B. 2012, \apj, 757, 42

\bibitem[{{Foreman-Mackey} {et~al.}(2013){Foreman-Mackey}, {Hogg}, {Lang}, \&
  {Goodman}}]{Foreman-Mackey13}
{Foreman-Mackey}, D., {Hogg}, D.~W., {Lang}, D., \& {Goodman}, J. 2013, \pasp,
  125, 306

\bibitem[{{Foreman-Mackey} {et~al.}(2015){Foreman-Mackey}, {Montet}, {Hogg},
  {Morton}, {Wang}, \& {Sch{\"o}lkopf}}]{Foreman-Mackey15}
{Foreman-Mackey}, D., {Montet}, B.~T., {Hogg}, D.~W., {Morton}, T.~D., {Wang},
  D., \& {Sch{\"o}lkopf}, B. 2015, \apj, 806, 215

\bibitem[{{Fressin} {et~al.}(2013){Fressin}, {Torres}, {Charbonneau}, {Bryson},
  {Christiansen}, {Dressing}, {Jenkins}, {Walkowicz}, \& {Batalha}}]{Fressin13}
{Fressin}, F., {et~al.} 2013, \apj, 766, 81

\bibitem[{Gelman \& Rubin(1992)}]{Gelman92}
Gelman, A., \& Rubin, D.~B. 1992, Statist. Sci., 7, 457

\bibitem[{{Gladman}(1993)}]{Gladman93}
{Gladman}, B. 1993, \icarus, 106, 247

\bibitem[{{Hansen} \& {Murray}(2012)}]{Hansen12}
{Hansen}, B.~M.~S., \& {Murray}, N. 2012, \apj, 751, 158

\bibitem[{{Holman} \& {Murray}(2005)}]{Holman05}
{Holman}, M.~J., \& {Murray}, N.~W. 2005, Science, 307, 1288

\bibitem[{{Howard} {et~al.}(2009){Howard}, {Johnson}, {Marcy}, {Fischer},
  {Wright}, {Henry}, {Giguere}, {Isaacson}, {Valenti}, {Anderson}, \&
  {Piskunov}}]{Howard09}
{Howard}, A.~W., {et~al.} 2009, \apj, 696, 75

\bibitem[{{Howard} {et~al.}(2010){Howard}, {Johnson}, {Marcy}, {Fischer},
  {Wright}, {Bernat}, {Henry}, {Peek}, {Isaacson}, {Apps}, {Endl}, {Cochran},
  {Valenti}, {Anderson}, \& {Piskunov}}]{Howard10}
---. 2010, \apj, 721, 1467

\bibitem[{{Howard} {et~al.}(2012){Howard}, {Marcy}, {Bryson}, {Jenkins},
  {Rowe}, {Batalha}, {Borucki}, {Koch}, {Dunham}, {Gautier}, {Van Cleve},
  {Cochran}, {Latham}, {Lissauer}, {Torres}, {Brown}, {Gilliland}, {Buchhave},
  {Caldwell}, {Christensen-Dalsgaard}, {Ciardi}, {Fressin}, {Haas}, {Howell},
  {Kjeldsen}, {Seager}, {Rogers}, {Sasselov}, {Steffen}, {Basri},
  {Charbonneau}, {Christiansen}, {Clarke}, {Dupree}, {Fabrycky}, {Fischer},
  {Ford}, {Fortney}, {Tarter}, {Girouard}, {Holman}, {Johnson}, {Klaus},
  {Machalek}, {Moorhead}, {Morehead}, {Ragozzine}, {Tenenbaum}, {Twicken},
  {Quinn}, {Isaacson}, {Shporer}, {Lucas}, {Walkowicz}, {Welsh}, {Boss},
  {Devore}, {Gould}, {Smith}, {Morris}, {Prsa}, {Morton}, {Still}, {Thompson},
  {Mullally}, {Endl}, \& {MacQueen}}]{Howard12}
---. 2012, \apjs, 201, 15

\bibitem[{{Howard} {et~al.}(2014){Howard}, {Marcy}, {Fischer}, {Isaacson},
  {Muirhead}, {Henry}, {Boyajian}, {von Braun}, {Becker}, {Wright}, \&
  {Johnson}}]{Howard14}
---. 2014, \apj, 794, 51

\bibitem[{{Howell} {et~al.}(2014){Howell}, {Sobeck}, {Haas}, {Still},
  {Barclay}, {Mullally}, {Troeltzsch}, {Aigrain}, {Bryson}, {Caldwell},
  {Chaplin}, {Cochran}, {Huber}, {Marcy}, {Miglio}, {Najita}, {Smith},
  {Twicken}, \& {Fortney}}]{Howell14}
{Howell}, S.~B., {et~al.} 2014, \pasp, 126, 398

\bibitem[{{Isaacson} \& {Fischer}(2010)}]{Isaacson10}
{Isaacson}, H., \& {Fischer}, D. 2010, \apj, 725, 875

\bibitem[{{Kaiser} {et~al.}(2010){Kaiser}, {Burgett}, {Chambers}, {Denneau},
  {Heasley}, {Jedicke}, {Magnier}, {Morgan}, {Onaka}, \& {Tonry}}]{Kaiser10}
{Kaiser}, N., {et~al.} 2010, in Society of Photo-Optical Instrumentation
  Engineers (SPIE) Conference Series, Vol. 7733, Society of Photo-Optical
  Instrumentation Engineers (SPIE) Conference Series, 0

\bibitem[{{Kolbl} {et~al.}(2015){Kolbl}, {Marcy}, {Isaacson}, \&
  {Howard}}]{Kolbl15}
{Kolbl}, R., {Marcy}, G.~W., {Isaacson}, H., \& {Howard}, A.~W. 2015, \aj, 149,
  18

\bibitem[{{Kraus} \& {Hillenbrand}(2007)}]{Kraus07}
{Kraus}, A.~L., \& {Hillenbrand}, L.~A. 2007, \aj, 134, 2340

\bibitem[{{Law} {et~al.}(2014){Law}, {Morton}, {Baranec}, {Riddle},
  {Ravichandran}, {Ziegler}, {Johnson}, {Tendulkar}, {Bui}, {Burse}, {Das},
  {Dekany}, {Kulkarni}, {Punnadi}, \& {Ramaprakash}}]{Law14}
{Law}, N.~M., {et~al.} 2014, \apj, 791, 35

\bibitem[{{Liddle}(2007)}]{Liddle07}
{Liddle}, A.~R. 2007, \mnras, 377, L74

\bibitem[{{Lissauer} {et~al.}(2011{\natexlab{a}}){Lissauer}, {Fabrycky},
  {Ford}, {Borucki}, {Fressin}, {Marcy}, {Orosz}, {Rowe}, {Torres}, {Welsh},
  {Batalha}, {Bryson}, {Buchhave}, {Caldwell}, {Carter}, {Charbonneau},
  {Christiansen}, {Cochran}, {Desert}, {Dunham}, {Fanelli}, {Fortney},
  {Gautier}, {Geary}, {Gilliland}, {Haas}, {Hall}, {Holman}, {Koch}, {Latham},
  {Lopez}, {McCauliff}, {Miller}, {Morehead}, {Quintana}, {Ragozzine},
  {Sasselov}, {Short}, \& {Steffen}}]{Lissauer11}
{Lissauer}, J.~J., {et~al.} 2011{\natexlab{a}}, \nat, 470, 53

\bibitem[{{Lissauer} {et~al.}(2011{\natexlab{b}}){Lissauer}, {Ragozzine},
  {Fabrycky}, {Steffen}, {Ford}, {Jenkins}, {Shporer}, {Holman}, {Rowe},
  {Quintana}, {Batalha}, {Borucki}, {Bryson}, {Caldwell}, {Carter}, {Ciardi},
  {Dunham}, {Fortney}, {Gautier}, {Howell}, {Koch}, {Latham}, {Marcy},
  {Morehead}, \& {Sasselov}}]{Lissauer11b}
---. 2011{\natexlab{b}}, \apjs, 197, 8

\bibitem[{{Lissauer} {et~al.}(2012){Lissauer}, {Marcy}, {Rowe}, {Bryson},
  {Adams}, {Buchhave}, {Ciardi}, {Cochran}, {Fabrycky}, {Ford}, {Fressin},
  {Geary}, {Gilliland}, {Holman}, {Howell}, {Jenkins}, {Kinemuchi}, {Koch},
  {Morehead}, {Ragozzine}, {Seader}, {Tanenbaum}, {Torres}, \&
  {Twicken}}]{Lissauer12}
---. 2012, \apj, 750, 112

\bibitem[{{Lissauer} {et~al.}(2014){Lissauer}, {Marcy}, {Bryson}, {Rowe},
  {Jontof-Hutter}, {Agol}, {Borucki}, {Carter}, {Ford}, {Gilliland}, {Kolbl},
  {Star}, {Steffen}, \& {Torres}}]{Lissauer14}
---. 2014, \apj, 784, 44

\bibitem[{{Lithwick} \& {Wu}(2012)}]{Lithwick12}
{Lithwick}, Y., \& {Wu}, Y. 2012, \apjl, 756, L11

\bibitem[{{Mann} {et~al.}(2013{\natexlab{a}}){Mann}, {Gaidos}, \&
  {Ansdell}}]{Mann13_temp}
{Mann}, A.~W., {Gaidos}, E., \& {Ansdell}, M. 2013{\natexlab{a}}, \apj, 779,
  188

\bibitem[{{Mann} {et~al.}(2013{\natexlab{b}}){Mann}, {Gaidos}, {Kraus}, \&
  {Hilton}}]{Mann13_metal}
{Mann}, A.~W., {Gaidos}, E., {Kraus}, A., \& {Hilton}, E.~J.
  2013{\natexlab{b}}, \apj, 770, 43

\bibitem[{{Marcus} {et~al.}(2010){Marcus}, {Sasselov}, {Hernquist}, \&
  {Stewart}}]{Marcus10}
{Marcus}, R.~A., {Sasselov}, D., {Hernquist}, L., \& {Stewart}, S.~T. 2010,
  \apjl, 712, L73

\bibitem[{{Marcy} \& {Butler}(1992)}]{Marcy92}
{Marcy}, G.~W., \& {Butler}, R.~P. 1992, \pasp, 104, 270

\bibitem[{{Meunier} {et~al.}(2010){Meunier}, {Desort}, \&
  {Lagrange}}]{Meunier10}
{Meunier}, N., {Desort}, M., \& {Lagrange}, A.-M. 2010, \aap, 512, A39

\bibitem[{{Middelkoop}(1982)}]{Middelkoop82}
{Middelkoop}, F. 1982, \aap, 107, 31

\bibitem[{{Montet} {et~al.}(2015{\natexlab{a}}){Montet}, {Johnson}, {Muirhead},
  {Villar}, {Vassallo}, {Baranec}, {Law}, {Riddle}, {Marcy}, {Howard}, \&
  {Isaacson}}]{Montet15b}
{Montet}, B.~T., {et~al.} 2015{\natexlab{a}}, \apj, 800, 134

\bibitem[{{Montet} {et~al.}(2015{\natexlab{b}}){Montet}, {Morton},
  {Foreman-Mackey}, {Johnson}, {Hogg}, {Bowler}, {Latham}, {Bieryla}, \&
  {Mann}}]{Montet15}
---. 2015{\natexlab{b}}, \apj, 809, 25

\bibitem[{{Morton}(2015)}]{Morton15}
{Morton}, T.~D. 2015, {isochrones: Stellar model grid package}, Astrophysics
  Source Code Library

\bibitem[{{Morton} \& {Johnson}(2011)}]{Morton11}
{Morton}, T.~D., \& {Johnson}, J.~A. 2011, \apj, 738, 170

\bibitem[{{Mullally} {et~al.}(2015){Mullally}, {Coughlin}, {Thompson}, {Rowe},
  {Burke}, {Latham}, {Batalha}, {Bryson}, {Christiansen}, {Henze}, {Ofir},
  {Quarles}, {Shporer}, {Van Eylen}, {Van Laerhoven}, {Shah}, {Wolfgang},
  {Chaplin}, {Xie}, {Akeson}, {Argabright}, {Bachtell}, {Barclay}, {Borucki},
  {Caldwell}, {Campbell}, {Catanzarite}, {Cochran}, {Duren}, {Fleming},
  {Fraquelli}, {Girouard}, {Haas}, {He{\l}miniak}, {Howell}, {Huber}, {Larson},
  {Gautier}, {Jenkins}, {Li}, {Lissauer}, {McArthur}, {Miller}, {Morris},
  {Patil-Sabale}, {Plavchan}, {Putnam}, {Quintana}, {Ramirez}, {Silva Aguirre},
  {Seader}, {Smith}, {Steffen}, {Stewart}, {Stober}, {Still}, {Tenenbaum},
  {Troeltzsch}, {Twicken}, \& {Zamudio}}]{Mullally15}
{Mullally}, F., {et~al.} 2015, \apjs, 217, 31

\bibitem[{{Narita} {et~al.}(2015){Narita}, {Hirano}, {Fukui}, {Hori},
  {Sanchis-Ojeda}, {Winn}, {Ryu}, {Kusakabe}, {Kudo}, {Onitsuka}, {Delrez},
  {Gillon}, {Jehin}, {McCormac}, {Holman}, {Izumiura}, {Takeda}, {Tamura}, \&
  {Yanagisawa}}]{Narita15}
{Narita}, N., {et~al.} 2015, \apj, 815, 47

\bibitem[{{Newton} {et~al.}(2015){Newton}, {Charbonneau}, {Irwin}, \&
  {Mann}}]{Newton15}
{Newton}, E.~R., {Charbonneau}, D., {Irwin}, J., \& {Mann}, A.~W. 2015, \apj,
  800, 85

\bibitem[{{Noyes} {et~al.}(1984){Noyes}, {Hartmann}, {Baliunas}, {Duncan}, \&
  {Vaughan}}]{Noyes84}
{Noyes}, R.~W., {Hartmann}, L.~W., {Baliunas}, S.~L., {Duncan}, D.~K., \&
  {Vaughan}, A.~H. 1984, \apj, 279, 763

\bibitem[{{Petigura}(2015)}]{Petigura15b}
{Petigura}, E.~A. 2015, PhD thesis, University of California, Berkeley,
  arXiv:1510.03902

\bibitem[{{Petigura} {et~al.}(2013){Petigura}, {Howard}, \&
  {Marcy}}]{Petigura13}
{Petigura}, E.~A., {Howard}, A.~W., \& {Marcy}, G.~W. 2013, Proceedings of the
  National Academy of Science, 110, 19273

\bibitem[{{Petigura} \& {Marcy}(2012)}]{Petigura12}
{Petigura}, E.~A., \& {Marcy}, G.~W. 2012, \pasp, 124, 1073

\bibitem[{{Petigura} {et~al.}(2015){Petigura}, {Schlieder}, {Crossfield},
  {Howard}, {Deck}, {Ciardi}, {Sinukoff}, {Allers}, {Best}, {Liu}, {Beichman},
  {Isaacson}, {Hansen}, \& {L{\'e}pine}}]{Petigura15}
{Petigura}, E.~A., {et~al.} 2015, \apj, 811, 102

\bibitem[{{Petigura} {et~al.}(2016){Petigura}, {Howard}, {Lopez}, {Deck},
  {Fulton}, {Crossfield}, {Ciardi}, {Chiang}, {Lee}, {Isaacson}, {Beichman},
  {Hansen}, {Schlieder}, \& {Sinukoff}}]{Petigura2037}
---. 2016, \apj, 818, 36

\bibitem[{{Rayner} {et~al.}(2009){Rayner}, {Cushing}, \& {Vacca}}]{Rayner09}
{Rayner}, J.~T., {Cushing}, M.~C., \& {Vacca}, W.~D. 2009, \apjs, 185, 289

\bibitem[{{Rayner} {et~al.}(2003){Rayner}, {Toomey}, {Onaka}, {Denault},
  {Stahlberger}, {Vacca}, {Cushing}, \& {Wang}}]{Rayner03}
{Rayner}, J.~T., {Toomey}, D.~W., {Onaka}, P.~M., {Denault}, A.~J.,
  {Stahlberger}, W.~E., {Vacca}, W.~D., {Cushing}, M.~C., \& {Wang}, S. 2003,
  \pasp, 115, 362

\bibitem[{{Rojas-Ayala} {et~al.}(2012){Rojas-Ayala}, {Covey}, {Muirhead}, \&
  {Lloyd}}]{Rojas12}
{Rojas-Ayala}, B., {Covey}, K.~R., {Muirhead}, P.~S., \& {Lloyd}, J.~P. 2012,
  \apj, 748, 93

\bibitem[{{Rowe} {et~al.}(2014){Rowe}, {Bryson}, {Marcy}, {Lissauer},
  {Jontof-Hutter}, {Mullally}, {Gilliland}, {Issacson}, {Ford}, {Howell},
  {Borucki}, {Haas}, {Huber}, {Steffen}, {Thompson}, {Quintana}, {Barclay},
  {Still}, {Fortney}, {Gautier}, {Hunter}, {Caldwell}, {Ciardi}, {Devore},
  {Cochran}, {Jenkins}, {Agol}, {Carter}, \& {Geary}}]{Rowe14}
{Rowe}, J.~F., {et~al.} 2014, \apj, 784, 45

\bibitem[{{Rowe} {et~al.}(2015){Rowe}, {Coughlin}, {Antoci}, {Barclay},
  {Batalha}, {Borucki}, {Burke}, {Bryson}, {Caldwell}, {Campbell},
  {Catanzarite}, {Christiansen}, {Cochran}, {Gilliland}, {Girouard}, {Haas},
  {He{\l}miniak}, {Henze}, {Hoffman}, {Howell}, {Huber}, {Hunter},
  {Jang-Condell}, {Jenkins}, {Klaus}, {Latham}, {Li}, {Lissauer}, {McCauliff},
  {Morris}, {Mullally}, {Ofir}, {Quarles}, {Quintana}, {Sabale}, {Seader},
  {Shporer}, {Smith}, {Steffen}, {Still}, {Tenenbaum}, {Thompson}, {Twicken},
  {Van Laerhoven}, {Wolfgang}, \& {Zamudio}}]{Rowe15}
---. 2015, \apjs, 217, 16

\bibitem[{{Schmitt} {et~al.}(2014){Schmitt}, {Wang}, {Fischer}, {Jek},
  {Moriarty}, {Boyajian}, {Schwamb}, {Lintott}, {Lynn}, {Smith}, {Parrish},
  {Schawinski}, {Simpson}, {LaCourse}, {Omohundro}, {Winarski}, {Goodman},
  {Jebson}, {Schwengeler}, {Paterson}, {Sejpka}, {Terentev}, {Jacobs},
  {Alsaadi}, {Bailey}, {Ginman}, {Granado}, {Vonstad Guttormsen}, {Mallia},
  {Papillon}, {Rossi}, \& {Socolovsky}}]{Schmitt14}
{Schmitt}, J.~R., {et~al.} 2014, \aj, 148, 28

\bibitem[{Schwarz(1978)}]{Schwarz78}
Schwarz, G. 1978, Ann. Statist., 6, 461

\bibitem[{{Seager} {et~al.}(2007){Seager}, {Kuchner}, {Hier-Majumder}, \&
  {Militzer}}]{Seager07}
{Seager}, S., {Kuchner}, M., {Hier-Majumder}, C.~A., \& {Militzer}, B. 2007,
  \apj, 669, 1279

\bibitem[{{Southworth}(2011)}]{Southworth11}
{Southworth}, J. 2011, \mnras, 417, 2166

\bibitem[{{Southworth} {et~al.}(2004){Southworth}, {Maxted}, \&
  {Smalley}}]{Southworth04}
{Southworth}, J., {Maxted}, P.~F.~L., \& {Smalley}, B. 2004, \mnras, 351, 1277

\bibitem[{{Terrien} {et~al.}(2012){Terrien}, {Mahadevan}, {Bender},
  {Deshpande}, {Ramsey}, \& {Bochanski}}]{Terrien12}
{Terrien}, R.~C., {Mahadevan}, S., {Bender}, C.~F., {Deshpande}, R., {Ramsey},
  L.~W., \& {Bochanski}, J.~J. 2012, \apjl, 747, L38

\bibitem[{{Vacca} {et~al.}(2003){Vacca}, {Cushing}, \& {Rayner}}]{Vacca03}
{Vacca}, W.~D., {Cushing}, M.~C., \& {Rayner}, J.~T. 2003, \pasp, 115, 389

\bibitem[{{Valencia} {et~al.}(2013){Valencia}, {Guillot}, {Parmentier}, \&
  {Freedman}}]{Valencia13}
{Valencia}, D., {Guillot}, T., {Parmentier}, V., \& {Freedman}, R.~S. 2013,
  \apj, 775, 10

\bibitem[{{Valenti} {et~al.}(1995){Valenti}, {Butler}, \& {Marcy}}]{Valenti95}
{Valenti}, J.~A., {Butler}, R.~P., \& {Marcy}, G.~W. 1995, \pasp, 107, 966

\bibitem[{{Vanderburg} {et~al.}(2015){Vanderburg}, {Montet}, {Johnson},
  {Buchhave}, {Zeng}, {Pepe}, {Collier Cameron}, {Latham}, {Molinari}, {Udry},
  {Lovis}, {Matthews}, {Cameron}, {Law}, {Bowler}, {Angus}, {Baranec},
  {Bieryla}, {Boschin}, {Charbonneau}, {Cosentino}, {Dumusque}, {Figueira},
  {Guenther}, {Harutyunyan}, {Hellier}, {Kuschnig}, {Lopez-Morales}, {Mayor},
  {Micela}, {Moffat}, {Pedani}, {Phillips}, {Piotto}, {Pollacco}, {Queloz},
  {Rice}, {Riddle}, {Rowe}, {Rucinski}, {Sasselov}, {S{\'e}gransan},
  {Sozzetti}, {Szentgyorgyi}, {Watson}, \& {Weiss}}]{Vanderburg15}
{Vanderburg}, A., {et~al.} 2015, \apj, 800, 59

\bibitem[{{Vanderburg} {et~al.}(2016){Vanderburg}, {Latham}, {Buchhave},
  {Bieryla}, {Berlind}, {Calkins}, {Esquerdo}, {Welsh}, \&
  {Johnson}}]{Vanderburg15b}
---. 2016, \apjs, 222, 14

\bibitem[{{Vogt} {et~al.}(1994){Vogt}, {Allen}, {Bigelow}, {Bresee}, {Brown},
  {Cantrall}, {Conrad}, {Couture}, {Delaney}, {Epps}, {Hilyard}, {Hilyard},
  {Horn}, {Jern}, {Kanto}, {Keane}, {Kibrick}, {Lewis}, {Osborne},
  {Pardeilhan}, {Pfister}, {Ricketts}, {Robinson}, {Stover}, {Tucker}, {Ward},
  \& {Wei}}]{Vogt94}
{Vogt}, S.~S., {et~al.} 1994, in Society of Photo-Optical Instrumentation
  Engineers (SPIE) Conference Series, Vol. 2198, Instrumentation in Astronomy
  VIII, ed. D.~L. {Crawford} \& E.~R. {Craine}, 362

\bibitem[{{Weiss} \& {Marcy}(2014)}]{Weiss14}
{Weiss}, L.~M., \& {Marcy}, G.~W. 2014, \apjl, 783, L6

\bibitem[{{Wilson}(1968)}]{Wilson68}
{Wilson}, O.~C. 1968, \apj, 153, 221

\bibitem[{{Winn} {et~al.}(2010){Winn}, {Johnson}, {Howard}, {Marcy}, {Bakos},
  {Hartman}, {Torres}, {Albrecht}, \& {Narita}}]{Winn10}
{Winn}, J.~N., {et~al.} 2010, \apj, 718, 575

\bibitem[{{Wolfgang} {et~al.}(2015){Wolfgang}, {Rogers}, \&
  {Ford}}]{Wolfgang15}
{Wolfgang}, A., {Rogers}, L.~A., \& {Ford}, E.~B. 2015, arXiv:1504.07557

\bibitem[{{Wright}(2005)}]{Wright05}
{Wright}, J.~T. 2005, \pasp, 117, 657

\bibitem[{{Wright} \& {Howard}(2009)}]{Wright09}
{Wright}, J.~T., \& {Howard}, A.~W. 2009, \apjs, 182, 205

\bibitem[{{Wurm} {et~al.}(2013){Wurm}, {Trieloff}, \& {Rauer}}]{Wurm13}
{Wurm}, G., {Trieloff}, M., \& {Rauer}, H. 2013, \apj, 769, 78

\bibitem[{{Zeng} \& {Sasselov}(2013)}]{Zeng13}
{Zeng}, L., \& {Sasselov}, D. 2013, \pasp, 125, 227

\end{thebibliography}



\appendix
\label{sec:appendix}

This appendix provides light curves and parameters for each planetary system.  We show \ktwo aperture photometry corrected for spacecraft systematics, de-trended photometry with planet transits identified by colored vertical ticks, and photometry phased to the orbital period of each planet.  We also provide a table of physical, orbital, and model parameters for each planetary system.  The most important stellar and planetary properties are summarized in Table~\ref{tb:highlights}.

\clearpage

\begin{figure*}
\centering
\subfigure{
\includegraphics[trim={0 0 0 7.4cm},clip, width=\textwidth]{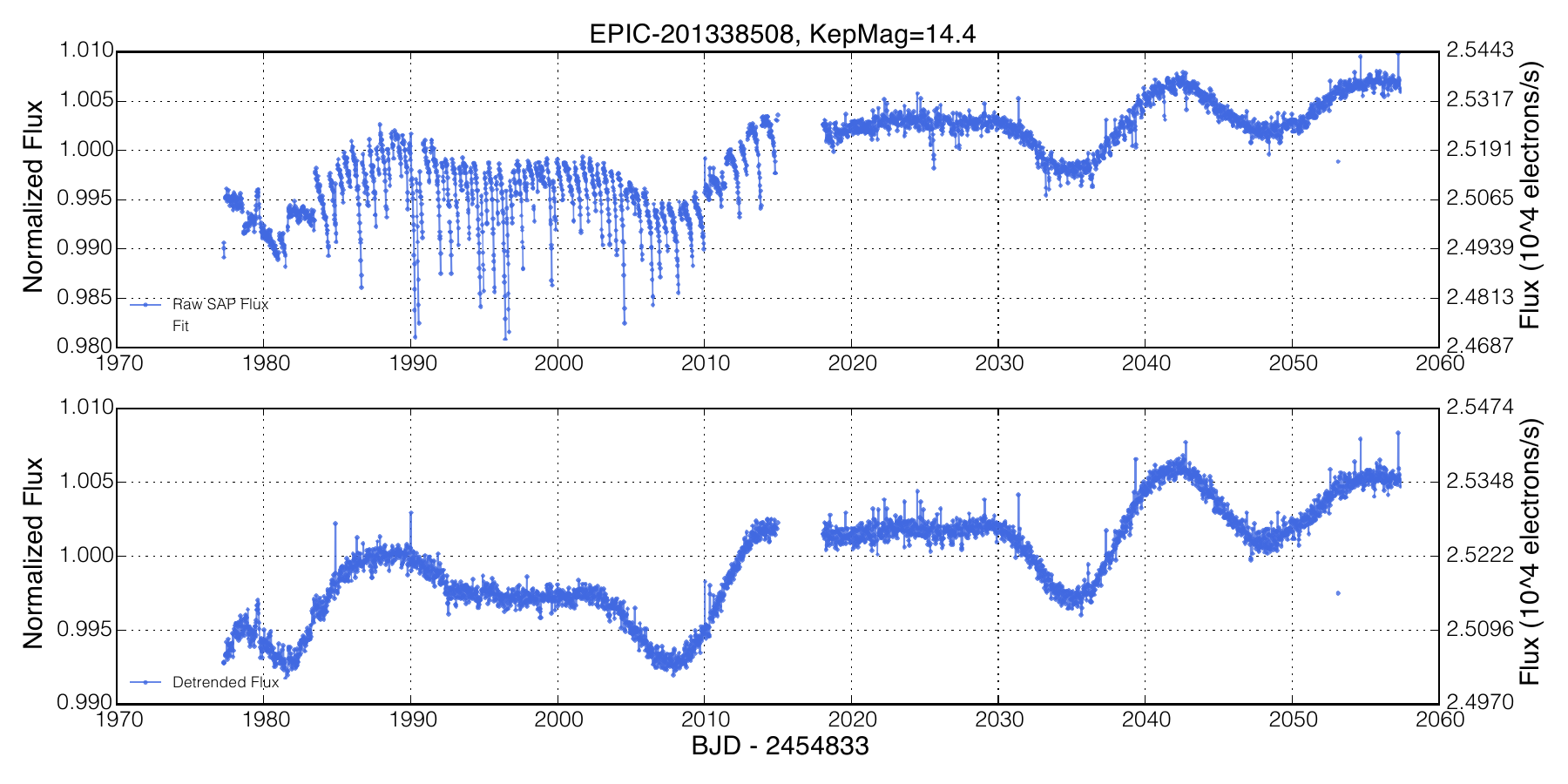}
}\\
\subfigure{
\includegraphics[width=\textwidth]{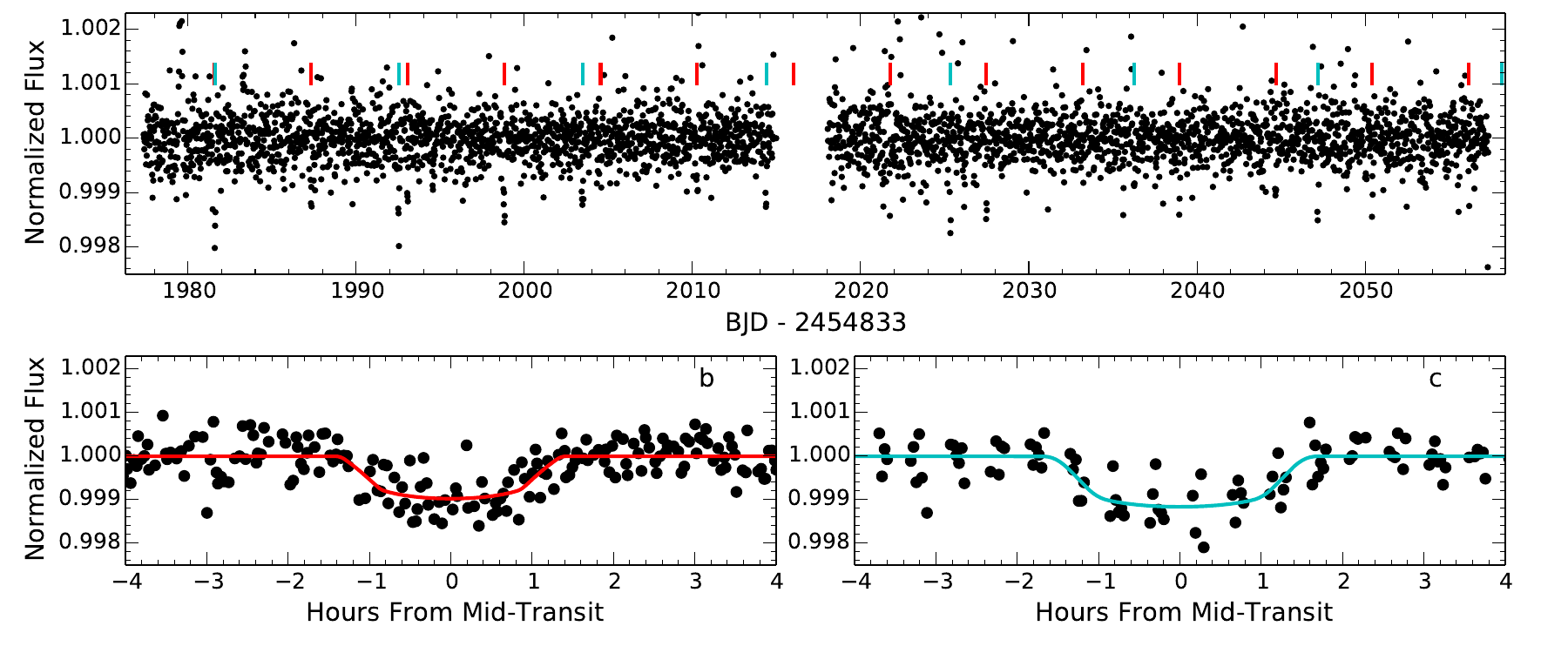}
}
\caption{\textit{Top}: \ktwo photometry for \ktwoeoh after subtracting variations caused by telescope roll. \textit{Middle}: Calibrated \ktwo photometry.  Vertical ticks indicate times of planet transits. \textit{Bottom}: Phase-folded
photometry and best fitting light curves for each planet.}  
\label{fig:eoh_LC}
\end{figure*}

\begin{deluxetable*}{lrrr}[h]
\centering
\tabletypesize{\footnotesize}
\tablecaption{Planet properties, \ktwoeoh}
\tablehead{Parameter & ~~~~~~~~Planet b & Planet c & Units}
\startdata
\sidehead{\textbf{Transit Model}}
$T_0$ & \eohbTt & \eohcTt & BJD$_{\mathrm{TDB}}$--2456000 \\
$P$ & \eohbPer & \eohcPer & d \\
$i$ & \eohbincdeg & \eohcincdeg & deg \\
$R_p/R_\star$ & \eohbRpRs & \eohcRpRs & -- \\
$R_\star/a$ & \eohbRsa & \eohcRsa & -- \\
$u$ & \eohbu & \eohcu & -- \\
$b$ & \eohbb & \eohcb & -- \\
$t_{\mathrm{14}}$ & \eohbtdur & \eohctdur & hrs \\
$R_p$ & \eohbRp & \eohcRp & $R_{\oplus}$ \\
$\rho_{\star,\mathrm{circ}}$ & \eohbrhostar & \eohcrhostar & g\,cm$^{-3}$ \\
\sidehead{\textbf{Derived Properties}}
$a$ & \eohba & \eohca & AU \\
$S_{\mathrm{inc}}$ & \eohbSinc & \eohcSinc & $S_{\oplus}$ \\
$T_{\mathrm{eq}}$ & \eohbTeq & \eohcTeq & $K$
\enddata
\tablenotetext{}{$T_0$ = mid-transit time, $i$ = orbital inclination, $a$ = orbital semi-major axis, $u$ = linear limb-darkening coefficient, $b$ = impact parameter, $t_{\mathrm{14}}$ = transit duration, $\rho_{\star,\mathrm{circ}}$ = stellar density, $S_{\mathrm{inc}}$ = incident stellar flux, $T_{\mathrm{eq}}$ = equilibrium temperature}
\label{tb:eoh}
\end{deluxetable*}

\begin{figure*}[h!]
\centering
\subfigure{
\includegraphics[trim={0 0 0 7.4cm},clip, width=\textwidth]{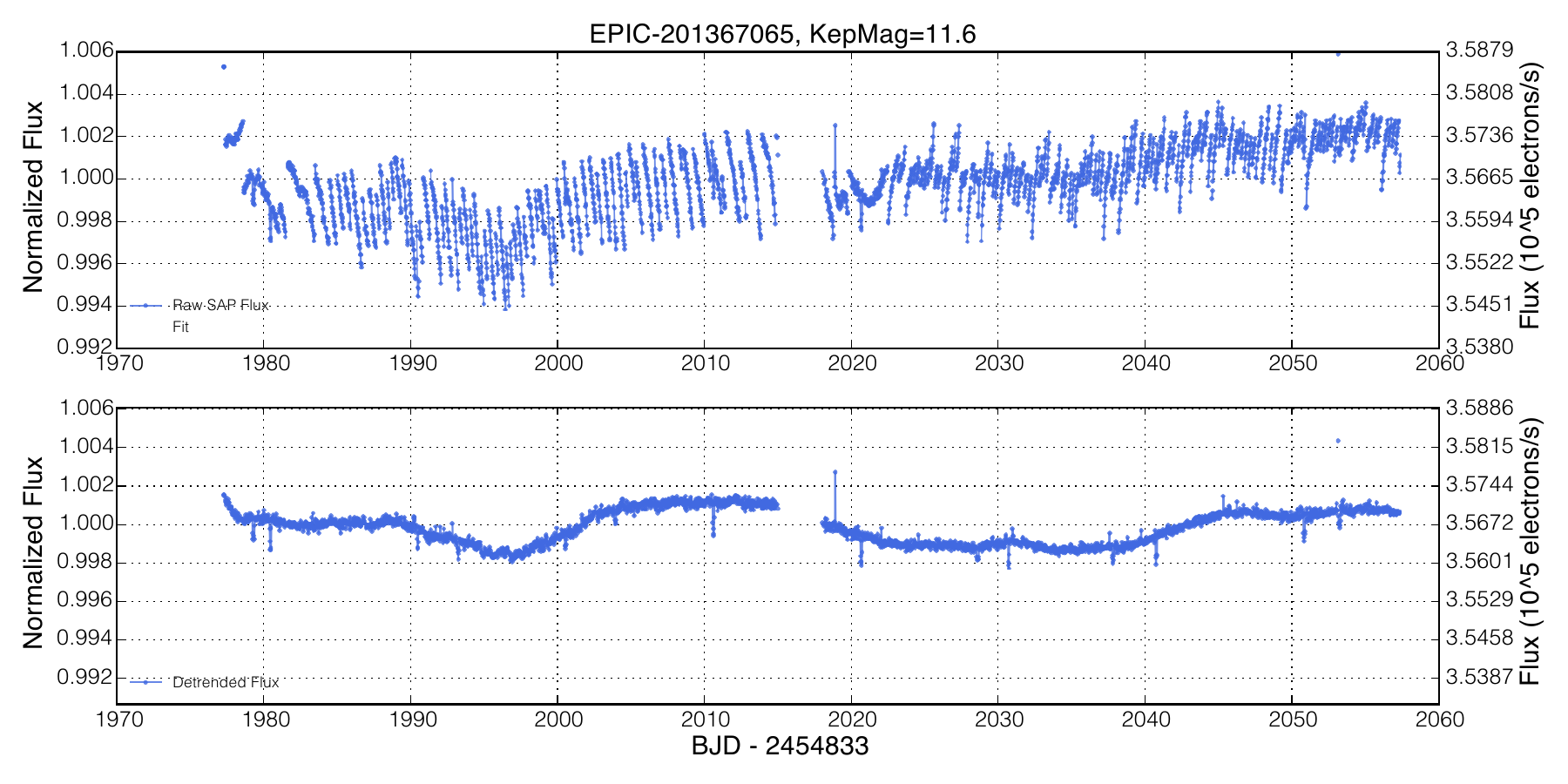}
}\\
\subfigure{
\includegraphics[width=\textwidth]{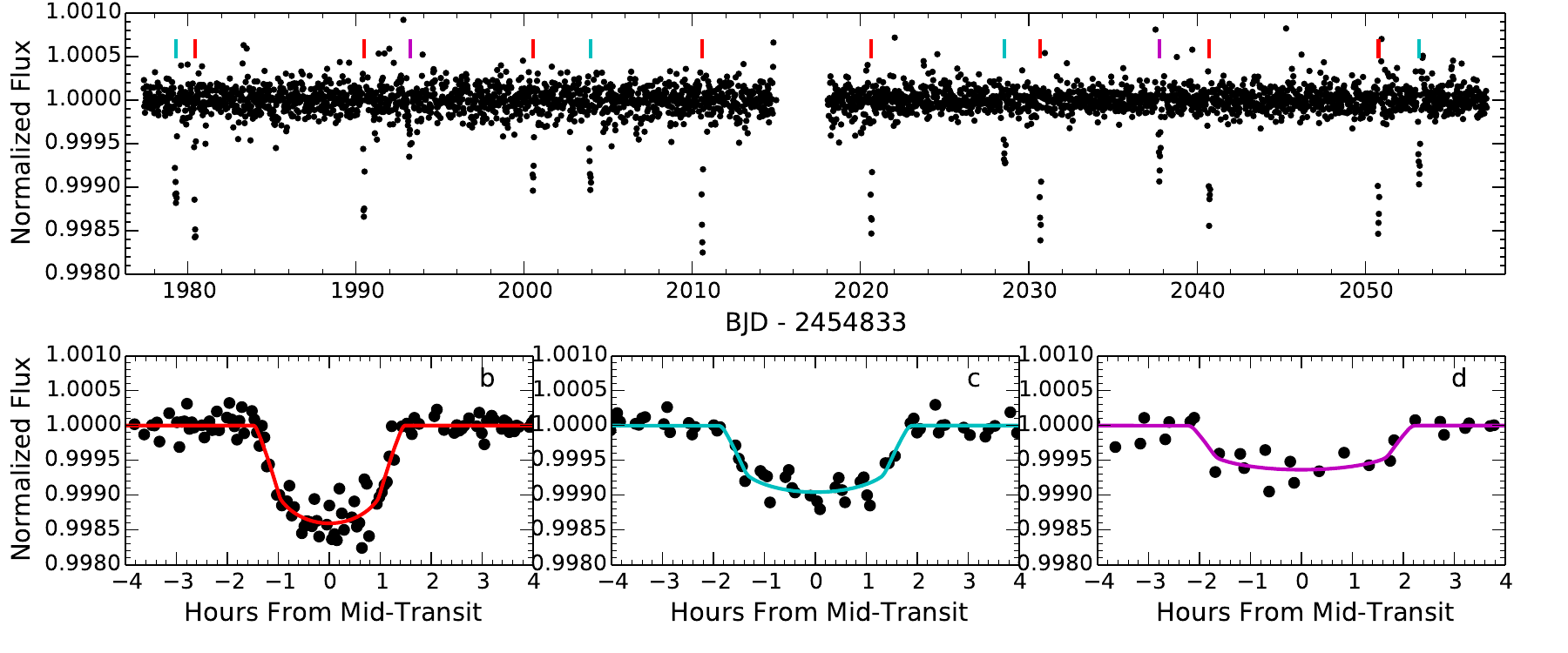}
}
\caption{\textit{Top}: \ktwo photometry for \ktwoofe after subtracting variations caused by telescope roll. \textit{Middle}: Calibrated \ktwo photometry.  Vertical ticks indicate times of planet transits. \textit{Bottom}: Phase-folded
photometry and best fitting light curves for each planet.}  
\label{fig:ofe_LC}
\end{figure*}

\begin{deluxetable*}{lrrrr}[h!]
\centering
\tabletypesize{\footnotesize}
\tablecaption{Planet properties, \ktwoofe}
\tablehead{Parameter & ~~~~~~~~Planet b & Planet c & Planet d & Units}
\startdata
\sidehead{\textbf{Transit Model}}
$T_0$ & \ofebTt & \ofecTt & \ofedTt & BJD$_{\mathrm{TDB}}$--2456000\\
$P$ & \ofebPer & \ofecPer & \ofedPer & d \\
$i$ & \ofebincdeg & \ofecincdeg & \ofedincdeg & deg \\
$R_p/R_\star$ & \ofebRpRs & \ofecRpRs & \ofedRpRs & -- \\
$R_\star/a$ & \ofebRsa & \ofecRsa & \ofedRsa & -- \\
$u$ & \ofebu & \ofecu & \ofedu & -- \\
$b$ & \ofebb & \ofecb & \ofedb & -- \\
$t_{\mathrm{14}}$ & \ofebtdur & \ofectdur & \ofedtdur & hrs \\
$R_p$ & \ofebRp & \ofecRp & \ofedRp & $R_{\oplus}$ \\
$\rho_{\star,\mathrm{circ}}$ & \ofebrhostar & \ofecrhostar & \ofedrhostar & g\,cm$^{-3}$ \\
\sidehead{\textbf{Derived Properties}}
$a$ & \ofeba & \ofeca & \ofeda & AU \\
$S_{\mathrm{inc}}$ & \ofebSinc & \ofecSinc & \ofedSinc & $S_{\oplus}$ \\
$T_{\mathrm{eq}}$ & \ofebTeq & \ofecTeq & \ofedTeq & $K$ 
\enddata
\tablenotetext{}{Same footnotes as Table \ref{tb:eoh}}
\label{tb:ofe}
\end{deluxetable*}

\begin{figure*}[h]
\centering
\subfigure{
\includegraphics[trim={0 0 0 7.4cm},clip, width=\textwidth]{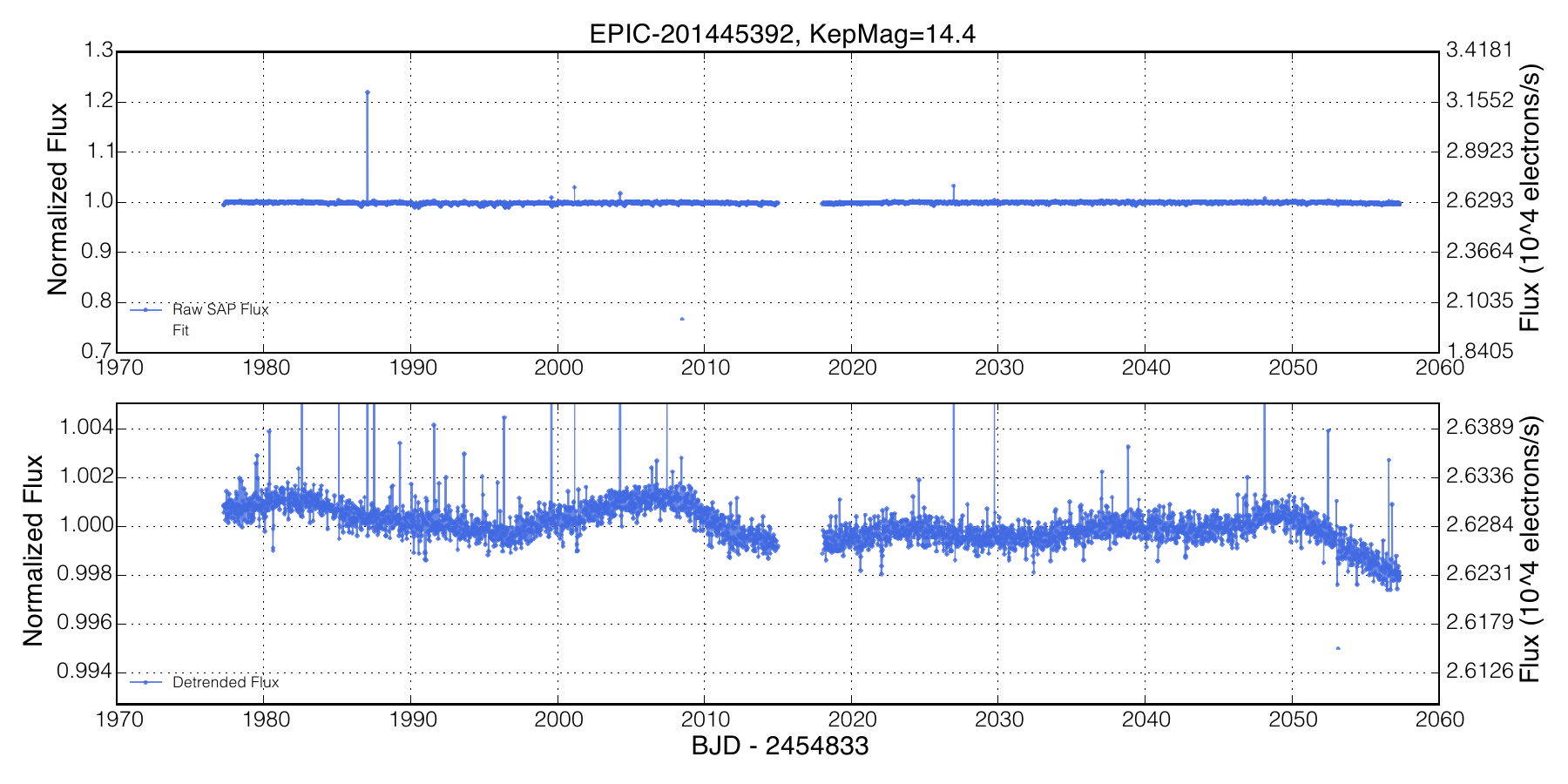}
}\\
\subfigure{
\includegraphics[width=\textwidth]{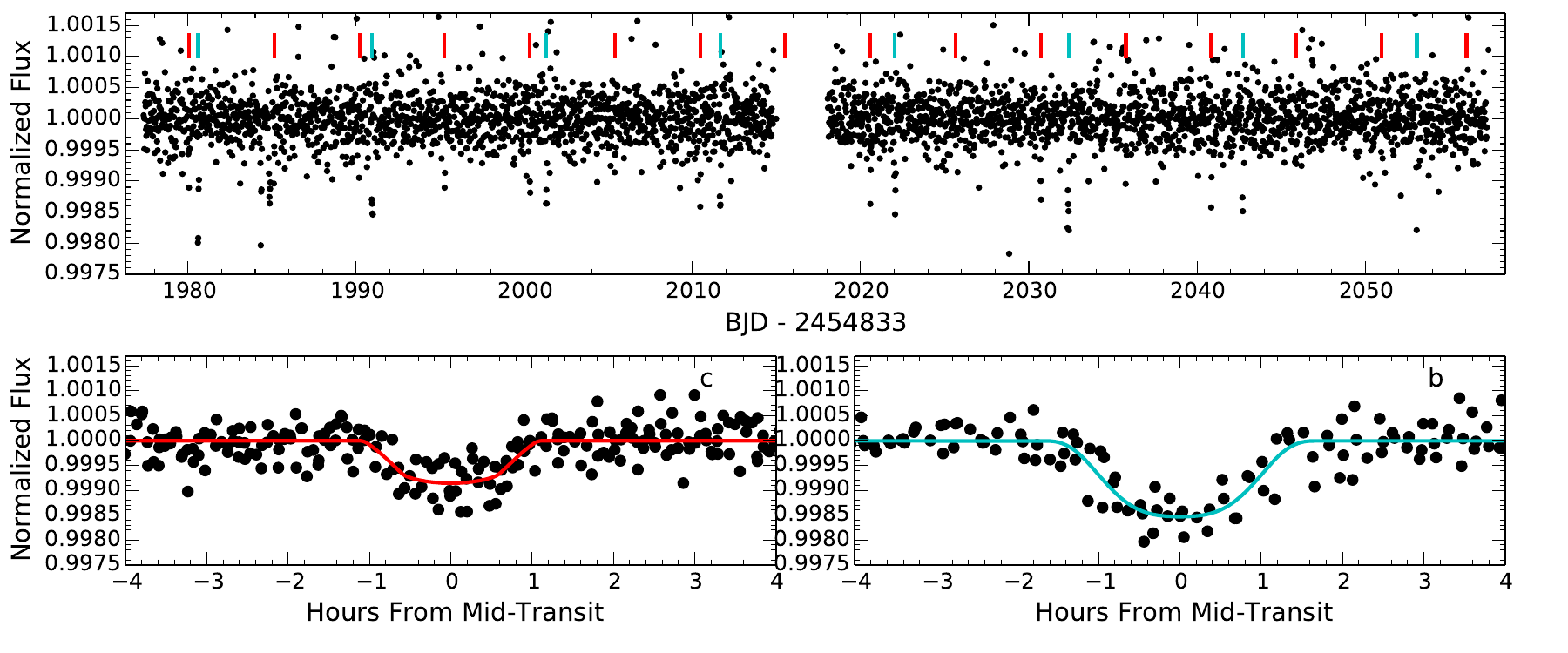}
}
\caption{\textit{Top}: \ktwo photometry for \ktwocib after subtracting variations caused by telescope roll. \textit{Middle}: Calibrated \ktwo photometry.  Vertical ticks indicate times of planet transits. \textit{Bottom}: Phase-folded
photometry and best fitting light curves for each planet.}  
\label{fig:cib_LC}
\end{figure*}

\begin{deluxetable*}{lrrr}[h]
\centering
\tabletypesize{\footnotesize}
\tablecaption{Planet properties, \ktwocib}
\tablehead{Parameter & ~~~~~~~~Planet b & Planet c & Units}
\startdata
\sidehead{\textbf{Transit Model}}
$T_0$ & \cibbTt & \cibcTt & BJD$_{\mathrm{TDB}}$--2456000\\
$P$ & \cibbPer & \cibcPer & d \\
$i$ & \cibbincdeg & \cibcincdeg & deg \\
$R_p/R_\star$ & \cibbRpRs & \cibcRpRs & -- \\
$R_\star/a$ & \cibbRsa & \cibcRsa & -- \\
$u$ & \cibbu & \cibcu & -- \\
$b$ & \cibbb & \cibcb & -- \\
$t_{\mathrm{14}}$ & \cibbtdur & \cibctdur & hrs \\
$R_p$ & \cibbRp & \cibcRp & $R_{\oplus}$ \\
$\rho_{\star,\mathrm{circ}}$ & \cibbrhostar & \cibcrhostar & g\,cm$^{-3}$ \\
\sidehead{\textbf{Derived Properties}}
$a$ & \cibba & \cibca & AU \\
$S_{\mathrm{inc}}$ & \cibbSinc & \cibcSinc & $S_{\oplus}$ \\
$T_{\mathrm{eq}}$ & \cibbTeq & \cibcTeq & $K$ 
\enddata
\tablenotetext{}{Same footnotes as Table \ref{tb:eoh}}
\label{tb:cib}
\end{deluxetable*}

\begin{figure*}[h]
\centering
\subfigure{
\includegraphics[trim={0 0 0 7.4cm},clip, width=\textwidth]{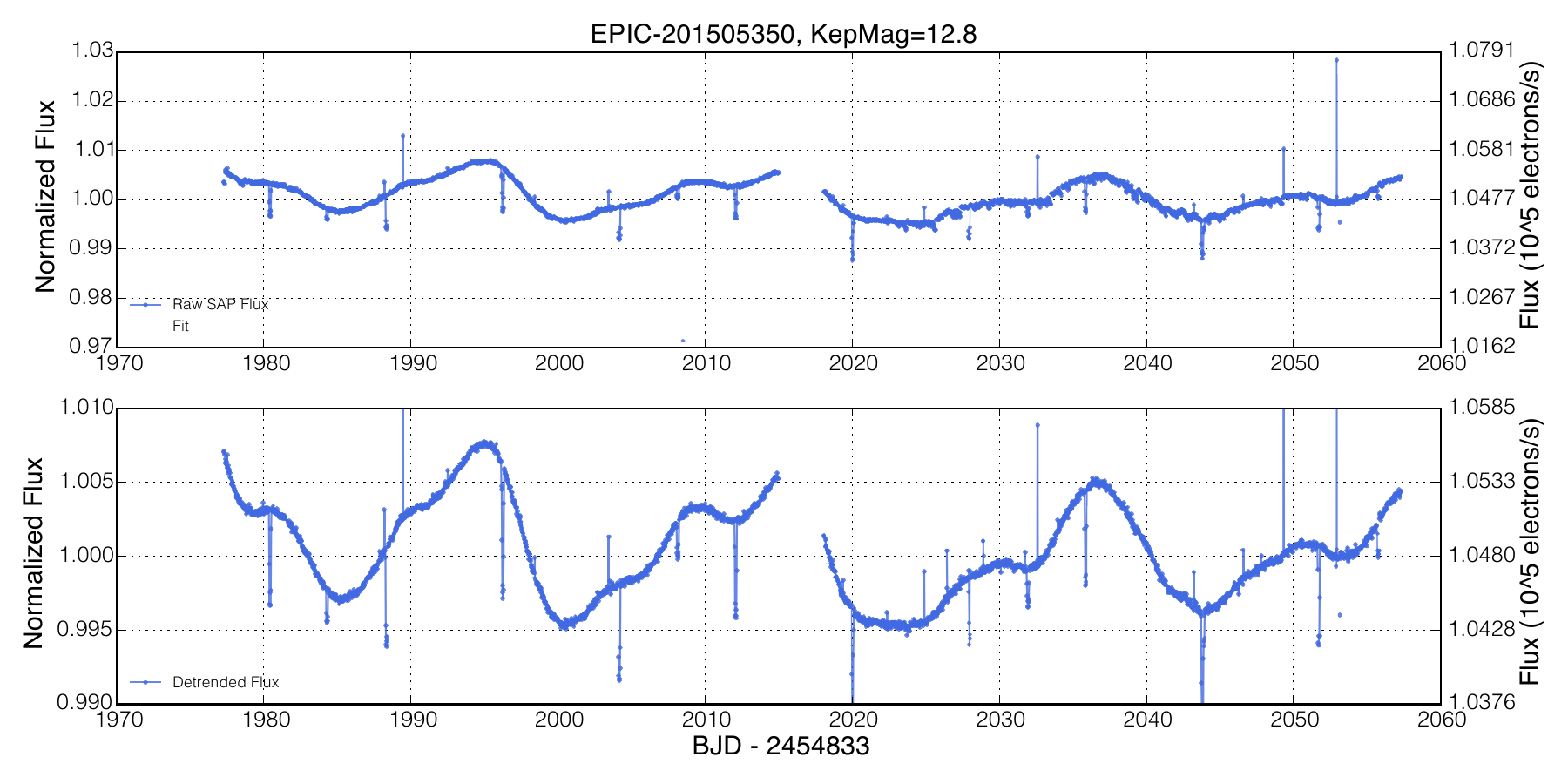}
}\\
\subfigure{
\includegraphics[width=\textwidth]{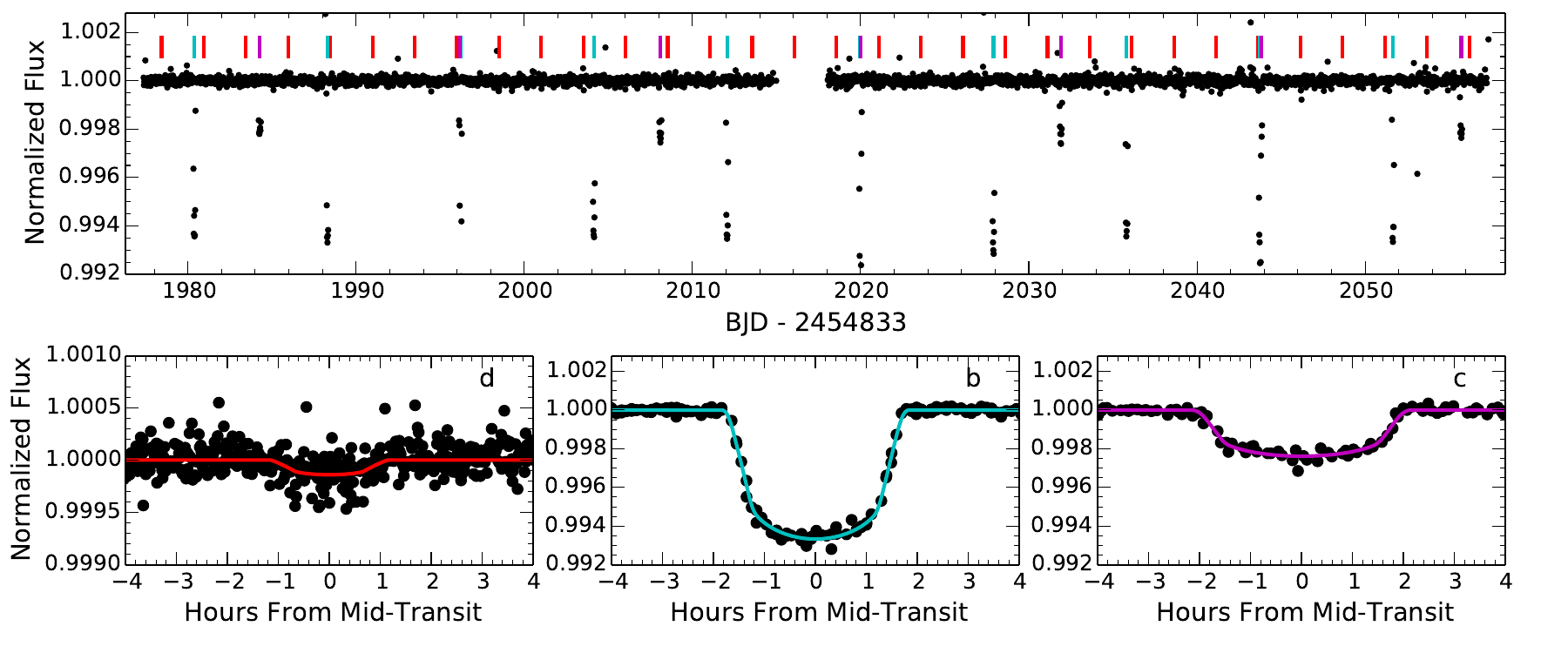}
}
\caption{\textit{Top}: \ktwo photometry for \ktwoceo after subtracting variations caused by telescope roll. \textit{Middle}: Calibrated \ktwo photometry.  Vertical ticks indicate times of planet transits. \textit{Bottom}: Phase-folded
photometry and best fitting light curves for each planet.}  
\label{fig:ceo_LC}
\end{figure*}

\begin{deluxetable*}{lrrrr}[h]
\centering
\tabletypesize{\footnotesize}
\tablecaption{Planet properties, \ktwoceo}
\tablehead{Parameter & ~~~~~~~~Planet b & Planet c & Planet d & Units}
\startdata
\sidehead{\textbf{Transit Model}}
$T_0$ & \ceobTt & \ceocTt & \ceodTt & BJD$_{\mathrm{TDB}}$--2456000\\
$P$ & \ceobPer & \ceocPer & \ceodPer & d \\
$i$ & \ceobincdeg & \ceocincdeg & \ceodincdeg & deg \\
$R_p/R_\star$ & \ceobRpRs & \ceocRpRs & \ceodRpRs & -- \\
$R_\star/a$ & \ceobRsa & \ceocRsa & \ceodRsa & -- \\
$u$ & \ceobu & \ceocu & \ceodu & -- \\
$b$ & \ceobb & \ceocb & \ceodb & -- \\
$t_{\mathrm{14}}$ & \ceobtdur & \ceoctdur & \ceodtdur & hrs \\
$R_p$ & \ceobRp & \ceocRp & \ceodRp & $R_{\oplus}$ \\
$\rho_{\star,\mathrm{circ}}$ & \ceobrhostar & \ceocrhostar & \ceodrhostar & g\,cm$^{-3}$ \\
\sidehead{\textbf{Derived Properties}}
$a$ & \ceoba & \ceoca & \ceoda & AU \\
$S_{\mathrm{inc}}$ & \ceobSinc & \ceocSinc & \ceodSinc & $S_{\oplus}$ \\
$T_{\mathrm{eq}}$ & \ceobTeq & \ceocTeq & \ceodTeq & $K$ 
\enddata
\tablenotetext{}{Same footnotes as Table \ref{tb:eoh}}
\label{tb:ceo}
\end{deluxetable*}

\begin{figure*}[h]
\centering
\subfigure{
\includegraphics[trim={0 0 0 7.4cm},clip, width=\textwidth]{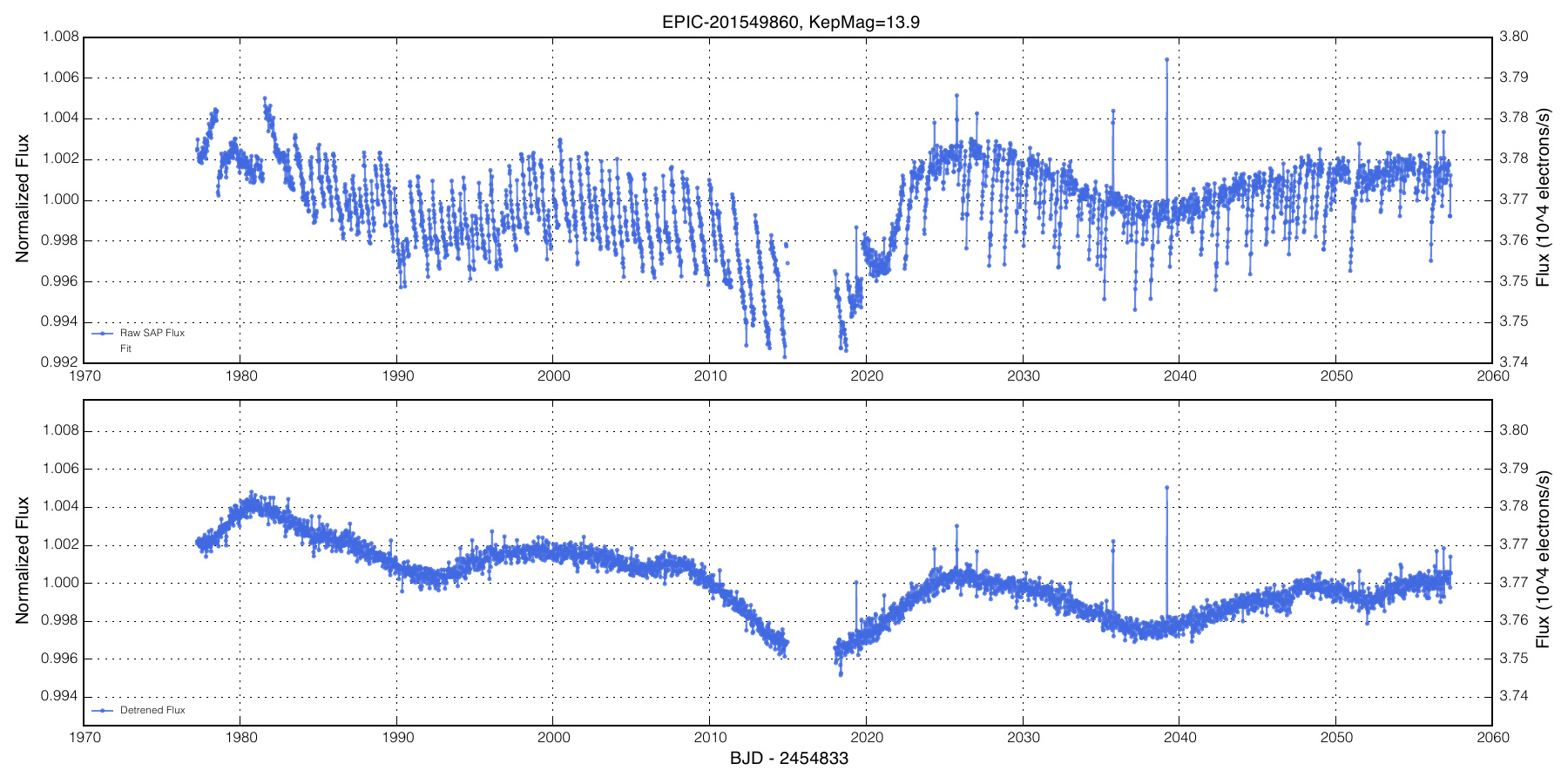}
}\\
\subfigure{
\includegraphics[width=\textwidth]{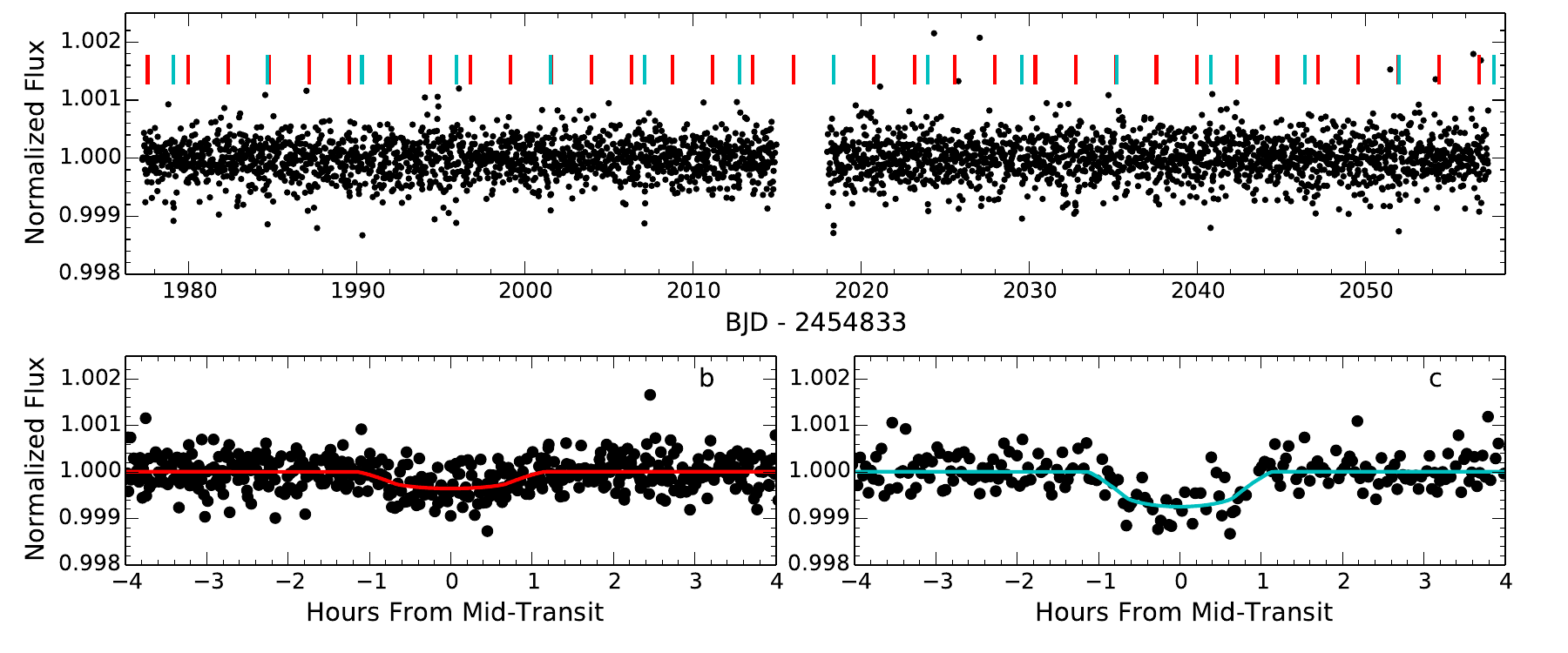}
}
\caption{\textit{Top}: \ktwo photometry for \ktwohfo after subtracting variations caused by telescope roll. \textit{Middle}: Calibrated \ktwo photometry.  Vertical ticks indicate times of planet transits. \textit{Bottom}: Phase-folded
photometry and best fitting light curves for each planet.}  
\label{fig:hfo_LC}
\end{figure*}

\begin{deluxetable*}{lrrr}[h]
\centering
\tabletypesize{\footnotesize}
\tablecaption{Planet properties, \ktwohfo}
\tablehead{Parameter & ~~~~~~~~Planet b & Planet c & Units}
\startdata
\sidehead{\textbf{Transit Model}}
$T_0$ & \hfobTt & \hfocTt & BJD$_{\mathrm{TDB}}$--2456000\\
$P$ & \hfobPer & \hfocPer & d \\
$i$ & \hfobincdeg & \hfocincdeg & deg \\
$R_p/R_\star$ & \hfobRpRs & \hfocRpRs & -- \\
$R_\star/a$ & \hfobRsa & \hfocRsa & -- \\
$u$ & \hfobu & \hfocu & -- \\
$b$ & \hfobb & \hfocb & -- \\
$t_{\mathrm{14}}$ & \hfobtdur & \hfoctdur & hrs \\
$R_p$ & \hfobRp & \hfocRp & $R_{\oplus}$ \\
$\rho_{\star,\mathrm{circ}}$ & \hfobrhostar & \hfocrhostar & g\,cm$^{-3}$ \\
\sidehead{\textbf{Derived Properties}}
$a$ & \hfoba & \hfoca & AU \\
$S_{\mathrm{inc}}$ & \hfobSinc & \hfocSinc & $S_{\oplus}$ \\
$T_{\mathrm{eq}}$ & \hfobTeq & \hfocTeq & $K$ 
\enddata
\tablenotetext{}{Same footnotes as Table \ref{tb:eoh}}
\label{tb:hfo}
\end{deluxetable*}

\begin{figure*}[h]
\centering
\subfigure{
\includegraphics[trim={0 0 0 7.4cm},clip, width=\textwidth]{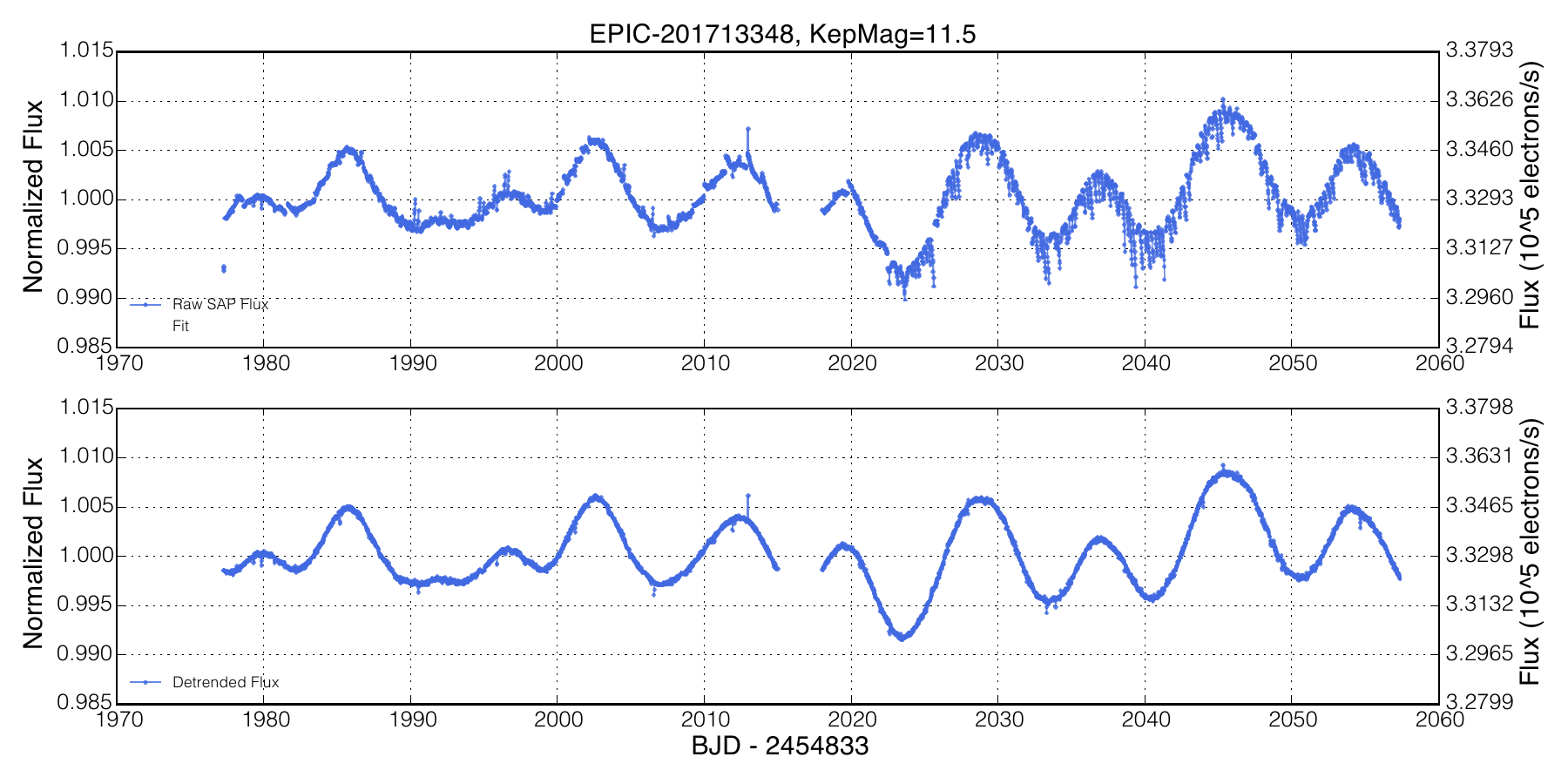}
}\\
\subfigure{
\includegraphics[width=\textwidth]{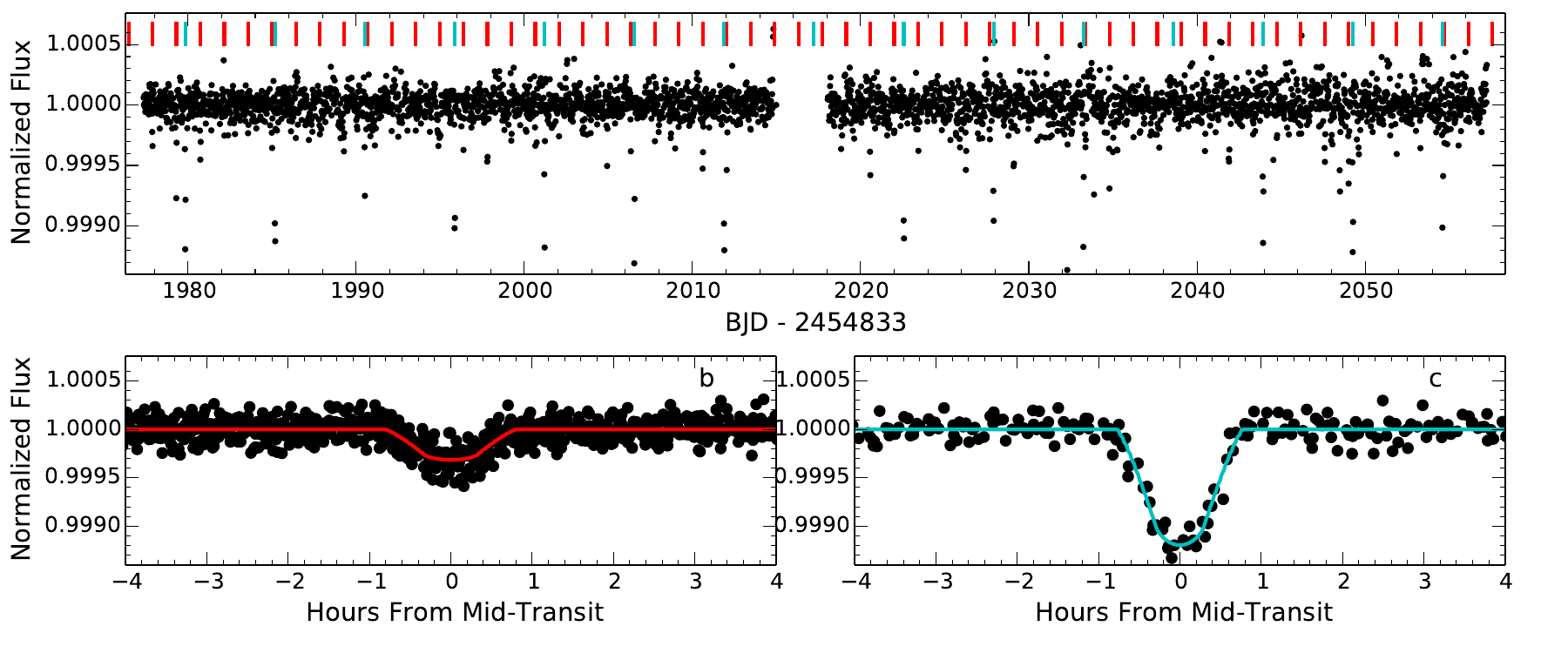}
}
\caption{\textit{Top}: \ktwo photometry for \ktwocdh after subtracting variations caused by telescope roll. \textit{Middle}: Calibrated \ktwo photometry.  Vertical ticks indicate times of planet transits. \textit{Bottom}: Phase-folded
photometry and best fitting light curves for each planet.}  
\label{fig:cdh_LC}
\end{figure*}

\begin{deluxetable*}{lrrr}[h]
\centering
\tabletypesize{\footnotesize}
\tablecaption{Planet properties, \ktwocdh}
\tablehead{Parameter & ~~~~~~~~Planet b & Planet c & Units}
\startdata
\sidehead{\textbf{Transit Model}}
$T_0$ & \cdhbTt & \cdhcTt & BJD$_{\mathrm{TDB}}$--2456000\\
$P$ & \cdhbPer & \cdhcPer & d \\
$i$ & \cdhbincdeg & \cdhcincdeg & deg \\
$R_p/R_\star$ & \cdhbRpRs & \cdhcRpRs & -- \\
$R_\star/a$ & \cdhbRsa & \cdhcRsa & -- \\
$u$ & \cdhbu & \cdhcu & -- \\
$b$ & \cdhbb & \cdhcb & -- \\
$t_{\mathrm{14}}$ & \cdhbtdur & \cdhctdur & hrs \\
$R_p$ & \cdhbRp & \cdhcRp & $R_{\oplus}$ \\
$\rho_{\star,\mathrm{circ}}$ & \cdhbrhostar & \cdhcrhostar & g\,cm$^{-3}$ \\
\sidehead{\textbf{Derived Properties}}
$a$ & \cdhba & \cdhca & AU \\
$S_{\mathrm{inc}}$ & \cdhbSinc & \cdhcSinc & $S_{\oplus}$ \\
$T_{\mathrm{eq}}$ & \cdhbTeq & \cdhcTeq & $K$ 
\enddata
\tablenotetext{}{Same footnotes as Table \ref{tb:eoh}}
\label{tb:cdh}
\end{deluxetable*}

\begin{figure*}[h]
\centering
\subfigure{
\includegraphics[trim={0 0 0 7.4cm},clip, width=\textwidth]{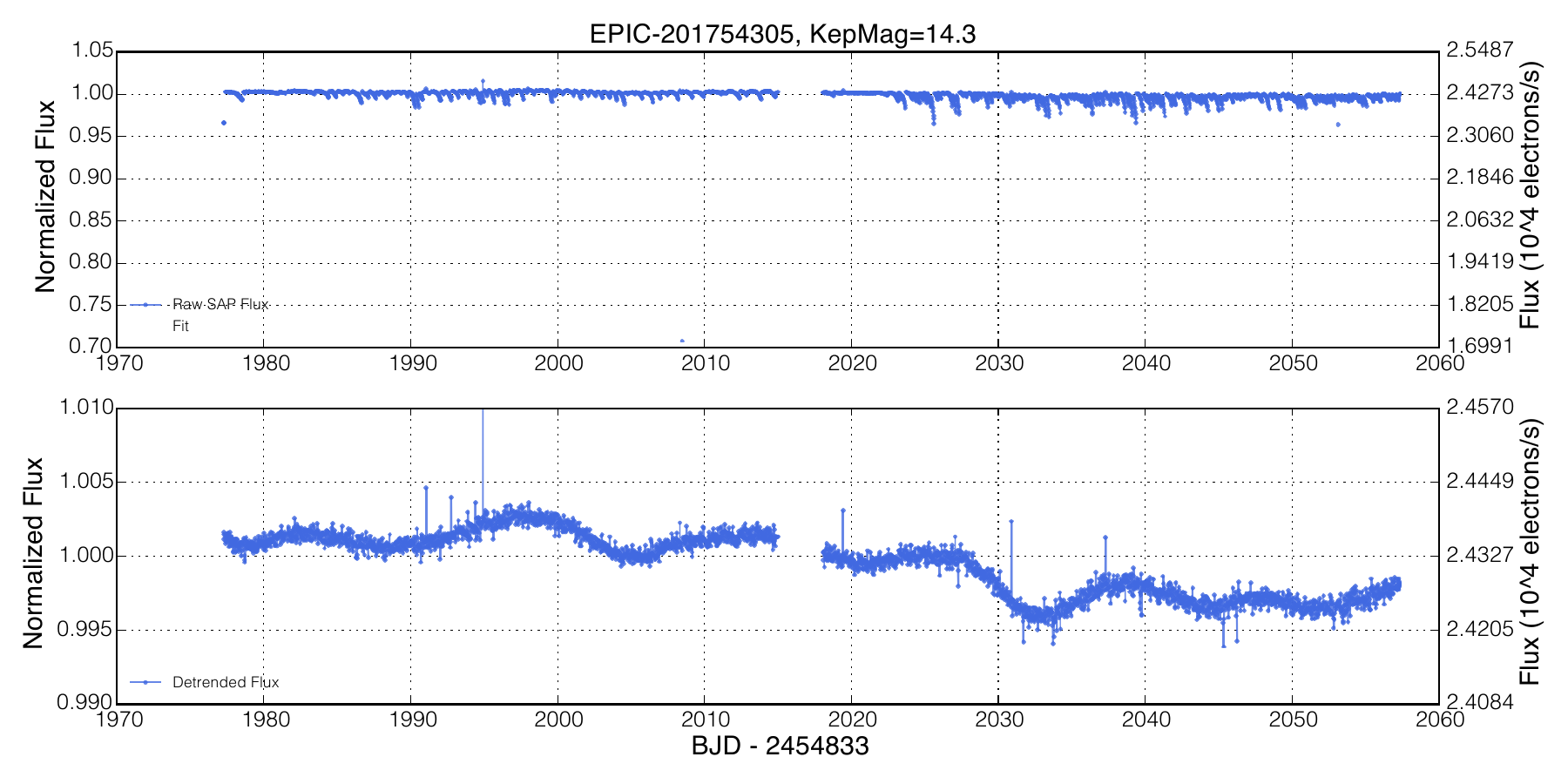}
}\\
\subfigure{
\includegraphics[width=\textwidth]{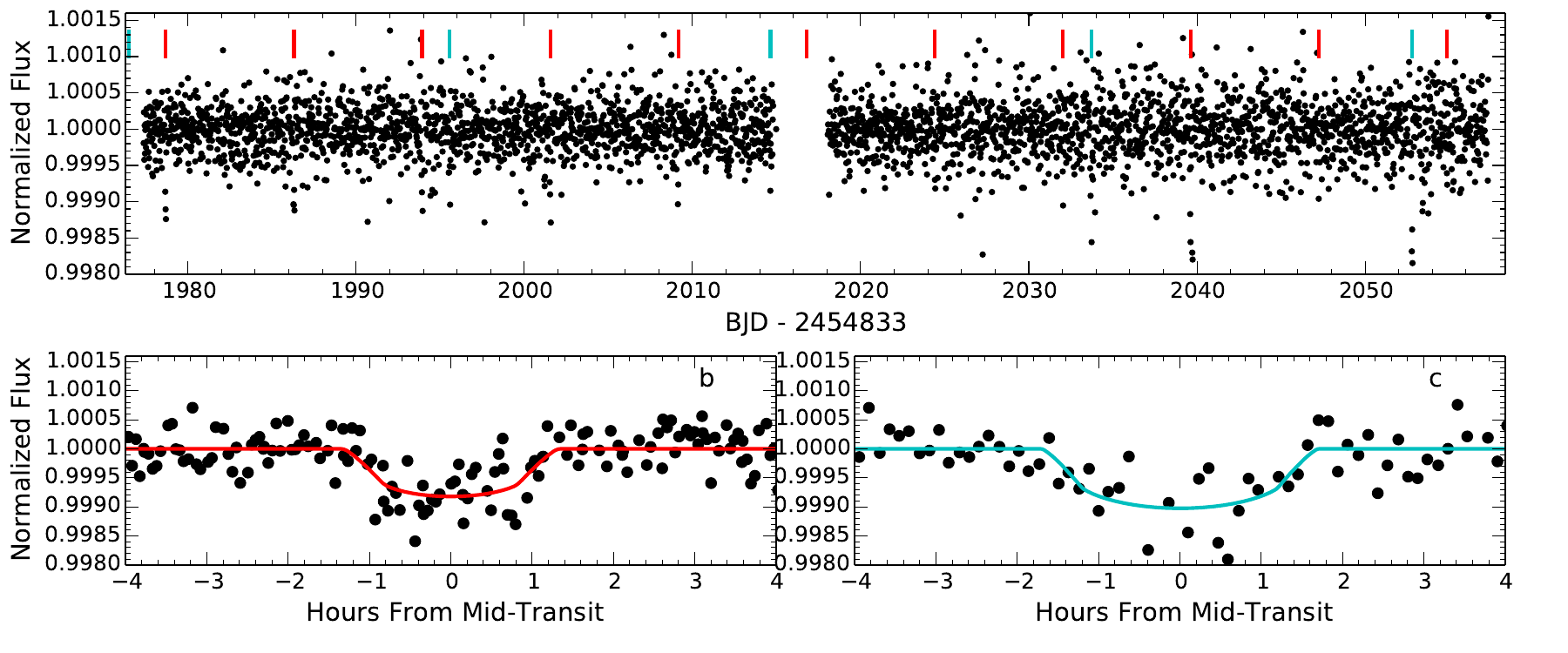}
}
\caption{\textit{Top}: \ktwo photometry for \ktwocoe after subtracting variations caused by telescope roll. \textit{Middle}: Calibrated \ktwo photometry.  Vertical ticks indicate times of planet transits. \textit{Bottom}: Phase-folded
photometry and best fitting light curves for each planet.}  
\label{fig:coe_LC}
\end{figure*}

\begin{deluxetable*}{lrrr}[h]
\centering
\tabletypesize{\footnotesize}
\tablecaption{Planet properties, \ktwocoe}
\tablehead{Parameter & ~~~~~~~~Planet b & Planet c & Units}
\startdata
\sidehead{\textbf{Transit Model}}
$T_0$ & \coebTt & \coecTt & BJD$_{\mathrm{TDB}}$--2456000\\
$P$ & \coebPer & \coecPer & d \\
$i$ & \coebincdeg & \coecincdeg & deg \\
$R_p/R_\star$ & \coebRpRs & \coecRpRs & -- \\
$R_\star/a$ & \coebRsa & \coecRsa & -- \\
$u$ & \coebu & \coecu & -- \\
$b$ & \coebb & \coecb & -- \\
$t_{\mathrm{14}}$ & \coebtdur & \coectdur & hrs \\
$R_p$ & \coebRp & \coecRp & $R_{\oplus}$ \\
$\rho_{\star,\mathrm{circ}}$ & \coebrhostar & \coecrhostar & g\,cm$^{-3}$ \\
\sidehead{\textbf{Derived Properties}}
$a$ & \coeba & \coeca & AU \\
$S_{\mathrm{inc}}$ & \coebSinc & \coecSinc & $S_{\oplus}$ \\
$T_{\mathrm{eq}}$ & \coebTeq & \coecTeq & $K$ 
\enddata
\tablenotetext{}{Same footnotes as Table \ref{tb:eoh}}
\label{tb:coe}
\end{deluxetable*}

\begin{figure*}[h]
\centering
\subfigure{
\includegraphics[trim={0 0 0 7.4cm},clip, width=\textwidth]{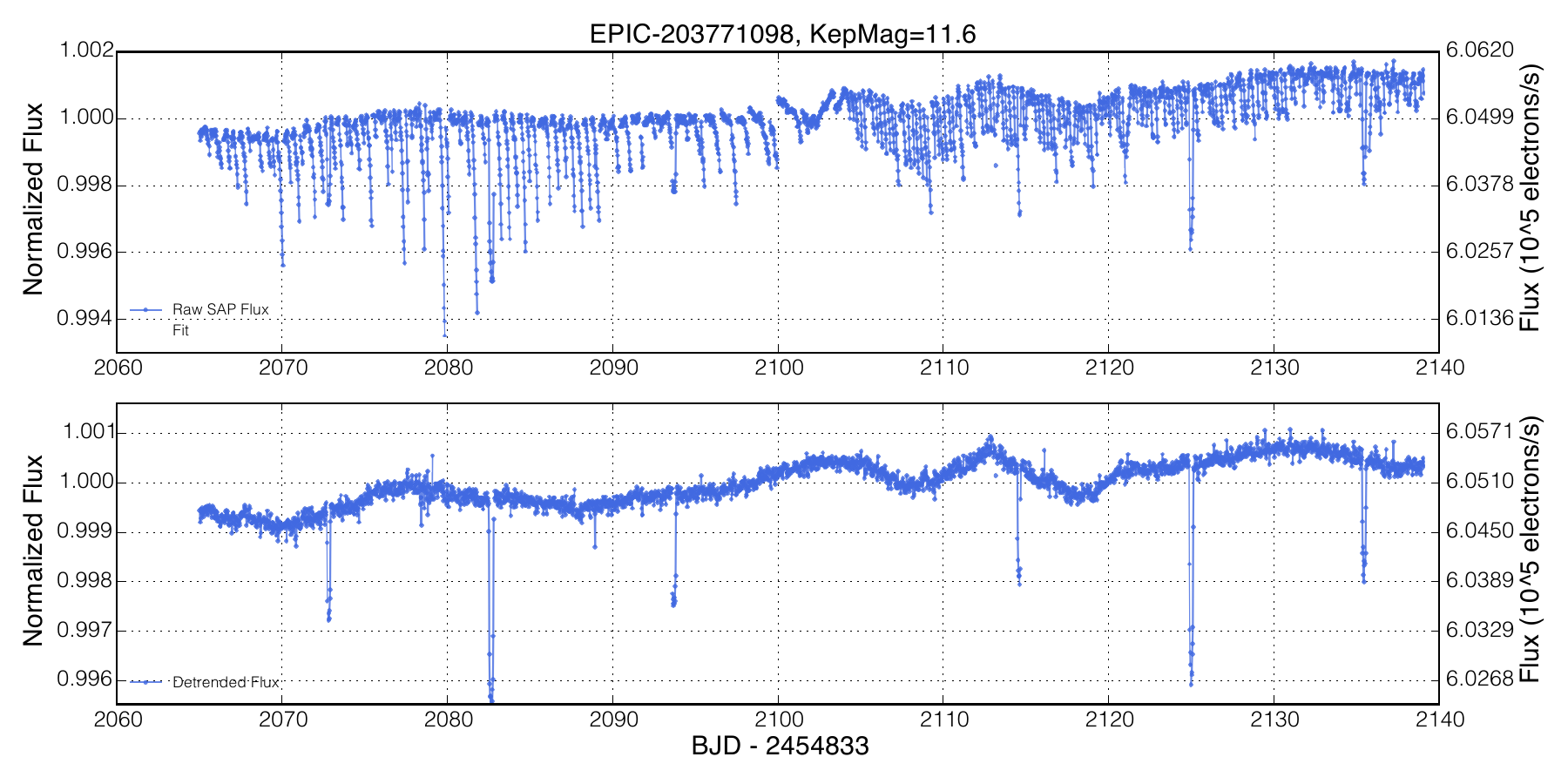}
}\\
\subfigure{
\includegraphics[width=\textwidth]{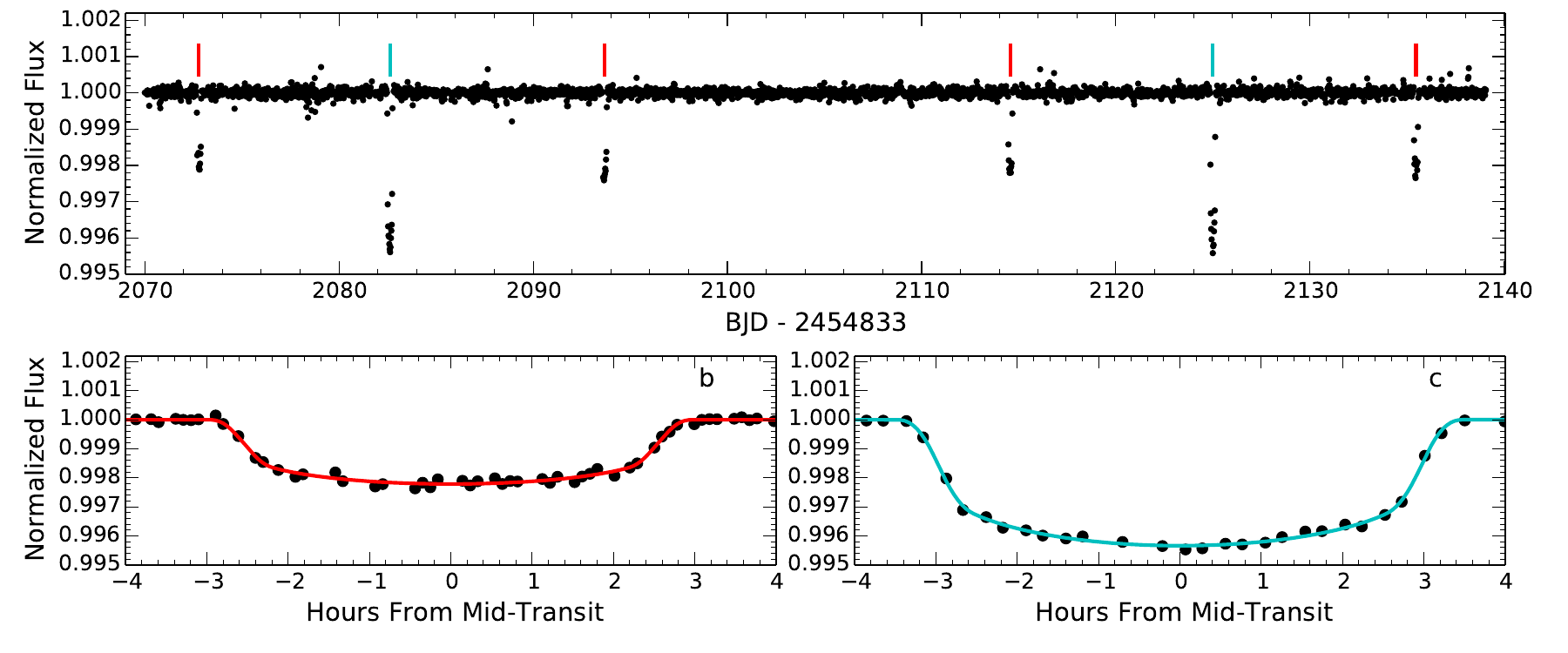}
}
\caption{\textit{Top}: \ktwo photometry for \ktwooih after subtracting variations caused by telescope roll. \textit{Middle}: Calibrated \ktwo photometry.  Vertical ticks indicate times of planet transits. \textit{Bottom}: Phase-folded
photometry and best fitting light curves for each planet.}  
\label{fig:oih_LC}
\end{figure*}

\begin{deluxetable*}{lrrr}[h]
\centering
\tabletypesize{\footnotesize}
\tablecaption{Planet properties, \ktwooih}
\tablehead{Parameter & ~~~~~~~~Planet b & Planet c & Units}
\startdata
\sidehead{\textbf{Transit Model}}
$T_0$ & \oihbTt & \oihcTt & BJD$_{\mathrm{TDB}}$--2456000\\
$P$ & \oihbPer & \oihcPer & d \\
$i$ & \oihbincdeg & \oihcincdeg & deg \\
$R_p/R_\star$ & \oihbRpRs & \oihcRpRs & -- \\
$R_\star/a$ & \oihbRsa & \oihcRsa & -- \\
$u$ & \oihbu & \oihcu & -- \\
$b$ & \oihbb & \oihcb & -- \\
$t_{\mathrm{14}}$ & \oihbtdur & \oihctdur & hrs \\
$R_p$ & \oihbRp & \oihcRp & $R_{\oplus}$ \\
$\rho_{\star,\mathrm{circ}}$ & \oihbrhostar & \oihcrhostar & g\,cm$^{-3}$ \\
\sidehead{\textbf{Derived Properties}}
$a$ & \oihba & \oihca & AU \\
$S_{\mathrm{inc}}$ & \oihbSinc & \oihcSinc & $S_{\oplus}$ \\
$T_{\mathrm{eq}}$ & \oihbTeq & \oihcTeq & $K$ 
\enddata
\tablenotetext{}{Same footnotes as Table \ref{tb:eoh}}
\label{tb:oih}
\end{deluxetable*}

\begin{figure*}[h]
\centering
\subfigure{
\includegraphics[trim={0 0 0 7.4cm},clip, width=\textwidth]{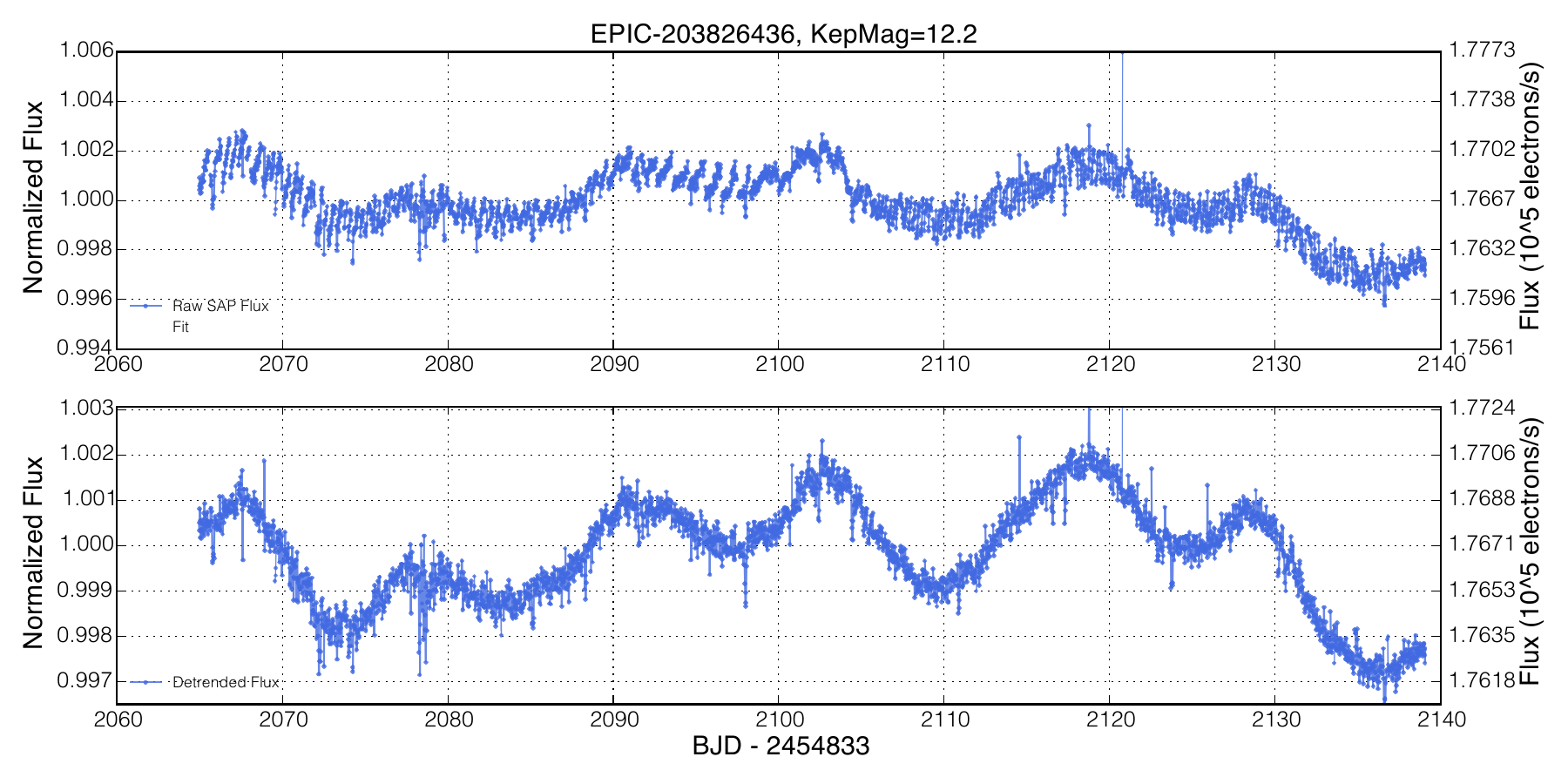}
}\\
\subfigure{
\includegraphics[width=\textwidth]{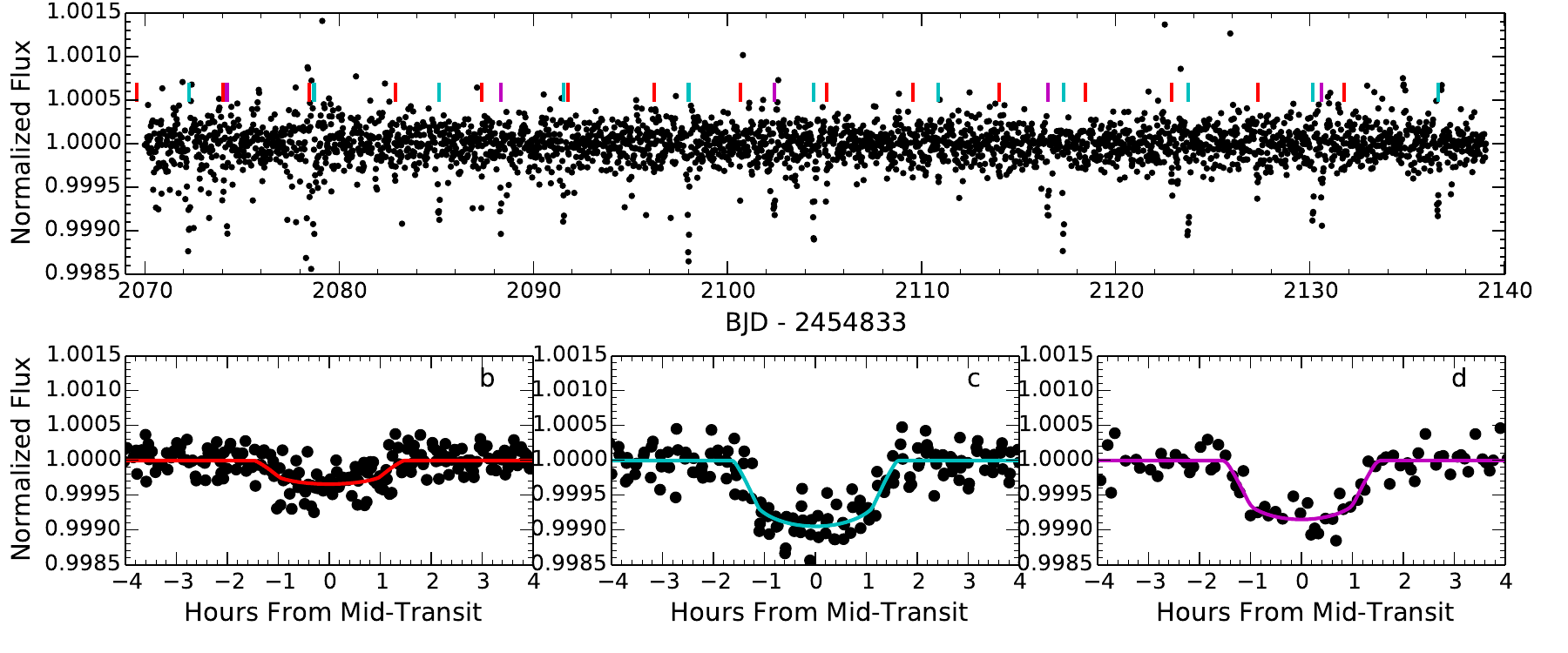}
}
\caption{\textit{Top}: \ktwo photometry for \ktwodcf after subtracting variations caused by telescope roll. \textit{Middle}: Calibrated \ktwo photometry.  Vertical ticks indicate times of planet transits. \textit{Bottom}: Phase-folded
photometry and best fitting light curves for each planet.}  
\label{fig:dcf_LC}
\end{figure*}

\begin{deluxetable*}{lrrrr}[h]
\centering
\tabletypesize{\footnotesize}
\tablecaption{Planet properties, \ktwodcf}
\tablehead{Parameter & ~~~~~~~~Planet b & Planet c & Planet d & Units}
\startdata
\sidehead{\textbf{Transit Model}}
$T_0$ & \dcfbTt & \dcfcTt & \dcfdTt & BJD$_{\mathrm{TDB}}$--2456000\\
$P$ & \dcfbPer & \dcfcPer & \dcfdPer & d \\
$i$ & \dcfbincdeg & \dcfcincdeg & \dcfdincdeg & deg \\
$R_p/R_\star$ & \dcfbRpRs & \dcfcRpRs & \dcfdRpRs & -- \\
$R_\star/a$ & \dcfbRsa & \dcfcRsa & \dcfdRsa & -- \\
$u$ & \dcfbu & \dcfcu & \dcfdu & -- \\
$b$ & \dcfbb & \dcfcb & \dcfdb & -- \\
$t_{\mathrm{14}}$ & \dcfbtdur & \dcfctdur & \dcfdtdur & hrs \\
$R_p$ & \dcfbRp & \dcfcRp & \dcfdRp & $R_{\oplus}$ \\
$\rho_{\star,\mathrm{circ}}$ & \dcfbrhostar & \dcfcrhostar & \dcfdrhostar & g\,cm$^{-3}$ \\
\sidehead{\textbf{Derived Properties}}
$a$ & \dcfba & \dcfca & \dcfda & AU \\
$S_{\mathrm{inc}}$ & \dcfbSinc & \dcfcSinc & \dcfdSinc & $S_{\oplus}$ \\
$T_{\mathrm{eq}}$ & \dcfbTeq & \dcfcTeq & \dcfdTeq & $K$ 
\enddata
\tablenotetext{}{Same footnotes as Table \ref{tb:eoh}}
\label{tb:dcf}
\end{deluxetable*}

\begin{figure*}[h]
\centering
\vspace{-0.1cm}
\subfigure{
\includegraphics[trim={0 0 0 7.4cm},clip, width=\textwidth]{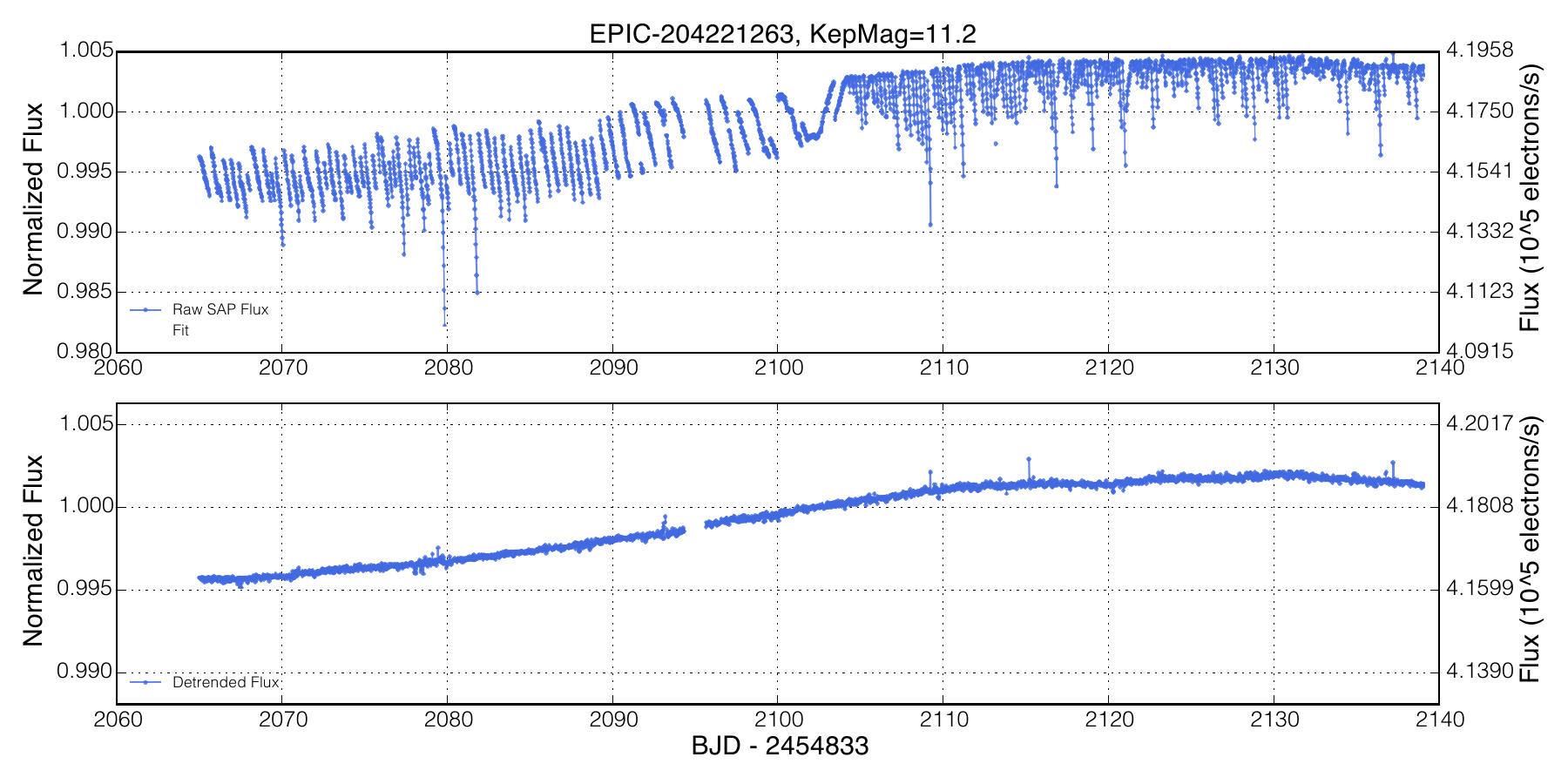}
}\\
\vspace{-0.6cm}
\subfigure{
\includegraphics[width=\textwidth]{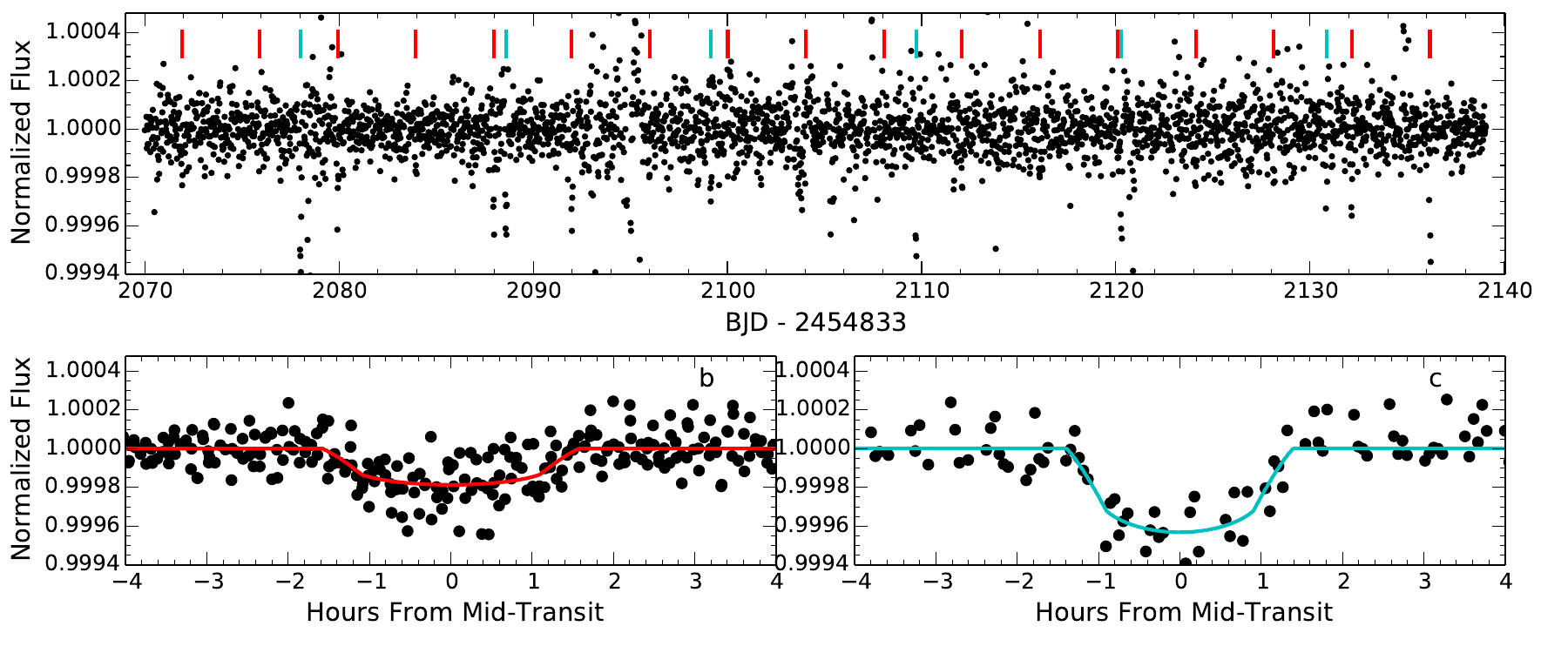}
}
\vspace{-0.5cm}
\caption{\textit{Top}: \ktwo photometry for \ktwobfc after subtracting variations caused by telescope roll. \textit{Middle}: Calibrated \ktwo photometry.  Vertical ticks indicate times of planet transits. \textit{Bottom}: Phase-folded
photometry and best fitting light curves for each planet.}  
\label{fig:bfc_LC}
\end{figure*}

\begin{deluxetable*}{lrrr}[h]
\centering
\tabletypesize{\footnotesize}
\tablecaption{Planet properties, \ktwobfc}
\tablehead{Parameter & ~~~~~~~~Planet b & Planet c & Units}
\startdata
\sidehead{\textbf{Transit Model}}
$T_0$ & \bfcbTt & \bfccTt & BJD$_{\mathrm{TDB}}$--2456000\\
$P$ & \bfcbPer & \bfccPer & d \\
$i$ & \bfcbincdeg & \bfccincdeg & deg \\
$R_p/R_\star$ & \bfcbRpRs & \bfccRpRs & -- \\
$R_\star/a$ & \bfcbRsa & \bfccRsa & -- \\
$u$ & \bfcbu & \bfccu & -- \\
$b$ & \bfcbb & \bfccb & -- \\
$t_{\mathrm{14}}$ & \bfcbtdur & \bfcctdur & hrs \\
$R_p$ & \bfcbRp & \bfccRp & $R_{\oplus}$ \\
$\rho_{\star,\mathrm{circ}}$ & \bfcbrhostar & \bfccrhostar & g\,cm$^{-3}$ \\
\sidehead{\textbf{Derived Properties}}
$a$ & \bfcba & \bfcca & AU \\
$S_{\mathrm{inc}}$ & \bfcbSinc & \bfccSinc & $S_{\oplus}$ \\
$T_{\mathrm{eq}}$ & \bfcbTeq & \bfccTeq & $K$
\enddata
\tablenotetext{}{Same footnotes as Table \ref{tb:eoh}}
\label{tb:bfc}
\end{deluxetable*}

\begin{figure*}[h]
\centering
\subfigure{
\includegraphics[trim={0 0 0 7.4cm},clip, width=\textwidth]{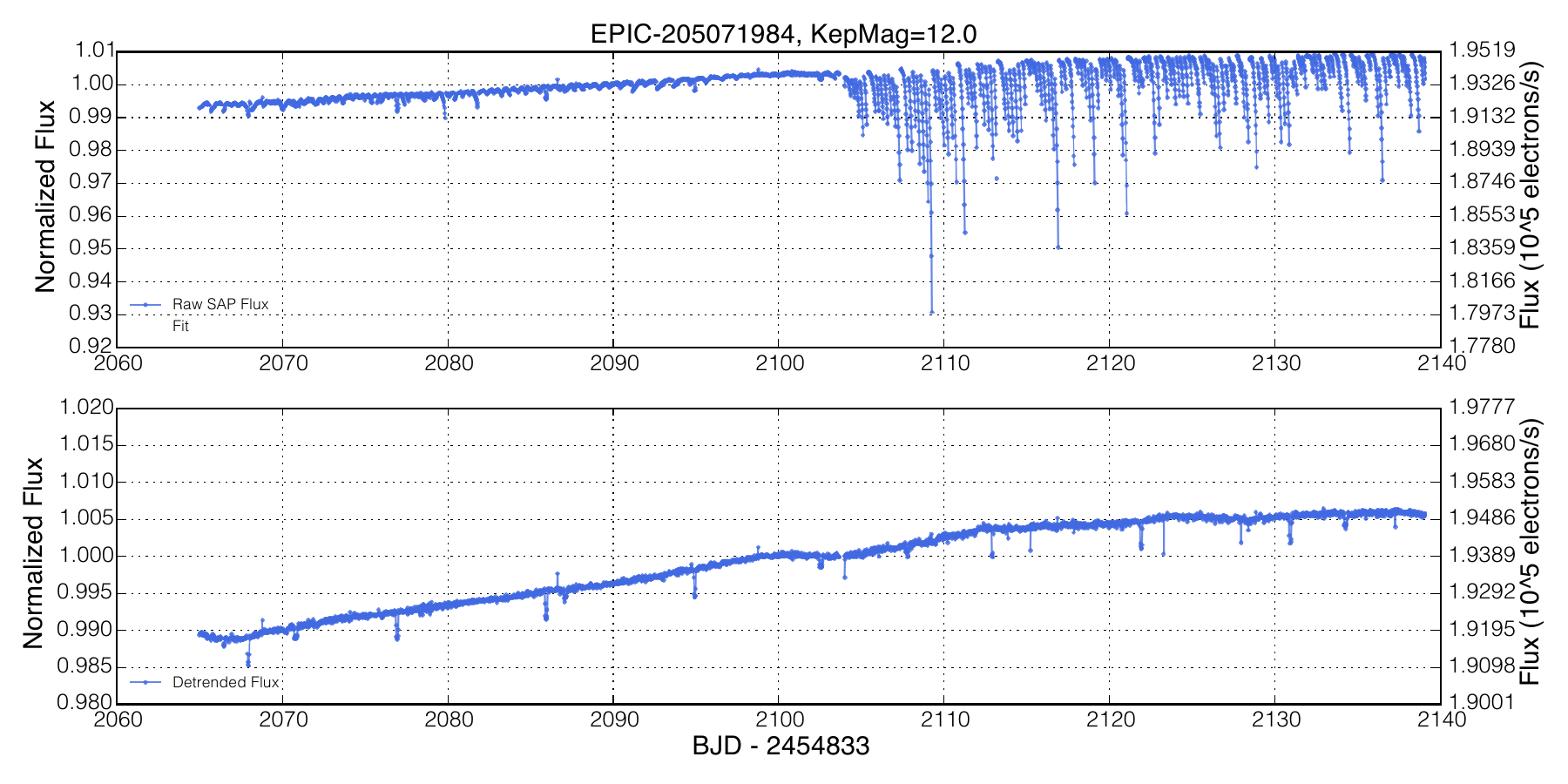}
}\\
\subfigure{
\includegraphics[width=\textwidth]{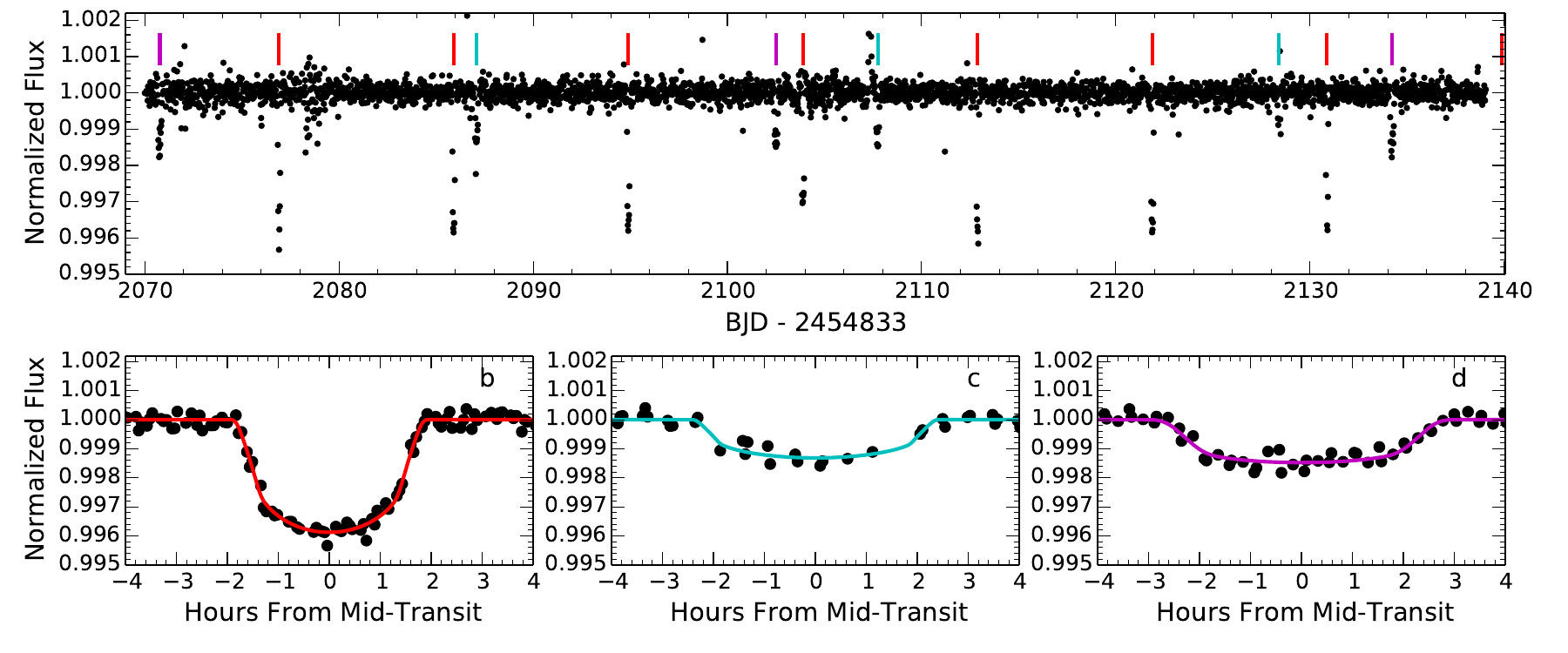}
}
\caption{\textit{Top}: \ktwo photometry for \ktwoihd after subtracting variations caused by telescope roll. \textit{Middle}: Calibrated \ktwo photometry.  Vertical ticks indicate times of planet transits. \textit{Bottom}: Phase-folded
photometry and best fitting light curves for each planet.}  
\label{fig:ihd_LC}
\end{figure*}

\begin{deluxetable*}{lrrrr}[h]
\centering
\tabletypesize{\footnotesize}
\tablecaption{Planet properties, \ktwoihd}
\tablehead{Parameter & ~~~~~~~~Planet b & Planet c & Planet d & Units}
\startdata
\sidehead{\textbf{Transit Model}}
$T_0$ & \ihdbTt & \ihdcTt & \ihddTt & BJD$_{\mathrm{TDB}}$--2456000 \\
$P$ & \ihdbPer & \ihdcPer & \ihddPer & d \\
$i$ & \ihdbincdeg & \ihdcincdeg & \ihddincdeg & deg \\
$R_p/R_\star$ & \ihdbRpRs & \ihdcRpRs & \ihddRpRs & -- \\
$R_\star/a$ & \ihdbRsa & \ihdcRsa & \ihddRsa & -- \\
$u$ & \ihdbu & \ihdcu & \ihddu & -- \\
$b$ & \ihdbb & \ihdcb & \ihddb & -- \\
$t_{\mathrm{14}}$ & \ihdbtdur & \ihdctdur & \ihddtdur & hrs \\
$R_p$ & \ihdbRp & \ihdcRp & \ihddRp & $R_{\oplus}$ \\
$\rho_{\star,\mathrm{circ}}$ & \ihdbrhostar & \ihdcrhostar & \ihddrhostar & g\,cm$^{-3}$ \\
\sidehead{\textbf{Derived Properties}}
$a$ & \ihdba & \ihdca & \ihdda & AU \\
$S_{\mathrm{inc}}$ & \ihdbSinc & \ihdcSinc & \ihddSinc & $S_{\oplus}$ \\
$T_{\mathrm{eq}}$ & \ihdbTeq & \ihdcTeq & \ihddTeq & $K$ 
\enddata
\tablenotetext{}{Same footnotes as Table \ref{tb:eoh}}
\label{tb:ihd}
\end{deluxetable*}

%
%
%
%
%
%
%
%
%

\end{document}